**С. В. Курапов**
**М. В. Давидовский**

# АЛГОРИТМИЧЕСКИЕ МЕТОДЫ КОНЕЧНЫХ ДИСКРЕТНЫХ СТРУКТУР

# ЗАДАЧА О КЛИКЕ ГРАФА

**(на правах рукописи)**

В работе представлен полиномиальный алгоритм выделения клики максимальной длины в несепарабельном графе. Алгоритм основан на свойствах представления клики как подмножества циклов длиной три, кольцевая сумма которых есть пустое множество. В результате выделения циклов длины три формируется вектор циклов, проходящих по ребрам, и вектор циклов, проходящих по вершинам. Числовые значения компонент этих векторов определяют веса вершин и ребер. Итерационный процесс построения множества векторов циклов, проходящих по ребрам, позволяет выявить основной вектор циклов, проходящих по ребрам. В свою очередь, выделение основного вектора позволяет строить клики графа. В работе указана вычислительная сложность представленного алгоритма.

 Для научных работников, преподавателей, студентов и аспирантов, специализирующихся на применении методов прикладной дискретной математики.

# Содержание

# Введение

Задача о выделении максимальной клики (в англоязычной литературе – MCP, от Maximum Clique Problem) была впервые сформулирована Ричардом Карпом в 1972 году. Тогда же был представлен переборный экспоненциальный алгоритм для решения этой задачи [2,5,7].

Основой алгебраического представления графа G(V,E;P) является линейное пространство суграфов £(G) [9]. В пространстве суграфов существует два подпространства – подпространство разрезов S(G) и подпространство циклов C(G). Таким образом, разрезы и циклы графа тесно связаны между собой. Эта связь отражается в преобразовании матрицы фундаментальных циклов в матрицу фундаментальных разрезов [1,2,5]. Поэтому, применение только одной матрицы смежностей графа отражает односторонний взгляд и ограничивает развитие прикладных методов теории графов. В качестве примера можно привести применение методов спектральной теории графов для решения задачи изоморфизма графов [11]. Таким образом, современное развитие прикладных методов теории графов требует комплексного подхода к разработке алгоритмов. Требуется создание более простых алгоритмов с одновременным использованием как матрицы смежностей, так и матрицы независимых циклов графа.

Хорошим примером для иллюстрации вычислительной сложности методов выделения максимальной клики являются графы Муна-Мозера. Данная задача имеет многочисленные приложения. В биоинформатике задача о максимальной клике используется при компьютерном анализе геномных баз данных, например, при поиске потенциальных регуляторных структур рибонуклеиновых кислот. В социальных сетях задача о максимальной клике применяется при кластеризации данных – при разделении различных сообществ на группы (кластеры), обладающие общими свойствами. Выделение кластеров позволяет обрабатывать каждый из них отдельным вспомогательным сервером. В химии задача о максимальной клике лежит в основе поиска «максимальной общей подструктуры» в графе, описывающем структуру химического соединения. Кроме того, задача о максимальной клике является математической моделью ряда прикладных задач, возникающих в процессе разработки систем автоматизированного проектирования.

# Глава 1. Простые графы и графы Муна-Мозера

## 1.1. Порождение клик графа Муна-Мозера

Представление о простоте переборных алгоритмов для решения задач справедливо только для небольших графов (с числом атрибутов до 20). Однако с увеличением количества атрибутов этот метод поиска становится громоздким с вычислительной точки зрения. Такой вывод можно получить, если исследовать зависимость количества клик от количества вершин на графах Муна–Мозера [13,14]. Количество вершин в клике будем называть *длиной клики*.

Число клик в графе может расти экспоненциально относительно числа вершин. Рассмотрим граф $M_n$ Муна–Мозера с $3n$ вершинами $\{1, 2, …, 3n\}$, в котором вершины разбиты на триады $\{1,2,3\}$, $\{4,5,6\}$, …, $\{3n – 2, 3n – 1, 3n\}$; $M_n$ не имеет ребер внутри любой триады, но вне их каждая вершина связана с каждой из остальных. Графы $M_1$, $M_2$, $M_3$ показаны на рис. 1.1.

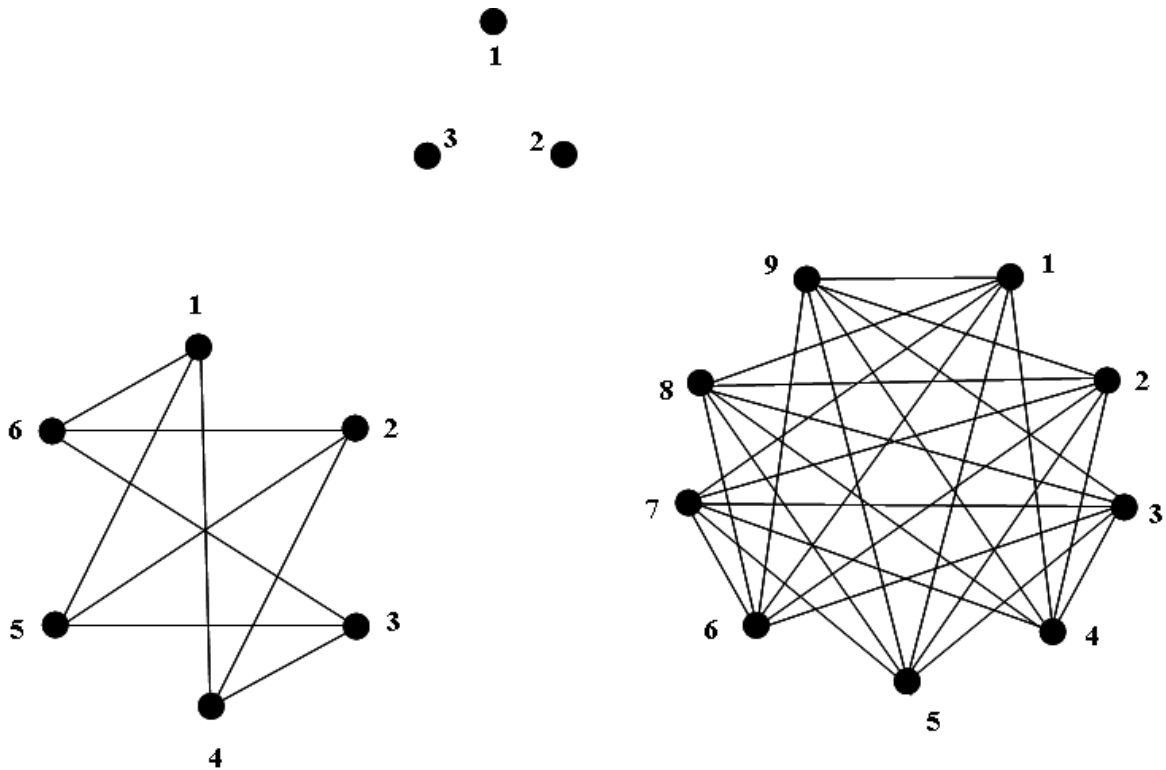

Рис.1.1. Первые три графа Муна-Мозера.

Легко доказать, что $M_n$ имеет $3^n$ клик, каждая из которых содержит $n$ вершин. Это верно для $M_1$, в котором кликами являются сами вершины. Если $M_{n-1}$ имеет $3^{(n-1)}$ клик, каждый из которых состоит из $(n – 1)$ вершин, то каждая из трех вершин, добавленных для построения $M_n$, формирует клику с каждой из $3^{n-1}$ клик $M_{n–1}$. Поскольку только они являются новыми кликами, $M_n$ имеет $3 \cdot 3^{n-1}$

= $3^n$ клик, каждая из которых состоит из *n* вершин. Таким образом, число клик в $M_n$ растет экспоненциально относительно числа вершин.

Рассмотрим граф Муна-Мозера на 6 вершин. Имеем 9 ребер (длина клики равна 2):

| | | |
|---|---|---|
| $v_1\ v_4$ | $v_2\ v_4$ | $v_3\ v_4$ |
| $v_1\ v_5$ | $v_2\ v_5$ | $v_3\ v_5$ |
| $v_1\ v_6$ | $v_2\ v_6$ | $v_3\ v_6$ |

Рассмотрим граф Муна-Мозера на 9 (3×3) вершин. Количество клик формируем путем добавления к каждой предыдущей клике поочередно трех новых вершин $v_7, v_8, v_9$. В результате имеем 27 ($3^3$) треугольных клик.

| | | |
|---|---|---|
| $v_1\ v_4\ v_7$ | $v_2\ v_4\ v_7$ | $v_3\ v_4\ v_7$ |
| $v_1\ v_4\ v_8$ | $v_2\ v_4\ v_8$ | $v_3\ v_4\ v_8$ |
| $v_1\ v_4\ v_9$ | $v_2\ v_4\ v_9$ | $v_3\ v_4\ v_9$ |
| $v_1\ v_5\ v_7$ | $v_2\ v_5\ v_7$ | $v_3\ v_5\ v_7$ |
| $v_1\ v_5\ v_8$ | $v_2\ v_5\ v_8$ | $v_3\ v_5\ v_8$ |
| $v_1\ v_5\ v_9$ | $v_2\ v_5\ v_9$ | $v_3\ v_5\ v_9$ |
| $v_1\ v_6\ v_7$ | $v_2\ v_6\ v_7$ | $v_3\ v_6\ v_7$ |
| $v_1\ v_6\ v_8$ | $v_2\ v_6\ v_8$ | $v_3\ v_6\ v_8$ |
| $v_1\ v_6\ v_9$ | $v_2\ v_6\ v_9$ | $v_3\ v_6\ v_9$ |

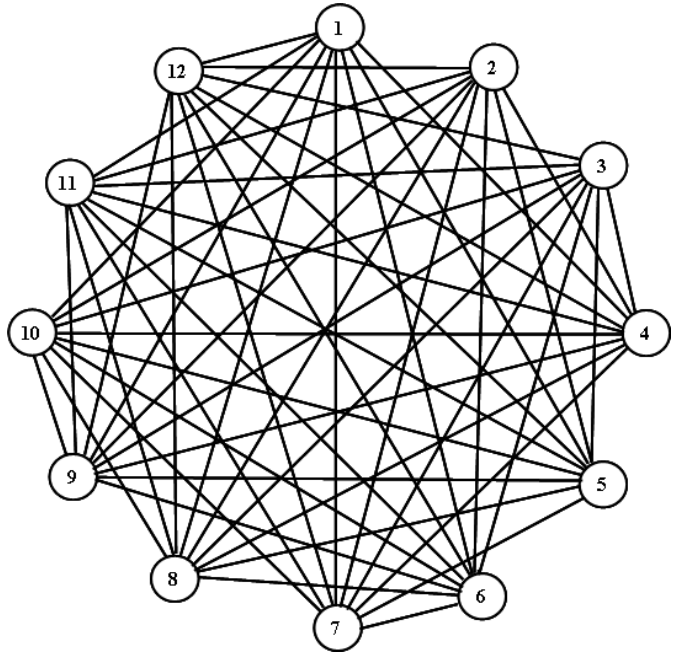

Рис. 1.2. Граф Муна-Мозера на 12 вершин.

Рассмотрим граф Муна-Мозера на 12 (3×4) вершин (рис. 1.2). Количество клик формируем путем добавления к каждой предыдущей клике поочерёдно трех новых вершин $v_{10}, v_{11}, v_{12}$. В результате имеем 81 ($3^4$) четырехвалентных клик.

| | | |
|---|---|---|
| $v_1\ v_4\ v_7\ v_{10}$ | $v_2\ v_4\ v_7\ v_{10}$ | $v_3\ v_4\ v_7\ v_{10}$ |
| $v_1\ v_4\ v_7\ v_{11}$ | $v_2\ v_4\ v_7\ v_{11}$ | $v_3\ v_4\ v_7\ v_{11}$ |
| $v_1\ v_4\ v_7\ v_{12}$ | $v_2\ v_4\ v_7\ v_{12}$ | $v_3\ v_4\ v_7\ v_{12}$ |
| $v_1\ v_4\ v_8\ v_{10}$ | $v_2\ v_4\ v_8\ v_{10}$ | $v_3\ v_4\ v_8\ v_{10}$ |
| $v_1\ v_4\ v_8\ v_{11}$ | $v_2\ v_4\ v_8\ v_{11}$ | $v_3\ v_4\ v_8\ v_{11}$ |
| $v_1\ v_4\ v_8\ v_{12}$ | $v_2\ v_4\ v_8\ v_{12}$ | $v_3\ v_4\ v_8\ v_{12}$ |
| $v_1\ v_4\ v_9\ v_{10}$ | $v_2\ v_4\ v_9\ v_{10}$ | $v_3\ v_4\ v_9\ v_{10}$ |
| $v_1\ v_4\ v_9\ v_{11}$ | $v_2\ v_4\ v_9\ v_{11}$ | $v_3\ v_4\ v_9\ v_{11}$ |
| $v_1\ v_4\ v_9\ v_{12}$ | $v_2\ v_4\ v_9\ v_{12}$ | $v_3\ v_4\ v_9\ v_{12}$ |
| $v_1\ v_5\ v_7\ v_{10}$ | $v_2\ v_5\ v_7\ v_{10}$ | $v_3\ v_5\ v_7\ v_{10}$ |
| $v_1\ v_5\ v_7\ v_{11}$ | $v_2\ v_5\ v_7\ v_{11}$ | $v_3\ v_5\ v_7\ v_{11}$ |
| $v_1\ v_5\ v_7\ v_{12}$ | $v_2\ v_5\ v_7\ v_{12}$ | $v_3\ v_5\ v_7\ v_{12}$ |
| $v_1\ v_5\ v_8\ v_{10}$ | $v_2\ v_5\ v_8\ v_{10}$ | $v_3\ v_5\ v_8\ v_{10}$ |
| $v_1\ v_5\ v_8\ v_{11}$ | $v_2\ v_5\ v_8\ v_{11}$ | $v_3\ v_5\ v_8\ v_{11}$ |
| $v_1\ v_5\ v_8\ v_{12}$ | $v_2\ v_5\ v_8\ v_{12}$ | $v_3\ v_5\ v_8\ v_{12}$ |
| $v_1\ v_5\ v_9\ v_{10}$ | $v_2\ v_5\ v_9\ v_{10}$ | $v_3\ v_5\ v_9\ v_{10}$ |
| $v_1\ v_5\ v_9\ v_{11}$ | $v_2\ v_5\ v_9\ v_{11}$ | $v_3\ v_5\ v_9\ v_{11}$ |
| $v_1\ v_5\ v_9\ v_{12}$ | $v_2\ v_5\ v_9\ v_{12}$ | $v_3\ v_5\ v_9\ v_{12}$ |
| $v_1\ v_6\ v_7\ v_{10}$ | $v_2\ v_6\ v_7\ v_{10}$ | $v_3\ v_6\ v_7\ v_{10}$ |
| $v_1\ v_6\ v_7\ v_{11}$ | $v_2\ v_6\ v_7\ v_{11}$ | $v_3\ v_6\ v_7\ v_{11}$ |
| $v_1\ v_6\ v_7\ v_{12}$ | $v_2\ v_6\ v_7\ v_{12}$ | $v_3\ v_6\ v_7\ v_{12}$ |
| $v_1\ v_6\ v_8\ v_{10}$ | $v_2\ v_6\ v_8\ v_{10}$ | $v_3\ v_6\ v_8\ v_{10}$ |
| $v_1\ v_6\ v_8\ v_{11}$ | $v_2\ v_6\ v_8\ v_{11}$ | $v_3\ v_6\ v_8\ v_{11}$ |
| $v_1\ v_6\ v_8\ v_{13}$ | $v_2\ v_6\ v_8\ v_{13}$ | $v_3\ v_6\ v_8\ v_{13}$ |
| $v_1\ v_6\ v_9\ v_{10}$ | $v_2\ v_6\ v_9\ v_{10}$ | $v_3\ v_6\ v_9\ v_{10}$ |
| $v_1\ v_6\ v_9\ v_{11}$ | $v_2\ v_6\ v_9\ v_{11}$ | $v_3\ v_6\ v_9\ v_{11}$ |
| $v_1\ v_6\ v_9\ v_{12}$ | $v_2\ v_6\ v_9\ v_{12}$ | $v_3\ v_6\ v_9\ v_{12}$ |

Дальнейшая логика построения клик для графа Муна-Мозера очевидна.

Таким образом, задача перечисления всех клик в графе является экспоненциальной по сложности вычисления (так как существуют графы с количеством клик $3^n$).

### 1.2. Метод Магу

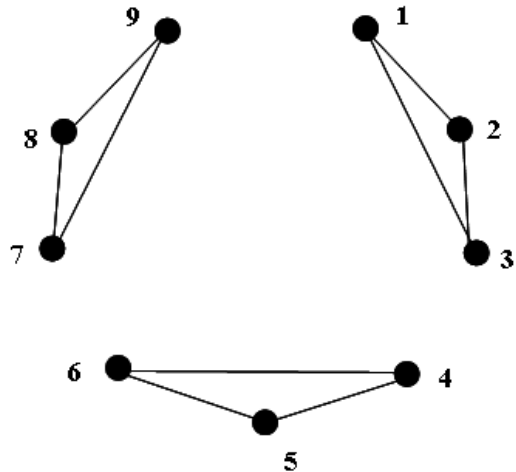

Рис. 1.3. $\overline{M}_3$ – дополнительный граф Муна-Мозера на 9 вершин.

Для решения задачи определения клик в графе Муна-Мозера применим метод Магу. Рассмотрим следующее представление графа $\overline{M}_3$:

$$\Pi = \prod_{i=1}^{m} e_i = (v_{i1} \vee v_{i2})_i = 1. \tag{1.1}$$

$\Pi = (v_1 + v_2)(v_1 + v_3)(v_2 + v_3)(v_4 + v_5)(v_4 + v_6)(v_5 + v_6)(v_7 + v_8)(v_7 + v_9)(v_8 + v_9) =$
$= (v_1 + v_2v_3)(v_2 + v_3)(v_4 + v_5v_6)(v_5 + v_6)(v_7 + v_8v_9)(v_8 + v_9) =$
$= (v_1v_2 + v_1v_3 + v_2v_3)(v_4v_5 + v_4v_6 + v_5v_6)(v_7v_8 + v_7v_9 + v_8v_9) =$
$= (v_1v_2v_4v_5 + v_1v_2v_4v_6 + v_1v_2v_5v_6 + v_1v_3v_4v_5 + v_1v_3v_4v_6 + v_1v_3v_5v_6 + v_2v_3v_4v_5 + v_2v_3v_4v_6 + v_2v_3v_5v_6) \times$
$\times (v_7v_8 + v_7v_9 + v_8v_9) = v_1v_2v_4v_5v_7v_8 + v_1v_2v_4v_6v_7v_8 + v_1v_2v_5v_6v_7v_8 + v_1v_3v_4v_5v_7v_8 +$
$+ v_1v_3v_4v_6v_7v_8 + v_1v_3v_5v_6v_7v_8 + v_2v_3v_4v_5v_7v_8 + v_2v_3v_4v_6v_7v_8 + v_2v_3v_5v_6v_7v_8 + v_1v_2v_4v_5v_7v_9 +$
$+ v_1v_2v_4v_6v_7v_9 + v_1v_2v_5v_6v_7v_9 + v_1v_3v_4v_5v_7v_9 + v_1v_3v_4v_6v_7v_9 + v_1v_3v_5v_6v_7v_9 + v_2v_3v_4v_5v_7v_9 +$
$+ v_2v_3v_4v_6v_7v_9 + v_2v_3v_5v_6v_7v_9 + v_1v_2v_4v_5v_8v_9 + v_1v_2v_4v_6v_8v_9 + v_1v_2v_5v_6v_8v_9 + v_1v_3v_4v_5v_8v_9 +$
$+ v_1v_3v_4v_6v_8v_9 + v_1v_3v_5v_6v_8v_9 + v_2v_3v_4v_5v_8v_9 + v_2v_3v_4v_6v_8v_9 + v_2v_3v_5v_6v_8v_9.$

Отсюда определим клики как дополнение:

$K_1 = \{v_1,v_2,v_3,v_4,v_5,v_6,v_7,v_8,v_9\} \setminus \{v_1,v_2,v_4,v_5,v_7,v_8\} = \{v_3,v_6,v_9\};$

$K_2 = \{v_1,v_2,v_3,v_4,v_5,v_6,v_7,v_8,v_9\} \setminus \{v_1,v_2,v_4,v_6,v_7,v_8\} = \{v_3,v_5,v_9\};$

$K_3 = \{v_1,v_2,v_3,v_4,v_5,v_6,v_7,v_8,v_9\} \setminus \{v_1,v_2,v_5,v_6,v_7,v_8\} = \{v_3,v_4,v_9\};$

$K_4 = \{v_1,v_2,v_3,v_4,v_5,v_6,v_7,v_8,v_9\} \setminus \{v_1,v_2,v_4,v_5,v_7,v_9\} = \{v_3,v_6,v_8\};$

$K_5 = \{v_1,v_2,v_3,v_4,v_5,v_6,v_7,v_8,v_9\} \setminus \{v_1,v_2,v_4,v_6,v_7,v_9\} = \{v_3,v_5,v_8\};$

$K_6 = \{v_1,v_2,v_3,v_4,v_5,v_6,v_7,v_8,v_9\} \setminus \{v_1,v_2,v_5,v_6,v_7,v_9\} = \{v_3,v_4,v_8\};$

$K_7 = \{v_1,v_2,v_3,v_4,v_5,v_6,v_7,v_8,v_9\} \setminus \{v_1,v_2,v_4,v_5,v_8,v_9\} = \{v_3,v_6,v_7\};$

$K_8 = \{v_1,v_2,v_3,v_4,v_5,v_6,v_7,v_8,v_9\} \setminus \{v_1,v_2,v_4,v_6,v_8,v_9\} = \{v_3,v_5,v_7\};$

$K_9 = \{v_1,v_2,v_3,v_4,v_5,v_6,v_7,v_8,v_9\} \setminus \{v_1,v_2,v_5,v_6,v_8,v_9\} = \{v_3,v_4,v_7\};$

$K_{10} = \{v_1,v_2,v_3,v_4,v_5,v_6,v_7,v_8,v_9\} \setminus \{v_1,v_3,v_4,v_5,v_7,v_8\} = \{v_2,v_6,v_9\};$

$K_{11} = \{v_1,v_2,v_3,v_4,v_5,v_6,v_7,v_8,v_9\} \setminus \{v_1,v_3,v_4,v_6,v_7,v_8\} = \{v_2,v_5,v_9\};$

$K_{12} = \{v_1,v_2,v_3,v_4,v_5,v_6,v_7,v_8,v_9\} \setminus \{v_1,v_3,v_5,v_6,v_7,v_8\} = \{v_2,v_4,v_9\};$

$K_{13} = \{v_1,v_2,v_3,v_4,v_5,v_6,v_7,v_8,v_9\} \setminus \{v_1,v_3,v_4,v_5,v_7,v_9\} = \{v_2,v_6,v_8\};$

$K_{14} = \{v_1,v_2,v_3,v_4,v_5,v_6,v_7,v_8,v_9\} \setminus \{v_1,v_3,v_4,v_6,v_7,v_9\} = \{v_2,v_5,v_8\};$

$K_{15} = \{v_1,v_2,v_3,v_4,v_5,v_6,v_7,v_8,v_9\} \setminus \{v_1,v_3,v_5,v_6,v_7,v_9\} = \{v_2,v_4,v_8\};$

$K_{16} = \{v_1,v_2,v_3,v_4,v_5,v_6,v_7,v_8,v_9\} \setminus \{v_1,v_3,v_4,v_5,v_8,v_9\} = \{v_2,v_6,v_7\};$

$K_{17} = \{v_1,v_2,v_3,v_4,v_5,v_6,v_7,v_8,v_9\} \setminus \{v_1,v_3,v_4,v_6,v_8,v_9\} = \{v_2,v_5,v_7\};$

$K_{18} = \{v_1,v_2,v_3,v_4,v_5,v_6,v_7,v_8,v_9\} \setminus \{v_1,v_3,v_5,v_6,v_8,v_9\} = \{v_2,v_4,v_7\};$

$K_{19} = \{v_1,v_2,v_3,v_4,v_5,v_6,v_7,v_8,v_9\} \setminus \{v_2,v_3,v_4,v_5,v_7,v_8\} = \{v_1,v_6,v_9\}$;

$K_{20} = \{v_1,v_2,v_3,v_4,v_5,v_6,v_7,v_8,v_9\} \setminus \{v_2,v_3,v_4,v_6,v_7,v_8\} = \{v_1,v_5,v_9\}$;

$K_{21} = \{v_1,v_2,v_3,v_4,v_5,v_6,v_7,v_8,v_9\} \setminus \{v_2,v_3,v_5,v_6,v_7,v_8\} = \{v_1,v_4,v_9\}$;

$K_{22} = \{v_1,v_2,v_3,v_4,v_5,v_6,v_7,v_8,v_9\} \setminus \{v_2,v_3,v_4,v_5,v_7,v_9\} = \{v_1,v_6,v_8\}$;

$K_{23} = \{v_1,v_2,v_3,v_4,v_5,v_6,v_7,v_8,v_9\} \setminus \{v_2,v_3,v_4,v_6,v_7,v_9\} = \{v_1,v_5,v_8\}$;

$K_{24} = \{v_1,v_2,v_3,v_4,v_5,v_6,v_7,v_8,v_9\} \setminus \{v_2,v_3,v_5,v_6,v_7,v_9\} = \{v_1,v_4,v_8\}$;

$K_{25} = \{v_1,v_2,v_3,v_4,v_5,v_6,v_7,v_8,v_9\} \setminus \{v_2,v_3,v_4,v_5,v_8,v_9\} = \{v_1,v_6,v_7\}$;

$K_{26} = \{v_1,v_2,v_3,v_4,v_5,v_6,v_7,v_8,v_9\} \setminus \{v_2,v_3,v_4,v_6,v_8,v_9\} = \{v_1,v_5,v_7\}$;

$K_{27} = \{v_1,v_2,v_3,v_4,v_5,v_6,v_7,v_8,v_9\} \setminus \{v_2,v_3,v_5,v_6,v_8,v_9\} = \{v_1,v_4,v_7\}$.

Как видно, для метода Магу количество клик длиной три в графе Муна-Мозера на 9 вершин определяется величиной $3^3 = 27$.

Количество ребер в графе Муна-Мозера можно определить как

$$m = \frac{n(n-3)}{2} \quad (1.2)$$

Тогда количество элементарных операций для метода Магу можно оценить как $2^m$.

### 1.3. Общая постановка задачи

Рассмотрим граф К$_4$.

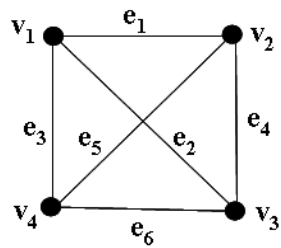

Рис. 1.4. Граф К$_4$

Будем представлять ребро концевыми вершинами. Обозначим вершины в треугольной клике как $v_{i1}, v_{i2}, v_{i3}$, тогда условие существования клики длиной 3 в графе К$_4$ можно записать в следующем виде:

$$F_1 = (v_{i1}, v_{i2}) \,\&\, (v_{i1}, v_{i3}) \,\&\, (v_{i2}, v_{i3}) = 1 \quad (1.3)$$

Обозначим вершины в клике длиной 4 как $(v_{i1}, v_{i2}, v_{i3}, v_{i4})$, тогда условие существование клики:

$$F_2 = (v_{i1}, v_{i2}) \,\&\, (v_{i1}, v_{i3}) \,\&\, (v_{i1}, v_{i4}) \,\&\, (v_{i2}, v_{i3}) \,\&\, (v_{i2}, v_{i4}) \,\&\, (v_{i3}, v_{i4}) = 1 \quad (1.4)$$

В общем виде, для клики длиной *n*, выражения (1.2 и 1.3) можно записать в виде следующего выражения:

$$F= \prod_{i=1}^{n(n-1)/2} e_i = e_1 \, \& \, e_2 \, \& \, ... \, \& \, e_{n(n-1)/2} = 1 \qquad (1.5)$$

Таким образом, можно рассматривать задачу о клике в представлении формулы (1.4).

|    | $e_1=(v_1,v_2)$ | $e_2=(v_1,v_3)$ | $e_3=(v_1,v_4)$ | $e_4=(v_2,v_3)$ | $e_5=(v_2,v_4)$ | $e_6=(v_3,v_4)$ | F |
|----|---|---|---|---|---|---|---|
|    | 1 | 2 | 3 | 4 | 5 | 6 | 7 |
| 0  | 0 | 0 | 0 | 0 | 0 | 0 | 0 |
| 1  | 0 | 0 | 0 | 0 | 0 | 1 | 0 |
| 2  | 0 | 0 | 0 | 0 | 1 | 0 | 0 |
| 3  | 0 | 0 | 0 | 1 | 0 | 0 | 0 |
| 4  | 0 | 0 | 1 | 0 | 0 | 0 | 0 |
|    |   |   | … |   |   |   | … |
| 60 | 1 | 1 | 1 | 1 | 0 | 0 | 0 |
| 61 | 1 | 1 | 1 | 1 | 0 | 1 | 0 |
| 62 | 1 | 1 | 1 | 1 | 1 | 0 | 0 |
| 63 | 1 | 1 | 1 | 1 | 1 | 1 | 1 |

В этом случае вычислительная сложность равна $2^m$.

Будем рассматривать решение задачи о максимальной клике для простого неориентированного графа G имеющего *n* вершин и *m* ребер в следующем виде:

1) Определяем существование в графе G клик длиной три.

2) Затем определяем существование в графе G клик длиной четыре.

…

n) Определяем существование в графе G клик длиной *n*.

Если на следующем этапе процесса поиска не существует клик длиной *n*+1, то поиск прекращается и клика длиной *n* объявляется максимальной.

**Вычислительная сложность.**

Данный процесс имеет экспоненциальную вычислительную сложность. Число сочетаний из *n* элементов по *k* равно биномиальному коэффициенту:

$$\binom{n}{k} = C_n^k = C_{n-1}^{k-1} + C_{n-1}^k = \frac{n!}{k!(n-k)!} \qquad (1.6)$$

Вычислительную сложность процессов, связанных с перестановками *n*! элементов можно оценить с помощью формулы Стирлинга. Так число перестановок длины *n* равно произведению всех натуральных чисел от 1 до *n* – *n*!. Асимптотическая формула Стирлинга позволяет быстро и довольно точно оценить эту величину для больших значений *n*:

$$n! = \sqrt{2\pi n}\left(\frac{n}{e}\right)^n \exp\frac{1}{12n+\theta_n} \tag{1.7}$$

где $0 < \theta_n < 1$, $n > 0$. В формуле (1.7) максимальное значение $\theta_n$ в действительности меньше 1 и примерно равно 0,7509.

Для определения характера вычислений воспользуемся следующим выражением:

$$n^x = e^{x\ln n}. \tag{1.8}$$

Если будем считать $x$ переменной, а $n$ константой, то процесс расчета носит экспоненциальный характер. Если же считать $x$ константой, а $n$ – переменной, то процесс расчета будет носить полиномиальный характер.

Следовательно, можно сделать вывод о том, что при факториальной зависимости процесс поиска клик будет носить экспоненциальный характер.

Преобразуем выражение (1.5):

$$C_n^k = \frac{n(n-1)(n-2)...(n-k+1)(n-k)...1}{k!(n-k)!} \tag{1.9}$$

Сократим числитель и знаменатель на величину $(n-k)!$, получим:

$$C_n^k = \frac{n(n-1)(n-2)...(n-k+1)}{k!} \text{ при } k \leq n-k \tag{1.10}$$

$$C_n^k = \frac{n(n-1)(n-2)...(k+1)}{(n-k)!} \text{ при } k \geq n-k \tag{1.11}$$

Данное выражение можно усилить:

$$C_n^k = \frac{n(n-1)(n-2)...(n-k+1)}{k!} \approx \frac{n^k}{k!} \text{ при } k \leq n-k \tag{1.12}$$

$$C_n^k = \frac{n(n-1)(n-2)...(k+1)}{(n-k)!} \approx \frac{n^{n-k}}{(n-k)!} \text{ при } k \geq n-k \tag{1.13}$$

Теперь можно сказать, что вычислительная сложность алгоритма построения сочетаний из $n$ элементов по $k$ равна $o(n^k)$ для $k \leq n-k$ и $o(n^{n-k})$ для $k \geq n-k$.

Таким образом, вычислительная сложность алгоритма построения сочетаний (операции сочетания $C_n^k$) носит экспоненциальный характер. Согласно выражению (1.8) $n^{n/2} = e^{n/2\ln n}$ [1,6].

# Глава 2. Рекуррентный алгоритм выделения максимальной клики графа

## 2.1. Множество треугольных циклов

Пространство суграфов графа включает в себя подпространство разрезов S(G) и подпространство циклов C(G). Поэтому соответствующие математические модели должны учитывать свойства указанных подпространств. Будем рассматривать следующую математическую модель для решения задачи выделения максимальной клики графа. Начнем с рассмотрения свойств клик несепарабельного неориентированного графа G(V,E) [8].

**Определение 1**. Связный неориентированный граф G, не имеющий мостов и точек сочленения, без петель, кратных ребер и вершин с локальной степенью меньшей или равной двум, называется *несепарабельным графом* G.

Введем некоторые определения и утверждения, характеризующие клики графа. Количество вершин в графе обозначим как *n*, а количество ребер – как *m* [3,4,10].

**Определение 2**. Количество вершин графа в клике называется *длиной клики* и обозначается латинской буквой $L_v$.

**Определение 3**. Количество ребер в цикле графа называется *длиной цикла* и обозначается латинской буквой $L_c$.

Клика состоит из циклов длиной три, таковых что: кольцевая сумма этих циклов равна пустому множеству [9], и/или суграф всех ребер клики и подграф, состоящий из этих циклов, являются полными подграфами.

Например, для графа $K_4$, циклы $\{e_1,e_2,e_4\} \oplus \{e_2,e_3,e_6\} \oplus \{e_1,e_3,e_5\} \oplus \{e_4,e_5,e_6\} = \emptyset$ и подграф, состоящий из этих циклов есть полный граф.

Циклы будем записывать как подмножество ребер и подмножество вершин.

**Утверждение 1**. Если ребро графа включено в состав клики длиной L, то данное ребро должно быть включено в L-2 циклов длиной три (треугольных циклов).

Количество циклов графа длиной три, можно рассчитать по формуле

$$k_c \leq \frac{n(n-1)(n-2)}{6} \qquad (2.1)$$

## 2.2. Итерационный процесс построения множества клик

Будем рассматривать следующий процесс исключения определённого количества циклов из множества циклов графа длиной три. Пусть данный процесс состоит из нескольких последовательно выполняемых действий (итераций). Каждая итерация характеризуется следующими параметрами:

i – номер итерации;

$C_i$ – множество треугольных графов для итерации;

$P_i = <b_1, b_2, \ldots, b_m>$ – кортеж количества треугольных циклов проходящих по ребру;

$b_j$ – количество циклов длиной три проходящих по ребру $e_j$ (вес ребра $e_j$);

$MAX_i$ – максимальное значение элемента кортежа $P_i$;

$MIN_i$ – минимальное значение элемента кортежа $P_i$ (исключая нули);

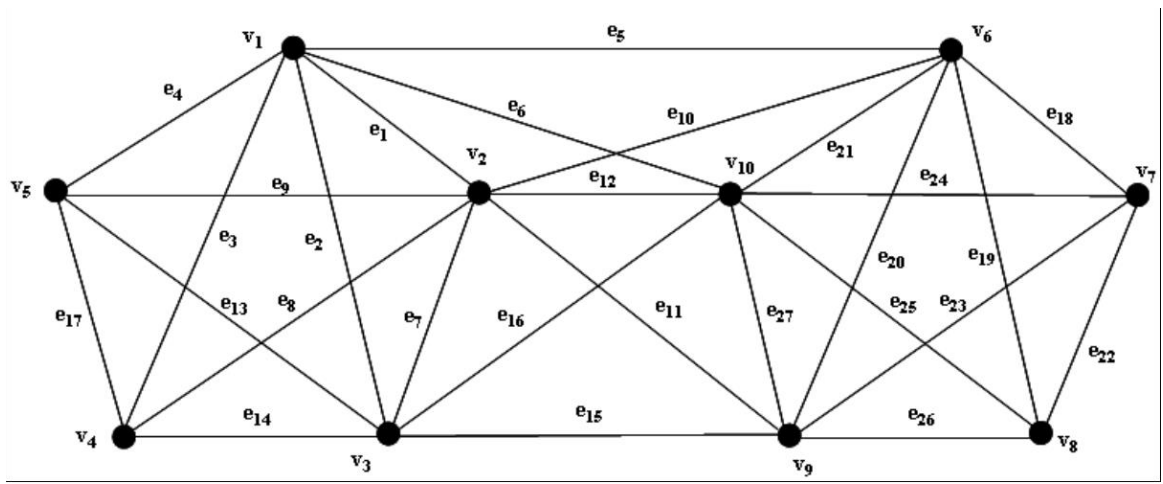

Рис. 2.1. Граф $G_1$.

Для получения элементов множества циклов $C_{i+1}$ последующей итерации, создаётся подмножество треугольных циклов $Q_i$, состоящее из циклов, имеющих рёбра с весом $MIN_i$.

$$C_{i+1} = C_i \setminus Q_i \qquad (2.2)$$

В качестве примера рассмотрим граф $G_1$ представленный на рис. 2.

Количество вершин графа = 10.
Количество рёбер графа = 27.
Количество треугольных циклов графа = 30.

Начальная итерация $i_0$.

Обозначим множество циклов длиной три в графе $G_1$ как $C_0$:

$c_1 = \{e_1, e_2, e_7\} \leftrightarrow \{v_2, v_3, v_1\}$;   $c_2 = \{e_1, e_3, e_8\} \leftrightarrow \{v_2, v_4, v_1\}$;
$c_3 = \{e_1, e_4, e_9\} \leftrightarrow \{v_2, v_5, v_1\}$;   $c_4 = \{e_1, e_5, e_{10}\} \leftrightarrow \{v_2, v_6, v_1\}$;
$c_5 = \{e_1, e_6, e_{12}\} \leftrightarrow \{v_2, v_{10}, v_1\}$;   $c_6 = \{e_2, e_3, e_{13}\} \leftrightarrow \{v_3, v_4, v_1\}$;
$c_7 = \{e_2, e_4, e_{14}\} \leftrightarrow \{v_3, v_5, v_1\}$;   $c_8 = \{e_2, e_6, e_{16}\} \leftrightarrow \{v_3, v_{10}, v_1\}$;
$c_9 = \{e_3, e_4, e_{17}\} \leftrightarrow \{v_4, v_5, v_1\}$;   $c_{10} = \{e_5, e_6, e_{21}\} \leftrightarrow \{v_6, v_{10}, v_1\}$;
$c_{11} = \{e_7, e_8, e_{13}\} \leftrightarrow \{v_3, v_4, v_2\}$;   $c_{12} = \{e_7, e_9, e_{14}\} \leftrightarrow \{v_3, v_5, v_2\}$;
$c_{13} = \{e_7, e_{11}, e_{15}\} \leftrightarrow \{v_3, v_9, v_2\}$;   $c_{14} = \{e_7, e_{12}, e_{16}\} \leftrightarrow \{v_3, v_{10}, v_2\}$;
$c_{15} = \{e_8, e_9, e_{17}\} \leftrightarrow \{v_4, v_5, v_2\}$;   $c_{16} = \{e_{10}, e_{11}, e_{20}\} \leftrightarrow \{v_6, v_9, v_2\}$;
$c_{17} = \{e_{10}, e_{12}, e_{21}\} \leftrightarrow \{v_6, v_{10}, v_2\}$;   $c_{18} = \{e_{11}, e_{12}, e_{27}\} \leftrightarrow \{v_9, v_{10}, v_2\}$;
$c_{19} = \{e_{13}, e_{14}, e_{17}\} \leftrightarrow \{v_4, v_5, v_3\}$;   $c_{20} = \{e_{15}, e_{16}, e_{27}\} \leftrightarrow \{v_9, v_{10}, v_3\}$;
$c_{21} = \{e_{18}, e_{19}, e_{22}\} \leftrightarrow \{v_7, v_8, v_6\}$;   $c_{22} = \{e_{18}, e_{20}, e_{23}\} \leftrightarrow \{v_7, v_9, v_6\}$;

$c_{23}=\{e_{18},e_{21},e_{24}\} \leftrightarrow \{v_7,v_{10},v_6\}$;  $c_{24}=\{e_{19},e_{20},e_{25}\} \leftrightarrow \{v_8,v_9,v_6\}$;
$c_{25}=\{e_{19},e_{21},e_{26}\} \leftrightarrow \{v_8,v_{10},v_6\}$;  $c_{26}=\{e_{20},e_{21},e_{27}\} \leftrightarrow \{v_9,v_{10},v_6\}$;
$c_{27}: \{e_{22},e_{23},e_{25}\} \leftrightarrow \{v_8,v_9,v_7\}$;  $c_{28}=\{e_{22},e_{24},e_{26}\} \leftrightarrow \{v_8,v_{10},v_7\}$;
$c_{29}=\{e_{23},e_{24},e_{27}\} \leftrightarrow \{v_9,v_{10},v_7\}$;  $c_{30}=\{e_{25},e_{26},e_{27}\} \leftrightarrow \{v_9,v_{10},v_8\}$.

Начальная конфигурация кортежа количества циклов проходящих по рёбрам $P_0$:

$P_0=<5,4,3,3,2,3,5,3,3,3,3,4,3,3,2,3,3,3,3,4,5,3,3,3,3,5>$;  $MIN_0=2$;  $MAX_0=5$.

Формируем множество $Q_0$ удаляя циклы, содержащие рёбра с минимальным весом. В данном случае это рёбра $e_5$ и $e_{15}$ с весом 2.

$c_4=\{e_1,e_5,e_{10}\} \leftrightarrow \{v_2,v_6,v_1\}$;  $c_{10}=\{e_5,e_6,e_{21}\} \leftrightarrow \{v_6,v_{10},v_1\}$;
$c_{13}=\{e_7,e_{11},e_{15}\} \leftrightarrow \{v_3,v_9,v_2\}$;  $c_{20}=\{e_{15},e_{16},e_{27}\} \leftrightarrow \{v_9,v_{10},v_3\}$;

Множество оставшихся циклов $C_1 = C_0 \backslash Q_0$:

$c_1=\{e_1,e_2,e_7\} \leftrightarrow \{v_2,v_3,v_1\}$;  $c_2=\{e_1,e_3,e_8\} \leftrightarrow \{v_2,v_4,v_1\}$;
$c_3=\{e_1,e_4,e_9\} \leftrightarrow \{v_2,v_5,v_1\}$;  $c_5=\{e_1,e_6,e_{12}\} \leftrightarrow \{v_2,v_{10},v_1\}$;
$c_6=\{e_2,e_3,e_{13}\} \leftrightarrow \{v_3,v_4,v_1\}$;  $c_7=\{e_2,e_4,e_{14}\} \leftrightarrow \{v_3,v_5,v_1\}$;
$c_8=\{e_2,e_6,e_{16}\} \leftrightarrow \{v_3,v_{10},v_1\}$;  $c_9=\{e_3,e_4,e_{17}\} \leftrightarrow \{v_4,v_5,v_1\}$;
$c_{11}=\{e_7,e_8,e_{13}\} \leftrightarrow \{v_3,v_4,v_2\}$;  $c_{12}=\{e_7,e_9,e_{14}\} \leftrightarrow \{v_3,v_5,v_2\}$;
$c_{14}=\{e_7,e_{12},e_{16}\} \leftrightarrow \{v_3,v_{10},v_2\}$;  $c_{15}=\{e_8,e_9,e_{17}\} \leftrightarrow \{v_4,v_5,v_2\}$;
$c_{16}=\{e_{10},e_{11},e_{20}\} \leftrightarrow \{v_6,v_9,v_2\}$;  $c_{17}=\{e_{10},e_{12},e_{21}\} \leftrightarrow \{v_6,v_{10},v_2\}$;
$c_{18}=\{e_{11},e_{12},e_{27}\} \leftrightarrow \{v_9,v_{10},v_2\}$;  $c_{19}=\{e_{13},e_{14},e_{17}\} \leftrightarrow \{v_4,v_5,v_3\}$;
$c_{21}=\{e_{18},e_{19},e_{22}\} \leftrightarrow \{v_7,v_8,v_6\}$;  $c_{22}=\{e_{18},e_{20},e_{23}\} \leftrightarrow \{v_7,v_9,v_6\}$;
$c_{23}=\{e_{18},e_{21},e_{24}\} \leftrightarrow \{v_7,v_{10},v_6\}$;  $c_{24}=\{e_{19},e_{20},e_{25}\} \leftrightarrow \{v_8,v_9,v_6\}$;
$c_{25}=\{e_{19},e_{21},e_{26}\} \leftrightarrow \{v_8,v_{10},v_6\}$;  $c_{26}=\{e_{20},e_{21},e_{27}\} \leftrightarrow \{v_9,v_{10},v_6\}$;
$c_{27}: \{e_{22},e_{23},e_{25}\} \leftrightarrow \{v_8,v_9,v_7\}$;  $c_{28}=\{e_{22},e_{24},e_{26}\} \leftrightarrow \{v_8,v_{10},v_7\}$;
$c_{29}=\{e_{23},e_{24},e_{27}\} \leftrightarrow \{v_9,v_{10},v_7\}$;  $c_{30}=\{e_{25},e_{26},e_{27}\} \leftrightarrow \{v_9,v_{10},v_8\}$.

Процесс порождения итераций производится до тех пор, пока не будет исчерпано все множество треугольных циклов графа. Отсюда следует, что максимальное количество итераций не может превосходить количество треугольных циклов в графе.

$$I_{\max} \leq \frac{n(n-1)(n-2)}{6} \qquad (2.3)$$

Среди всех порождённых итераций выбирается итерация с номером *i*, имеющая максимальное значение $MIN_i$. Такую итерацию будем называть *основной итерацией*.

Например, для графа $G_1$, представленного на рис. 2.1, кортеж количества циклов, проходящих по рёбрам, будет изменяться от одной итерации к другой:

$P_0=<5,4,3,3,2,3,5,3,3,3,3,4,3,3,2,3,3,3,3,4,5,3,3,3,3,5>$;  $MIN_0=2$;  $MAX_0=5$.
$P_1=<4,4,3,3,0,2,4,3,3,2,2,4,3,3,0,2,3,3,3,4,4,3,3,3,3,4>$;  $MIN_1=2$;  $MAX_1=4$.
$P_2=<3,3,3,3,0,0,3,3,3,0,0,0,3,3,0,0,3,3,3,3,3,3,3,3,3,3>$;  $MIN_2=3$;  $MAX_2=3$.
$P_3=<0,0,0,0,0,0,0,0,0,0,0,0,0,0,0,0,0,0,0,0,0,0,0,0,0,0>$;  $MIN_3=0$;  $MAX_3=0$.

Здесь вторая итерация является основной итерацией.

Множество $C_2$ имеет вид:

$c_1=\{e_1,e_2,e_7\} \leftrightarrow \{v_2,v_3,v_1\}$; $\qquad c_2=\{e_1,e_3,e_8\} \leftrightarrow \{v_2,v_4,v_1\}$;
$c_3=\{e_1,e_4,e_9\} \leftrightarrow \{v_2,v_5,v_1\}$; $\qquad c_6=\{e_2,e_3,e_{13}\} \leftrightarrow \{v_3,v_4,v_1\}$;
$c_7=\{e_2,e_4,e_{14}\} \leftrightarrow \{v_3,v_5,v_1\}$; $\qquad c_9=\{e_3,e_4,e_{17}\} \leftrightarrow \{v_4,v_5,v_1\}$;
$c_{11}=\{e_7,e_8,e_{13}\} \leftrightarrow \{v_3,v_4,v_2\}$; $\qquad c_{12}=\{e_7,e_9,e_{14}\} \leftrightarrow \{v_3,v_5,v_2\}$;
$c_{15}=\{e_8,e_9,e_{17}\} \leftrightarrow \{v_4,v_5,v_2\}$; $\qquad c_{19}=\{e_{13},e_{14},e_{17}\} \leftrightarrow \{v_4,v_5,v_3\}$;
$c_{21}=\{e_{18},e_{19},e_{22}\} \leftrightarrow \{v_7,v_8,v_6\}$; $\qquad c_{22}=\{e_{18},e_{20},e_{23}\} \leftrightarrow \{v_7,v_9,v_6\}$;
$c_{23}=\{e_{18},e_{21},e_{24}\} \leftrightarrow \{v_7,v_{10},v_6\}$; $\qquad c_{24}=\{e_{19},e_{20},e_{25}\} \leftrightarrow \{v_8,v_9,v_6\}$;
$c_{25}=\{e_{19},e_{21},e_{26}\} \leftrightarrow \{v_8,v_{10},v_6\}$; $\qquad c_{26}=\{e_{20},e_{21},e_{27}\} \leftrightarrow \{v_9,v_{10},v_6\}$;
$c_{27}: \{e_{22},e_{23},e_{25}\} \leftrightarrow \{v_8,v_9,v_7\}$; $\qquad c_{28}=\{e_{22},e_{24},e_{26}\} \leftrightarrow \{v_8,v_{10},v_7\}$;
$c_{29}=\{e_{23},e_{24},e_{27}\} \leftrightarrow \{v_9,v_{10},v_7\}$; $\qquad c_{30}=\{e_{25},e_{26},e_{27}\} \leftrightarrow \{v_9,v_{10},v_8\}$.

### 2.3. Выделение клики

В основной итерации произвольным образом выбираем ребро $e_j$, имеющее вес равный $MIN_i$. В подмножестве треугольных циклов $C_i$ выбираем циклы, имеющие в своем составе ребро $e_j$. Выбранное подмножество циклов образует подграф $H_j$. Если подграф $H_j$ соответствует полному графу, то подграф $H_j$ является *кликой*. Если подграф $H_j$ не является *кликой*, то к данному подграфу рекурсивно применяется процедура вычисления основной итерации и повторяется процесс определения клики.

В нашем примере выберем ребро $e_4$. В подмножестве $C_2$ выбираем циклы:

$c_3=\{e_1,e_4,e_9\} \leftrightarrow \{v_2,v_5,v_1\}$; $c_7=\{e_2,e_4,e_{14}\} \leftrightarrow \{v_3,v_5,v_1\}$; $c_9=\{e_3,e_4,e_{17}\} \leftrightarrow \{v_4,v_5,v_1\}$.

На основе выбранных циклов строим подграф $H_4=\{v_1,v_2,v_3,v_4,v_5\}$ как объединение всех вершин выбранных циклов. Формируем подмножество циклов, вершины которых включаются в подграф $H_4$:

$c_1=\{e_1,e_2,e_7\} \leftrightarrow \{v_2,v_3,v_1\}$; $\qquad c_2=\{e_1,e_3,e_8\} \leftrightarrow \{v_2,v_4,v_1\}$;
$c_3=\{e_1,e_4,e_9\} \leftrightarrow \{v_2,v_5,v_1\}$; $\qquad c_6=\{e_2,e_3,e_{13}\} \leftrightarrow \{v_3,v_4,v_1\}$;
$c_7=\{e_2,e_4,e_{14}\} \leftrightarrow \{v_3,v_5,v_1\}$; $\qquad c_9=\{e_3,e_4,e_{17}\} \leftrightarrow \{v_4,v_5,v_1\}$;
$c_{11}=\{e_7,e_8,e_{13}\} \leftrightarrow \{v_3,v_4,v_2\}$; $\qquad c_{12}=\{e_7,e_9,e_{14}\} \leftrightarrow \{v_3,v_5,v_2\}$;
$c_{15}=\{e_8,e_9,e_{17}\} \leftrightarrow \{v_4,v_5,v_2\}$; $\qquad c_{19}=\{e_{13},e_{14},e_{17}\} \leftrightarrow \{v_4,v_5,v_3\}$.

Это клика длиной 5, так как количество выбранных циклов соответствует полному графу, и кольцевая сумма циклов есть $\varnothing$.

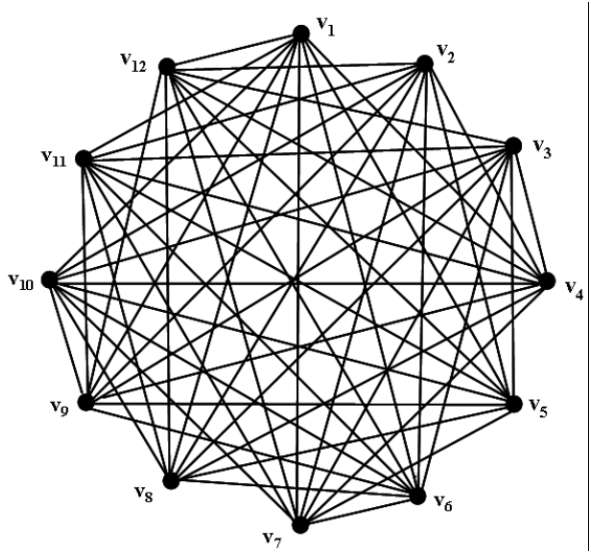

Рис. 2.2. Граф Муна-Мозера на 12 вершин.

Рассмотрим случай рекурсивного применения алгоритма выделения максимальной клики графа. В качестве примера рассмотрим граф Муна-Мозера на 12 вершин. Следует заметить, что графы Муна-Мозера являются частным случаем графов Турана.

Количество вершин графа = 12.

Количество ребер графа = 54.

Количество циклов графа длиной три = 108.

Смежность графа Муна-Мозера:

```
вершина  v₁:  v₄   v₅   v₆   v₇   v₈   v₉   v₁₀  v₁₁  v₁₂
вершина  v₂:  v₄   v₅   v₆   v₇   v₈   v₉   v₁₀  v₁₁  v₁₂
вершина  v₃:  v₄   v₅   v₆   v₇   v₈   v₉   v₁₀  v₁₁  v₁₂
вершина  v₄:  v₁   v₂   v₃   v₇   v₈   v₉   v₁₀  v₁₁  v₁₂
вершина  v₅:  v₁   v₂   v₃   v₇   v₈   v₉   v₁₀  v₁₁  v₁₂
вершина  v₆:  v₁   v₂   v₃   v₇   v₈   v₉   v₁₀  v₁₁  v₁₂
вершина  v₇:  v₁   v₂   v₃   v₄   v₅   v₆   v₁₀  v₁₁  v₁₂
вершина  v₈:  v₁   v₂   v₃   v₄   v₅   v₆   v₁₀  v₁₁  v₁₂
вершина  v₉:  v₁   v₂   v₃   v₄   v₅   v₆   v₁₀  v₁₁  v₁₂
вершина  v₁₀: v₁   v₂   v₃   v₄   v₅   v₆   v₇   v₈   v₉
вершина  v₁₁: v₁   v₂   v₃   v₄   v₅   v₆   v₇   v₈   v₉
вершина  v₁₂: v₁   v₂   v₃   v₄   v₅   v₆   v₇   v₈   v₉
```

Инцидентность графа Муна-Мозера:

```
вершина  v₁:  e₁   e₂   e₃   e₄   e₅   e₆   e₇   e₈   e₉
вершина  v₂:  e₁₀  e₁₁  e₁₂  e₁₃  e₁₄  e₁₅  e₁₆  e₁₇  e₁₈
вершина  v₃:  e₁₉  e₂₀  e₂₁  e₂₂  e₂₃  e₂₄  e₂₅  e₂₆  e₂₇
вершина  v₄:  e₁   e₁₀  e₁₉  e₂₈  e₂₉  e₃₀  e₃₁  e₃₂  e₃₃
вершина  v₅:  e₂   e₁₁  e₂₀  e₃₄  e₃₅  e₃₆  e₃₇  e₃₈  e₃₉
```

вершина $v_6$: $e_3$ $e_{12}$ $e_{21}$ $e_{40}$ $e_{41}$ $e_{42}$ $e_{43}$ $e_{44}$ $e_{45}$
вершина $v_7$: $e_4$ $e_{13}$ $e_{22}$ $e_{28}$ $e_{34}$ $e_{40}$ $e_{46}$ $e_{47}$ $e_{48}$
вершина $v_8$: $e_5$ $e_{14}$ $e_{23}$ $e_{29}$ $e_{35}$ $e_{41}$ $e_{49}$ $e_{50}$ $e_{51}$
вершина $v_9$: $e_6$ $e_{15}$ $e_{24}$ $e_{30}$ $e_{36}$ $e_{42}$ $e_{52}$ $e_{53}$ $e_{54}$
вершина $v_{10}$: $e_7$ $e_{16}$ $e_{25}$ $e_{31}$ $e_{37}$ $e_{43}$ $e_{46}$ $e_{49}$ $e_{52}$
вершина $v_{11}$: $e_8$ $e_{17}$ $e_{26}$ $e_{32}$ $e_{38}$ $e_{44}$ $e_{47}$ $e_{50}$ $e_{53}$
вершина $v_{12}$: $e_9$ $e_{18}$ $e_{27}$ $e_{33}$ $e_{39}$ $e_{45}$ $e_{48}$ $e_{51}$ $e_{54}$

Множество циклов длиной три:

$c_1 = \{e_1, e_4, e_{28}\} \leftrightarrow \{v_4, v_7, v_1\}$; $\quad c_2 = \{e_1, e_5, e_{29}\} \leftrightarrow \{v_4, v_8, v_1\}$;
$c_3 = \{e_1, e_6, e_{30}\} \leftrightarrow \{v_4, v_9, v_1\}$; $\quad c_4 = \{e_1, e_7, e_{31}\} \leftrightarrow \{v_4, v_{10}, v_1\}$;
$c_5 = \{e_1, e_8, e_{32}\} \leftrightarrow \{v_4, v_{11}, v_1\}$; $\quad c_6 = \{e_1, e_9, e_{33}\} \leftrightarrow \{v_4, v_{12}, v_1\}$;
$c_7 = \{e_2, e_4, e_{34}\} \leftrightarrow \{v_5, v_7, v_1\}$; $\quad c_8 = \{e_2, e_5, e_{35}\} \leftrightarrow \{v_5, v_8, v_1\}$;
$c_9 = \{e_2, e_6, e_{36}\} \leftrightarrow \{v_5, v_9, v_1\}$; $\quad c_{10} = \{e_2, e_7, e_{37}\} \leftrightarrow \{v_5, v_{10}, v_1\}$;
$c_{11} = \{e_2, e_8, e_{38}\} \leftrightarrow \{v_5, v_{11}, v_1\}$; $\quad c_{12} = \{e_2, e_9, e_{39}\} \leftrightarrow \{v_5, v_{12}, v_1\}$;
$c_{13} = \{e_3, e_4, e_{40}\} \leftrightarrow \{v_6, v_7, v_1\}$; $\quad c_{14} = \{e_3, e_5, e_{41}\} \leftrightarrow \{v_6, v_8, v_1\}$;
$c_{15} = \{e_3, e_6, e_{42}\} \leftrightarrow \{v_6, v_9, v_1\}$; $\quad c_{16} = \{e_3, e_7, e_{43}\} \leftrightarrow \{v_6, v_{10}, v_1\}$;
$c_{17} = \{e_3, e_8, e_{44}\} \leftrightarrow \{v_6, v_{11}, v_1\}$; $\quad c_{18} = \{e_3, e_9, e_{45}\} \leftrightarrow \{v_6, v_{12}, v_1\}$;
$c_{19} = \{e_4, e_7, e_{46}\} \leftrightarrow \{v_7, v_{10}, v_1\}$; $\quad c_{20} = \{e_4, e_8, e_{47}\} \leftrightarrow \{v_7, v_{11}, v_1\}$;
$c_{21} = \{e_4, e_9, e_{48}\} \leftrightarrow \{v_7, v_{12}, v_1\}$; $\quad c_{22} = \{e_5, e_7, e_{49}\} \leftrightarrow \{v_8, v_{10}, v_1\}$;
$c_{23} = \{e_5, e_8, e_{50}\} \leftrightarrow \{v_8, v_{11}, v_1\}$; $\quad c_{24} = \{e_5, e_9, e_{51}\} \leftrightarrow \{v_8, v_{12}, v_1\}$;
$c_{25} = \{e_6, e_7, e_{52}\} \leftrightarrow \{v_9, v_{10}, v_1\}$; $\quad c_{26} = \{e_6, e_8, e_{53}\} \leftrightarrow \{v_9, v_{11}, v_1\}$;
$c_{27} = \{e_6, e_9, e_{54}\} \leftrightarrow \{v_9, v_{12}, v_1\}$; $\quad c_{28} = \{e_{10}, e_{13}, e_{28}\} \leftrightarrow \{v_4, v_7, v_2\}$;
$c_{29} = \{e_{10}, e_{14}, e_{29}\} \leftrightarrow \{v_4, v_8, v_2\}$; $\quad c_{30} = \{e_{10}, e_{15}, e_{30}\} \leftrightarrow \{v_4, v_9, v_2\}$;
$c_{31} = \{e_{10}, e_{16}, e_{31}\} \leftrightarrow \{v_4, v_{10}, v_2\}$; $\quad c_{32} = \{e_{10}, e_{17}, e_{32}\} \leftrightarrow \{v_4, v_{11}, v_2\}$;
$c_{33} = \{e_{10}, e_{18}, e_{33}\} \leftrightarrow \{v_4, v_{12}, v_2\}$; $\quad c_{34} = \{e_{11}, e_{13}, e_{34}\} \leftrightarrow \{v_5, v_7, v_2\}$;
$c_{35} = \{e_{11}, e_{14}, e_{35}\} \leftrightarrow \{v_5, v_8, v_2\}$; $\quad c_{36} = \{e_{11}, e_{15}, e_{36}\} \leftrightarrow \{v_5, v_9, v_2\}$;
$c_{37} = \{e_{11}, e_{16}, e_{37}\} \leftrightarrow \{v_5, v_{10}, v_2\}$; $\quad c_{38} = \{e_{11}, e_{17}, e_{38}\} \leftrightarrow \{v_5, v_{11}, v_2\}$;
$c_{39} = \{e_{11}, e_{18}, e_{39}\} \leftrightarrow \{v_5, v_{12}, v_2\}$; $\quad c_{40} = \{e_{12}, e_{13}, e_{40}\} \leftrightarrow \{v_6, v_7, v_2\}$;
$c_{41} = \{e_{12}, e_{14}, e_{41}\} \leftrightarrow \{v_6, v_8, v_2\}$; $\quad c_{42} = \{e_{12}, e_{15}, e_{42}\} \leftrightarrow \{v_6, v_9, v_2\}$;
$c_{43} = \{e_{12}, e_{16}, e_{43}\} \leftrightarrow \{v_6, v_{10}, v_2\}$; $\quad c_{44} = \{e_{12}, e_{17}, e_{44}\} \leftrightarrow \{v_6, v_{11}, v_2\}$;
$c_{45} = \{e_{12}, e_{18}, e_{45}\} \leftrightarrow \{v_6, v_{12}, v_2\}$; $\quad c_{46} = \{e_{13}, e_{16}, e_{46}\} \leftrightarrow \{v_7, v_{10}, v_2\}$;
$c_{47} = \{e_{13}, e_{17}, e_{47}\} \leftrightarrow \{v_7, v_{11}, v_2\}$; $\quad c_{48} = \{e_{13}, e_{18}, e_{48}\} \leftrightarrow \{v_7, v_{12}, v_2\}$;
$c_{49} = \{e_{14}, e_{16}, e_{49}\} \leftrightarrow \{v_8, v_{10}, v_2\}$; $\quad c_{50} = \{e_{14}, e_{17}, e_{50}\} \leftrightarrow \{v_8, v_{11}, v_2\}$;
$c_{51} = \{e_{14}, e_{18}, e_{51}\} \leftrightarrow \{v_8, v_{12}, v_2\}$; $\quad c_{52} = \{e_{15}, e_{16}, e_{52}\} \leftrightarrow \{v_9, v_{10}, v_2\}$;
$c_{53} = \{e_{15}, e_{17}, e_{53}\} \leftrightarrow \{v_9, v_{11}, v_2\}$; $\quad c_{54} = \{e_{15}, e_{18}, e_{54}\} \leftrightarrow \{v_9, v_{12}, v_2\}$;
$c_{55} = \{e_{19}, e_{22}, e_{28}\} \leftrightarrow \{v_4, v_7, v_3\}$; $\quad c_{56} = \{e_{19}, e_{23}, e_{29}\} \leftrightarrow \{v_4, v_8, v_3\}$;
$c_{57} = \{e_{19}, e_{24}, e_{30}\} \leftrightarrow \{v_4, v_9, v_3\}$; $\quad c_{58} = \{e_{19}, e_{25}, e_{31}\} \leftrightarrow \{v_4, v_{10}, v_3\}$;
$c_{59} = \{e_{19}, e_{26}, e_{32}\} \leftrightarrow \{v_4, v_{11}, v_3\}$; $\quad c_{60} = \{e_{19}, e_{27}, e_{33}\} \leftrightarrow \{v_4, v_{12}, v_3\}$;
$c_{61} = \{e_{20}, e_{22}, e_{34}\} \leftrightarrow \{v_5, v_7, v_3\}$; $\quad c_{62} = \{e_{20}, e_{23}, e_{35}\} \leftrightarrow \{v_5, v_8, v_3\}$;
$c_{63} = \{e_{20}, e_{24}, e_{36}\} \leftrightarrow \{v_5, v_9, v_3\}$; $\quad c_{64} = \{e_{20}, e_{25}, e_{37}\} \leftrightarrow \{v_5, v_{10}, v_3\}$;
$c_{65} = \{e_{20}, e_{26}, e_{38}\} \leftrightarrow \{v_5, v_{11}, v_3\}$; $\quad c_{66} = \{e_{20}, e_{27}, e_{39}\} \leftrightarrow \{v_5, v_{12}, v_3\}$;

$c_{67} = \{e_{21},e_{22},e_{40}\} \leftrightarrow \{v_6,v_7,v_3\};$ $\quad c_{68} = \{e_{21},e_{23},e_{41}\} \leftrightarrow \{v_6,v_8,v_3\};$
$c_{69} = \{e_{21},e_{24},e_{42}\} \leftrightarrow \{v_6,v_9,v_3\};$ $\quad c_{70} = \{e_{21},e_{25},e_{43}\} \leftrightarrow \{v_6,v_{10},v_3\};$
$c_{71} = \{e_{21},e_{26},e_{44}\} \leftrightarrow \{v_6,v_{11},v_3\};$ $\quad c_{72} = \{e_{21},e_{27},e_{45}\} \leftrightarrow \{v_6,v_{12},v_3\};$
$c_{73} = \{e_{22},e_{25},e_{46}\} \leftrightarrow \{v_7,v_{10},v_3\};$ $\quad c_{74} = \{e_{22},e_{26},e_{47}\} \leftrightarrow \{v_7,v_{11},v_3\};$
$c_{75} = \{e_{22},e_{27},e_{48}\} \leftrightarrow \{v_7,v_{12},v_3\};$ $\quad c_{76} = \{e_{23},e_{25},e_{49}\} \leftrightarrow \{v_8,v_{10},v_3\};$
$c_{77} = \{e_{23},e_{26},e_{50}\} \leftrightarrow \{v_8,v_{11},v_3\};$ $\quad c_{78} = \{e_{23},e_{27},e_{51}\} \leftrightarrow \{v_8,v_{12},v_3\};$
$c_{79} = \{e_{24},e_{25},e_{52}\} \leftrightarrow \{v_9,v_{10},v_3\};$ $\quad c_{80} = \{e_{24},e_{26},e_{53}\} \leftrightarrow \{v_9,v_{11},v_3\};$
$c_{81} = \{e_{24},e_{27},e_{54}\} \leftrightarrow \{v_9,v_{12},v_3\};$ $\quad c_{82} = \{e_{28},e_{31},e_{46}\} \leftrightarrow \{v_7,v_{10},v_4\};$
$c_{83} = \{e_{28},e_{32},e_{47}\} \leftrightarrow \{v_7,v_{11},v_4\};$ $\quad c_{84} = \{e_{28},e_{33},e_{48}\} \leftrightarrow \{v_7,v_{12},v_4\};$
$c_{85} = \{e_{29},e_{31},e_{49}\} \leftrightarrow \{v_8,v_{10},v_4\};$ $\quad c_{86} = \{e_{29},e_{32},e_{50}\} \leftrightarrow \{v_8,v_{11},v_4\};$
$c_{87} = \{e_{29},e_{33},e_{51}\} \leftrightarrow \{v_8,v_{12},v_4\};$ $\quad c_{88} = \{e_{30},e_{31},e_{52}\} \leftrightarrow \{v_9,v_{10},v_4\};$
$c_{89} = \{e_{30},e_{32},e_{53}\} \leftrightarrow \{v_9,v_{11},v_4\};$ $\quad c_{90} = \{e_{30},e_{33},e_{54}\} \leftrightarrow \{v_9,v_{12},v_4\};$
$c_{91} = \{e_{34},e_{37},e_{46}\} \leftrightarrow \{v_7,v_{10},v_5\};$ $\quad c_{92} = \{e_{34},e_{38},e_{47}\} \leftrightarrow \{v_7,v_{11},v_5\};$
$c_{93} = \{e_{34},e_{39},e_{48}\} \leftrightarrow \{v_7,v_{12},v_5\};$ $\quad c_{94} = \{e_{35},e_{37},e_{49}\} \leftrightarrow \{v_8,v_{10},v_5\};$
$c_{95} = \{e_{35},e_{38},e_{50}\} \leftrightarrow \{v_8,v_{11},v_5\};$ $\quad c_{96} = \{e_{35},e_{39},e_{51}\} \leftrightarrow \{v_8,v_{12},v_5\};$
$c_{97} = \{e_{36},e_{37},e_{52}\} \leftrightarrow \{v_9,v_{10},v_5\};$ $\quad c_{98} = \{e_{36},e_{38},e_{53}\} \leftrightarrow \{v_9,v_{11},v_5\};$
$c_{99} = \{e_{36},e_{39},e_{54}\} \leftrightarrow \{v_9,v_{12},v_5\};$ $\quad c_{100} = \{e_{40},e_{43},e_{46}\} \leftrightarrow \{v_7,v_{10},v_6\};$
$c_{101} = \{e_{40},e_{44},e_{47}\} \leftrightarrow \{v_7,v_{11},v_6\};$ $\quad c_{102} = \{e_{40},e_{45},e_{48}\} \leftrightarrow \{v_7,v_{12},v_6\};$
$c_{103} = \{e_{41},e_{43},e_{49}\} \leftrightarrow \{v_8,v_{10},v_6\};$ $\quad c_{104} = \{e_{41},e_{44},e_{50}\} \leftrightarrow \{v_8,v_{11},v_6\};$
$c_{105} = \{e_{41},e_{45},e_{51}\} \leftrightarrow \{v_8,v_{12},v_6\};$ $\quad c_{106} = \{e_{42},e_{43},e_{52}\} \leftrightarrow \{v_9,v_{10},v_6\};$
$c_{107} = \{e_{42},e_{44},e_{53}\} \leftrightarrow \{v_9,v_{11},v_6\};$ $\quad c_{108} = \{e_{42},e_{45},e_{54}\} \leftrightarrow \{v_9,v_{12},v_6\}.$

Кортежи весов ребер для итераций процесса поиска максимальной клики графа имеют вид:

$P_0 = <6,6,6,6,6,6,6,6,6,6,6,6,6,6,6,6,6,6,6,6,6,6,6,6,6,6,6,6,6,6,6,6,6,6,6,6,6,6,6,$
$6,6,6,6,6,6,6,6,6,6,6,6,6,6>;$ $MIN_0 = 6;$ $MAX_0 = 6.$
$P_1 = <0,0,0,0,0,0,0,0,0,0,0,0,0,0,0,0,0,0,0,0,0,0,0,0,0,0,0,0,0,0,0,0,0,0,0,0,0,0,$
$0,0,0,0,0,0,0,0,0,0,0,0,0,0,0,0>;$ $MIN_1 = 0;$ $MAX_1 = 0.$

Основная итерация графа Муна-Мозера $i_0$ определяется кортежем, состоящим из равных весов ребер. Поэтому формирование подграфа максимальной длины можно начинать с любого ребра. Выберем ребро $e_1$.

$c_1 = \{e_1,e_4,e_{28}\} \leftrightarrow \{v_4,v_7,v_1\};$ $\quad c_2 = \{e_1,e_5,e_{29}\} \leftrightarrow \{v_4,v_8,v_1\};$
$c_3 = \{e_1,e_6,e_{30}\} \leftrightarrow \{v_4,v_9,v_1\};$ $\quad c_4 = \{e_1,e_7,e_{31}\} \leftrightarrow \{v_4,v_{10},v_1\};$
$c_5 = \{e_1,e_8,e_{32}\} \leftrightarrow \{v_4,v_{11},v_1\};$ $\quad c_6 = \{e_1,e_9,e_{33}\} \leftrightarrow \{v_4,v_{12},v_1\}.$

Объединив вершины циклов, получаем следующий подграф $H_1 = \{v_1,v_4,v_7,v_9,v_{10},v_{11},v_{12}\}$. Формируем подмножество циклов, вершины которых включаются в подграф $H_1$ (рис. 2.3).

$c_1 = \{e_1,e_4,e_{28}\} \leftrightarrow \{v_4,v_7,v_1\};$ $\quad c_2 = \{e_1,e_5,e_{29}\} \leftrightarrow \{v_4,v_8,v_1\};$
$c_3 = \{e_1,e_6,e_{30}\} \leftrightarrow \{v_4,v_9,v_1\};$ $\quad c_4 = \{e_1,e_7,e_{31}\} \leftrightarrow \{v_4,v_{10},v_1\};$
$c_5 = \{e_1,e_8,e_{32}\} \leftrightarrow \{v_4,v_{11},v_1\};$ $\quad c_6 = \{e_1,e_9,e_{33}\} \leftrightarrow \{v_4,v_{12},v_1\};$
$c_{19} = \{e_4,e_7,e_{46}\} \leftrightarrow \{v_7,v_{10},v_1\};$ $\quad c_{20} = \{e_4,e_8,e_{47}\} \leftrightarrow \{v_7,v_{11},v_1\};$
$c_{21} = \{e_4,e_9,e_{48}\} \leftrightarrow \{v_7,v_{12},v_1\};$ $\quad c_{22} = \{e_5,e_7,e_{49}\} \leftrightarrow \{v_8,v_{10},v_1\};$

c$_{23}$ = {e$_5$,e$_8$,e$_{50}$} ↔ {v$_8$,v$_{11}$,v$_1$};   c$_{24}$ = {e$_5$,e$_9$,e$_{51}$} ↔ {v$_8$,v$_{12}$,v$_1$};
c$_{25}$ = {e$_6$,e$_7$,e$_{52}$} ↔ {v$_9$,v$_{10}$,v$_1$};   c$_{26}$ = {e$_6$,e$_8$,e$_{53}$} ↔ {v$_9$,v$_{11}$,v$_1$};
c$_{27}$ = {e$_6$,e$_9$,e$_{54}$} ↔ {v$_9$,v$_{12}$,v$_1$};   c$_{82}$ = {e$_{28}$,e$_{31}$,e$_{46}$} ↔ {v$_7$,v$_{10}$,v$_4$};
c$_{83}$ = {e$_{28}$,e$_{32}$,e$_{47}$} ↔ {v$_7$,v$_{11}$,v$_4$};   c$_{84}$ = {e$_{28}$,e$_{33}$,e$_{48}$} ↔ {v$_7$,v$_{12}$,v$_4$};
c$_{85}$ = {e$_{29}$,e$_{31}$,e$_{49}$} ↔ {v$_8$,v$_{10}$,v$_4$};   c$_{86}$ = {e$_{29}$,e$_{32}$,e$_{50}$} ↔ {v$_8$,v$_{11}$,v$_4$};
c$_{87}$ = {e$_{29}$,e$_{33}$,e$_{51}$} ↔ {v$_8$,v$_{12}$,v$_4$};   c$_{88}$ = {e$_{30}$,e$_{31}$,e$_{52}$} ↔ {v$_9$,v$_{10}$,v$_4$};
c$_{89}$ = {e$_{30}$,e$_{32}$,e$_{53}$} ↔ {v$_9$,v$_{11}$,v$_4$};   c$_{90}$ = {e$_{30}$,e$_{33}$,e$_{54}$} ↔ {v$_9$,v$_{12}$,v$_4$}.

Рис. 2.3. Подграф для ребра e$_1$.

P$^*$ = <6,4,4,4,4,4,4,4,4,0,0,0,0,0,0,0,0,0,0,0,0,0,0,0,0,0,4,4,4,4,4,4,0,0,0,0,0,0,0,0,0,0,0,0,0,2,2,2,2,2,2,2,2,2>;  MIN$_0$ = 2;  MAX$_0$ = 6.

Выбираем любое ребро, имеющее вес равный MIN$_0$. Выделяем циклы, проходящие по ребру e$_{46}$:

c$_{19}$ = {e$_4$,e$_7$,e$_{46}$} ↔ {v$_7$,v$_{10}$,v$_1$};   c$_{82}$ = {e$_{28}$,e$_{31}$,e$_{46}$} ↔ {v$_7$,v$_{10}$,v$_4$}.

Объединив вершины циклов, получаем следующий подграф H$_{46}$ = {v$_1$,v$_4$,v$_7$,v$_{10}$}. Формируем подмножество циклов, вершины которых включаются в подграф H$_{46}$.

c$_1$ = {e$_1$,e$_4$,e$_{28}$} ↔ {v$_4$,v$_7$,v$_1$};   c$_4$ = {e$_1$,e$_7$,e$_{31}$} ↔ {v$_4$,v$_{10}$,v$_1$};
c$_{19}$ = {e$_4$,e$_7$,e$_{46}$} ↔ {v$_7$,v$_{10}$,v$_1$};   c$_{82}$ = {e$_{28}$,e$_{31}$,e$_{46}$} ↔ {v$_7$,v$_{10}$,v$_4$}.

Подграф H$_{46}$ – клика длиной 4.

Если мы выберем другое ребро, например e$_{47}$, то получим другую клику H$_{47}$={v$_1$,v$_4$,v$_7$,v$_{11}$} длиной 4. Количество таких клик с участием ребра e$_1$ в графе Муна-Мозера будет равно 9.

По составу кортежа весов ребер, принадлежащего основной итерации графа G, можно определить имеется ли одна максимальная клика в составе графа или их несколько, имеют ли максимальные клики точки сопряжения или нет.

Например, рассмотрим следующий кортеж весов ребер, имеющий веса MIN=MAX:

P=<3,3,3,3,0,0,3,3,3,0,0,0,3,3,0,0,3,3,3,3,3,3,3,3,3,3>; MIN=3; MAX=3.

Он характеризует граф, имеющий две клики максимальной длины, и клики не имеют точки сочленения (рис. 2.1).

Кортеж весов рёбер для графа $G_2$ (рис. 2.4), имеет следующие веса MIN ≠ MAX, и является основной итерацией:

P = <3,5,3,4,4,5,3,3,3,3,4,4,5,3,3,3,4,4>; MIN=3; MAX=5.

характеризует граф, имеющий клики длины 5 и точки их сочленения.

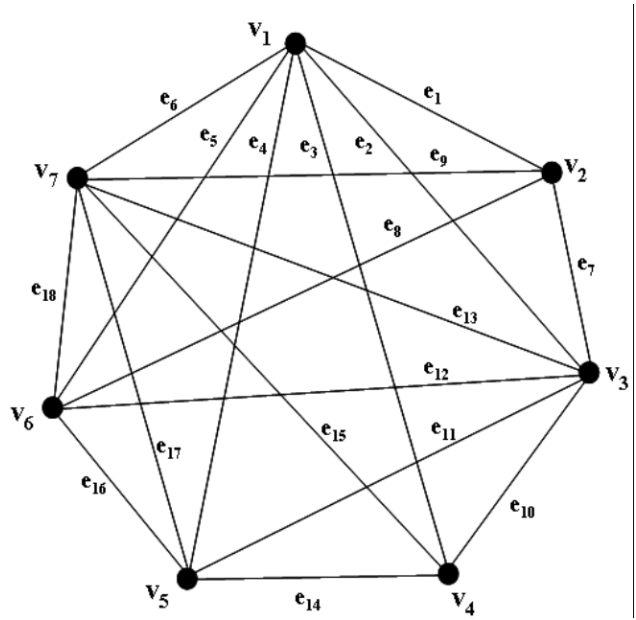

Рис. 2.4. Граф $G_2$.

### 2.4. Алгоритм вычисления максимальной клики графа.

Опишем алгоритм выделения максимальной клики графа.

**Инициализация.** Задан граф G.

**шаг 1**. [**Формирование множества циклов длины три**]. В графе G выделяем множество всех циклов длиной три.

**шаг 2**. [**Построение итерации с номером i**]. Если множество треугольных циклов $C_i$ не пусто, то строим для него кортеж весов рёбер $P_i$. Определяем переменные $MAX_i$ и $MIN_i$.

**шаг 3**. [**Удаление циклов с участием рёбер минимального веса**]. Используя кортеж циклов проходящих по рёбрам $P_i$, определяем рёбра с весом равным MIN. Формируем подмножество циклов $Q_i$ с участием рёбер минимального веса. Вычисляем подмножество $C_{i+1}=C_i \backslash Q_i$. Если подмножество $C_{i+1}$ не пусто, то идём на шаг 2. Иначе на шаг 4.

**шаг 4**. [**Определение основной итерации**]. Из множества итераций выбираем основную итерацию с максимальным значением величины MIN.

**шаг 5**. [**Выделение подграфа H**]. Определяем ребро с минимальным весом в основной итерации. Выделяем циклы, содержащие ребро $e_i$, и строим подграф, содержащий все вершины данного подмножества циклов. Определяем треугольные циклы, образующие подграф. Если этот подграф клика, то конец работы алгоритма, в противном случае к этому подграфу рекурсивно применяется алгоритм выделения клики графа.

Вычислительная сложность алгоритма определяется аддитивным сложением количества элементарных операций для последовательных шагов алгоритма в худшем случае [1]:

- количество циклов длиной три в графе в нашем случае приближённо равно $n^3$;
- количество операций для построения кортежа весов рёбер в итерации с номером $i$ можно рассчитать как количество рёбер во множестве треугольных циклов – приближенно $3n^3$;
- количество операций для определения основной итерации можно рассчитать, как произведение всех итераций (количество треугольных циклов) на величину количества операций для построения кортежа весов – примерно $n^3 3n^3 = 3n^6$;
- количество операций для поиска циклов, составляющих подграф относительно выбранного ребра во множестве треугольных циклов – приближенно $n^3$.

Таким образом, количество элементарных операций можно определить по формуле $n^3+3n^6+n^3$. С учётом рекурсии формулу можно представить как $2(n^3+3n^6+n^3)$. Следовательно, вычислительная сложность алгоритма определения максимальной клики графа $O(n^6)$.

Задача выделения максимальной клики графа является задачей с полиномиальной вычислительной сложностью. Количество максимальных клик в графах Муна-Мозера увеличивается экспоненциально с увеличением количества вершин. Поэтому задача перечисления всех максимальных клик графа является задачей с экспоненциальной вычислительной сложностью.

Основу алгоритма составляет выделение множества треугольных циклов графа, что позволяет осуществлять решение используя как подпространство циклов C(G), так и подпространство разрезов графа S(G).

### 2.5. Рекуррентный процесс построения суграфов

Удаление циклов длиной три порождает рекурсивный процесс построения суграфов (рис. 2.5 – 2.8).

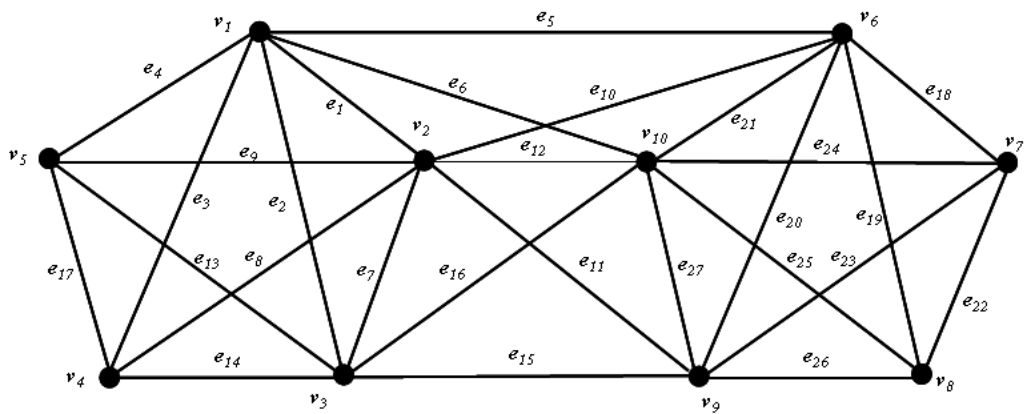
Рис. 2.5. Суграф 0–итерации.

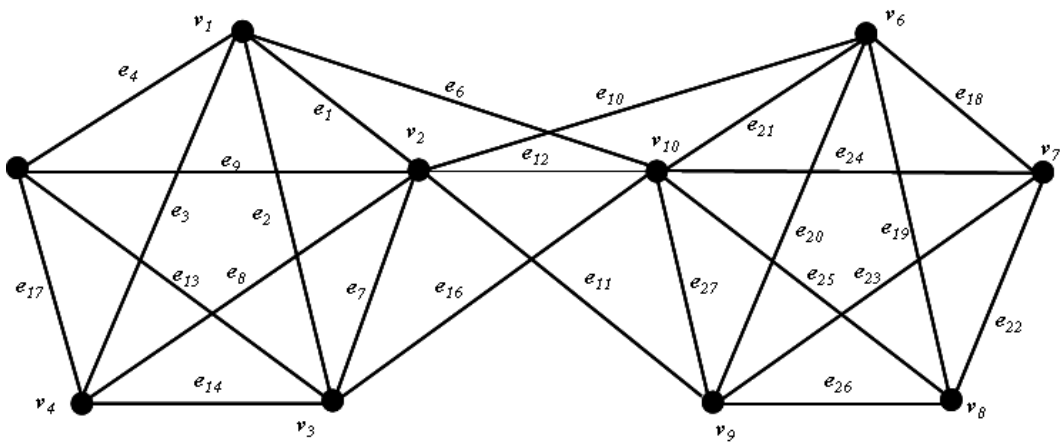
Рис. 2.6. Суграф 1-й итерации.

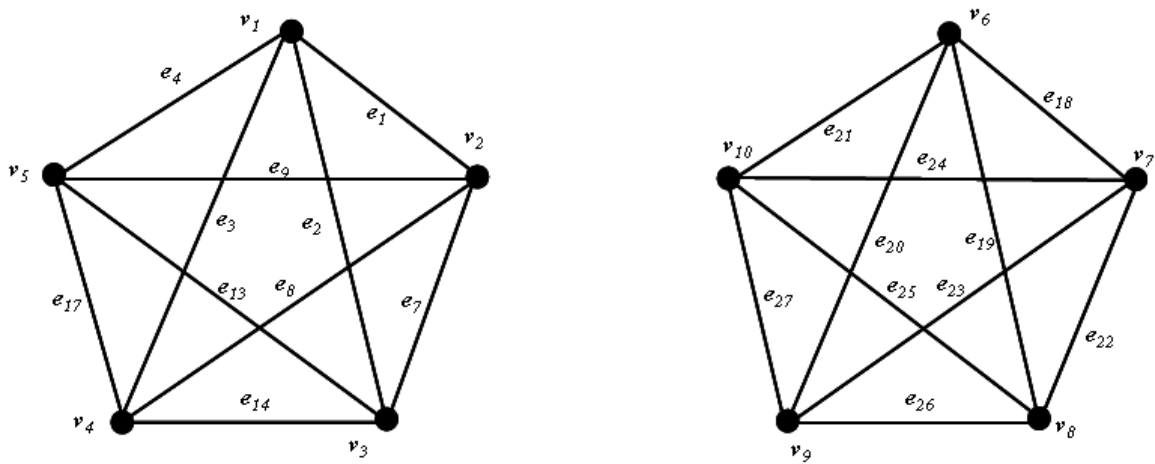
Рис. 2.7. Суграф 2-й итерации.

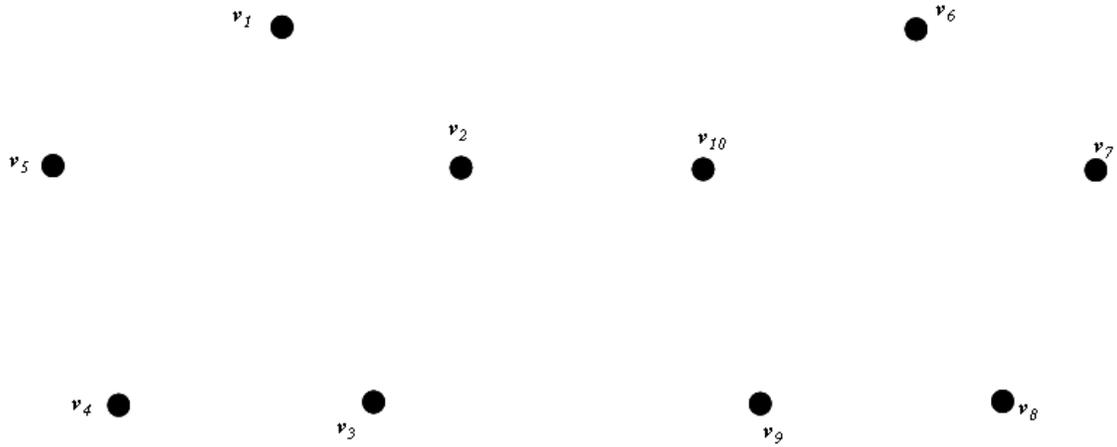

Рис. 2.8. Суграф 3-й итерации.

## 2.6. Рекуррентный процесс построения подграфов

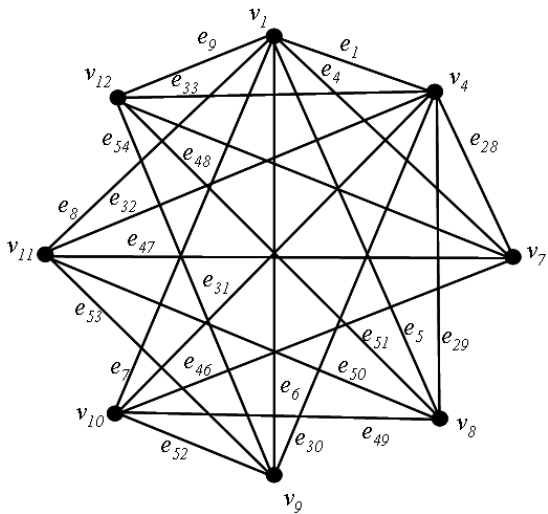
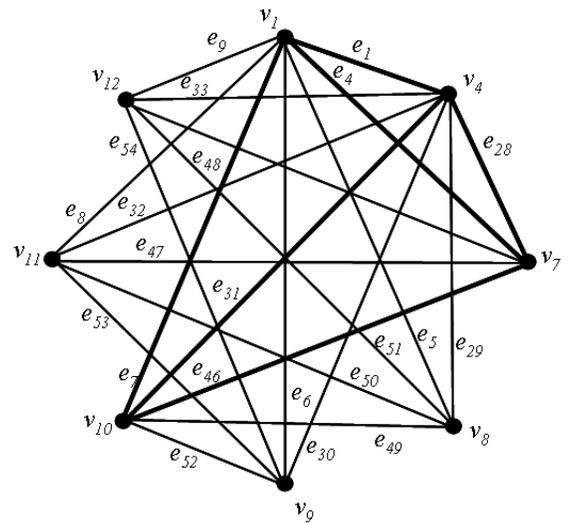

Рис. 2.9. Подграф графа Муна-Мозера относительно ребра $e_1$.

Рис. 2.10. Клика графа Муна-Мозера относительно рёбер $e_1$ и $e_{46}$.

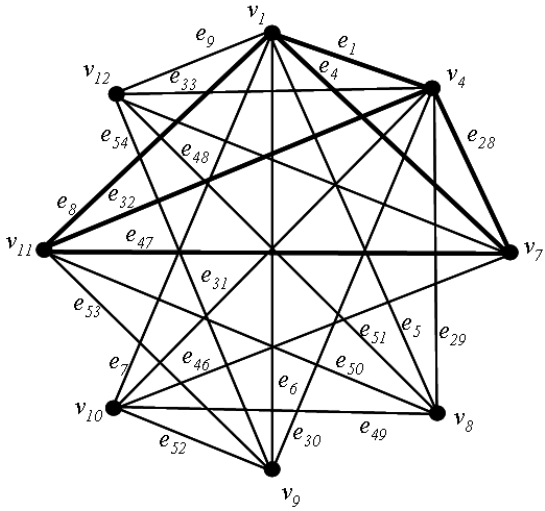
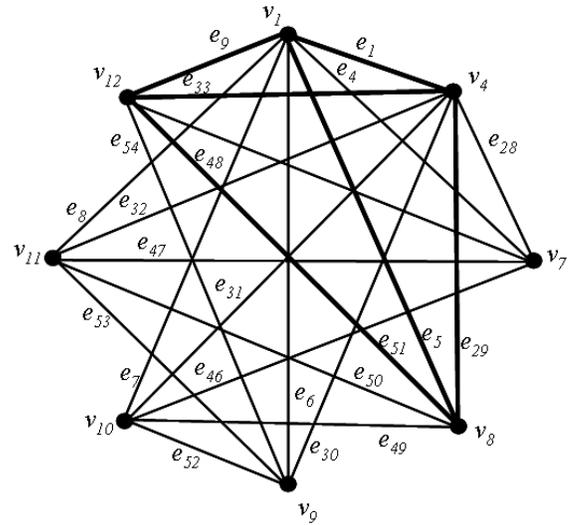

Рис. 2.11. Клика графа Муна-Мозера относительно ребер e₁ и e₄₇.

Рис. 2.12. Клика графа Муна-Мозера относительно ребер e₁ и e₄₈.

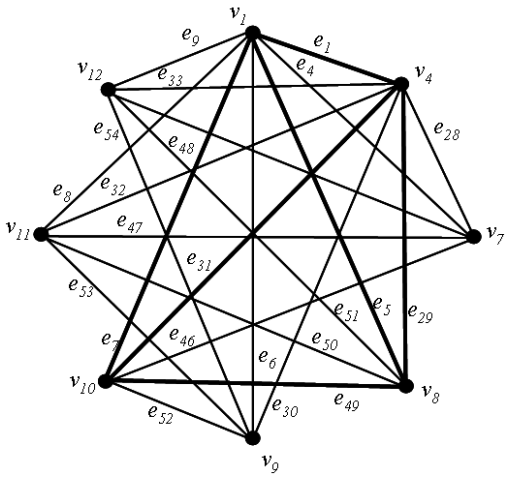
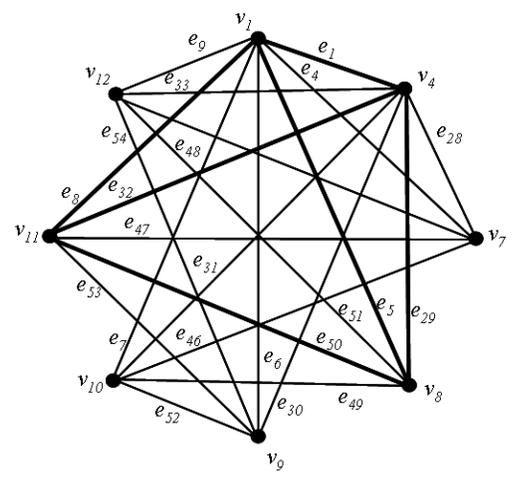

Рис. 2.13. Клика графа Муна-Мозера относительно ребер e₁ и e₄₉.

Рис. 2.14. Клика графа Муна-Мозера относительно ребер e₁ и e₅₀.

Рис. 2.15. Клика графа Муна-Мозера
относительно ребер $e_1$ и $e_{51}$.

Рис. 2.16. Клика графа Муна-Мозера
относительно ребер $e_1$ и $e_{52}$.

Вектор циклов по ребрам $P_e$, построенный относительно ребра $e_1$, порождает подграф, представленный на рис. 2.9:

$P^* =$
$=<6,4,4,4,4,4,4,4,4,0,0,0,0,0,0,0,0,0,0,0,0,0,0,0,0,0,4,4,4,4,4,4,0,0,0,0,0,0,0,0,0,0,0,0,0,0,2,2,2,2,2,2,2,2,2>$;
$MIN_0 = 2$; $MAX_0 = 6$.

Выбираем ребро с минимальным весом; пусть это ребро $e_{47}$. Выделяем циклы, проходящие по ребру $e_{47}$:

$c_{20} = \{e_4,e_8,e_{47}\} \leftrightarrow \{v_7,v_{11},v_1\}$;    $c_{83} = \{e_{28},e_{32},e_{47}\} \leftrightarrow \{v_7,v_{11},v_4\}$;

Объединив вершины циклов, получаем следующий подграф $H_{47} = \{v_1,v_4,v_7,v_{11}\}$ (рис. 2.11). Формируем подмножество циклов, вершины которых индуцируют подграф $H_{47}$.

$c_1 = \{e_1,e_4,e_{28}\} \leftrightarrow \{v_4,v_7,v_1\}$;    $c_5 = \{e_1,e_8,e_{32}\} \leftrightarrow \{v_4,v_{11},v_1\}$;

$c_{20} = \{e_4,e_8,e_{47}\} \leftrightarrow \{v_7,v_{11},v_1\}$;    $c_{83} = \{e_{28},e_{32},e_{47}\} \leftrightarrow \{v_7,v_{11},v_4\}$.

Порождение клики с ребром $e_{46}$ представлено на рис. 2.10, клики с ребром $e_{47}$ – на рис. 2.11, и так далее.

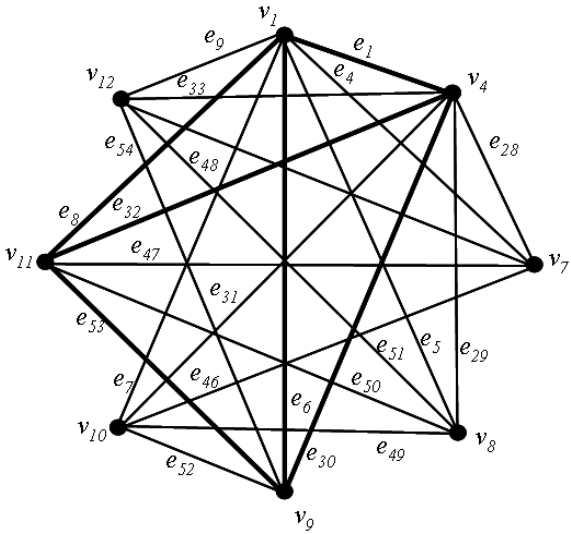
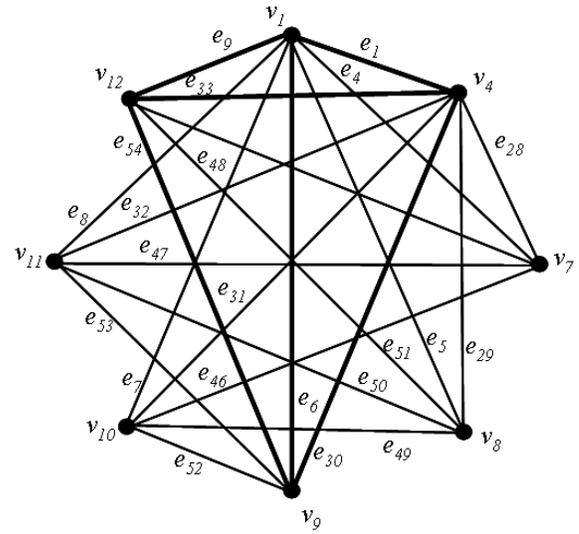

Рис. 2.17. Клика графа Муна-Мозера относительно ребер $e_1$ и $e_{53}$.

Рис. 2.18. Клика графа Муна-Мозера относительно ребер $e_1$ и $e_{54}$.

Рассмотрим граф $G_2$ (рис. 2.19), имеющий в своем составе следующие циклы длиной три:

$c_1 = \{e_1,e_2,e_7\} \leftrightarrow \{v_1,v_2,v_3\}$;   $c_2 = \{e_1,e_5,e_8\} \leftrightarrow \{v_1,v_2,v_6\}$;
$c_3 = \{e_1,e_6,e_9\} \leftrightarrow \{v_1,v_2,v_7\}$;   $c_4 = \{e_2,e_3,e_{10}\} \leftrightarrow \{v_1,v_3,v_4\}$;
$c_5 = \{e_2,e_4,e_{11}\} \leftrightarrow \{v_1,v_3,v_5\}$;   $c_6 = \{e_2,e_5,e_{12}\} \leftrightarrow \{v_1,v_3,v_6\}$;
$c_7 = \{e_2,e_6,e_{13}\} \leftrightarrow \{v_1,v_3,v_7\}$;   $c_8 = \{e_3,e_4,e_{14}\} \leftrightarrow \{v_1,v_4,v_5\}$;
$c_9 = \{e_3,e_6,e_{15}\} \leftrightarrow \{v_1,v_4,v_7\}$;   $c_{10} = \{e_4,e_5,e_{16}\} \leftrightarrow \{v_1,v_5,v_6\}$;
$c_{11} = \{e_4,e_6,e_{17}\} \leftrightarrow \{v_1,v_5,v_7\}$;   $c_{12} = \{e_5,e_6,e_{18}\} \leftrightarrow \{v_1,v_6,v_7\}$;
$c_{13} = \{e_7,e_8,e_{12}\} \leftrightarrow \{v_2,v_3,v_6\}$;   $c_{14} = \{e_7,e_9,e_{13}\} \leftrightarrow \{v_2,v_3,v_7\}$;
$c_{15} = \{e_8,e_9,e_{18}\} \leftrightarrow \{v_2,v_6,v_7\}$;   $c_{16} = \{e_{10},e_{11},e_{14}\} \leftrightarrow \{v_3,v_4,v_5\}$;
$c_{17} = \{e_{10},e_{13},e_{15}\} \leftrightarrow \{v_3,v_4,v_7\}$;   $c_{18} = \{e_{11},e_{12},e_{16}\} \leftrightarrow \{v_3,v_5,v_6\}$;
$c_{19} = \{e_{11},e_{13},e_{17}\} \leftrightarrow \{v_3,v_5,v_7\}$;   $c_{20} = \{e_{12},e_{13},e_{18}\} \leftrightarrow \{v_3,v_6,v_7\}$;
$c_{21} = \{e_{14},e_{15},e_{17}\} \leftrightarrow \{v_4,v_5,v_7\}$;   $c_{22} = \{e_{16},e_{17},e_{18}\} \leftrightarrow \{v_5,v_6,v_7\}$.

Возможны такие варианты:

а) по ребру $e_1$ проходят циклы:

$c_1 = \{e_1,e_2,e_7\} \leftrightarrow \{v_1,v_2,v_3\}$;   $c_2 = \{e_1,e_5,e_8\} \leftrightarrow \{v_1,v_2,v_6\}$;
$c_3 = \{e_1,e_6,e_9\} \leftrightarrow \{v_1,v_2,v_7\}$;

вершинами циклов порождается подграф $\{v_1,v_2,v_3,v_6,v_7\}$ (рис. 2.20);

b) по ребру $e_3$ проходят циклы:

$c_4 = \{e_2,e_3,e_{10}\} \leftrightarrow \{v_1,v_3,v_4\}$;   $c_8 = \{e_3,e_4,e_{14}\} \leftrightarrow \{v_1,v_4,v_5\}$;
$c_9 = \{e_3,e_6,e_{15}\} \leftrightarrow \{v_1,v_4,v_7\}$;

вершинами циклов порождается подграф $\{v_1,v_3,v_4,v_5,v_7\}$ (рис. 2.21);

c) по ребру $e_7$ проходят циклы:

$c_1 = \{e_1,e_2,e_7\} \leftrightarrow \{v_1,v_2,v_3\}$;   $c_{13} = \{e_7,e_8,e_{12}\} \leftrightarrow \{v_2,v_3,v_6\}$;
$c_{14} = \{e_7,e_9,e_{13}\} \leftrightarrow \{v_2,v_3,v_7\}$;

вершинами циклов порождается подграф $\{v_1,v_2,v_3,v_6,v_7\}$ (рис. 2.22);

d) по ребру $e_8$ проходят циклы:

$c_1 = \{e_1,e_5,e_8\} \leftrightarrow \{v_1,v_2,v_6\}$;     $c_{13} = \{e_7,e_8,e_{12}\} \leftrightarrow \{v_2,v_3,v_6\}$;
$c_{15} = \{e_8,e_9,e_{18}\} \leftrightarrow \{v_2,v_6,v_7\}$;

вершинами циклов порождается подграф $\{v_1,v_2,v_3,v_6,v_7\}$ (рис. 2.23);

e) по ребру $e_9$ проходят циклы:

$c_3 = \{e_1,e_6,e_9\} \leftrightarrow \{v_1,v_2,v_7\}$;     $c_{14} = \{e_7,e_9,e_{13}\} \leftrightarrow \{v_2,v_3,v_7\}$;
$c_{15} = \{e_8,e_9,e_{18}\} \leftrightarrow \{v_2,v_6,v_7\}$;

вершинами циклов порождается подграф $\{v_1,v_2,v_3,v_6,v_7\}$ (рис. 2.24);

f) по ребру $e_{10}$ проходят циклы:

$c_4 = \{e_2,e_3,e_{10}\} \leftrightarrow \{v_1,v_3,v_4\}$;     $c_{16} = \{e_{10},e_{11},e_{14}\} \leftrightarrow \{v_3,v_4,v_5\}$;
$c_{17} = \{e_{10},e_{13},e_{15}\} \leftrightarrow \{v_3,v_4,v_7\}$;

вершинами циклов порождается подграф $\{v_1,v_3,v_4,v_6,v_7\}$ (рис. 2.25);

g) по ребру $e_{14}$ проходят циклы:

$c_8 = \{e_3,e_4,e_{14}\} \leftrightarrow \{v_1,v_4,v_5\}$;     $c_{16} = \{e_{10},e_{11},e_{14}\} \leftrightarrow \{v_3,v_4,v_5\}$;
$c_{21} = \{e_{14},e_{15},e_{17}\} \leftrightarrow \{v_4,v_5,v_7\}$;

вершинами циклов порождается подграф $\{v_1,v_3,v_4,v_5,v_7\}$ (рис. 2.26);

h) по ребру $e_{15}$ проходят циклы:

$c_9 = \{e_3,e_6,e_{15}\} \leftrightarrow \{v_1,v_4,v_7\}$;     $c_{17} = \{e_{10},e_{13},e_{15}\} \leftrightarrow \{v_3,v_4,v_7\}$;
$c_{21} = \{e_{14},e_{15},e_{17}\} \leftrightarrow \{v_4,v_5,v_7\}$;

вершинами циклов порождается подграф $\{v_1,v_3,v_4,v_5,v_7\}$ (рис. 2.27);

i) по ребру $e_{16}$ проходят циклы:

$c_{10} = \{e_4,e_5,e_{16}\} \leftrightarrow \{v_1,v_5,v_6\}$;     $c_{18} = \{e_{11},e_{12},e_{16}\} \leftrightarrow \{v_3,v_5,v_6\}$;
$c_{22} = \{e_{16},e_{17},e_{18}\} \leftrightarrow \{v_5,v_6,v_7\}$;

вершинами циклов порождается подграф $\{v_1,v_3,v_5,v_6,v_7\}$ (рис. 2.28).

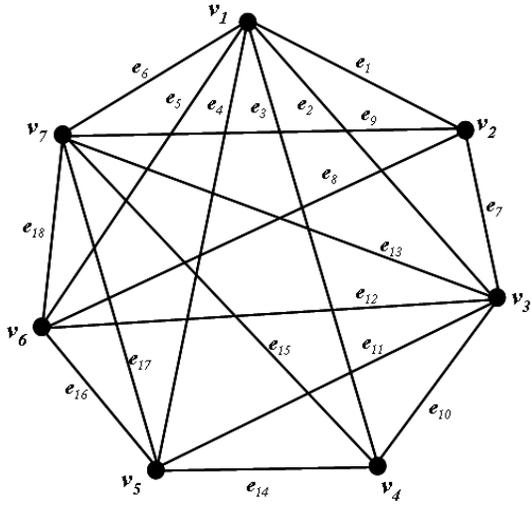
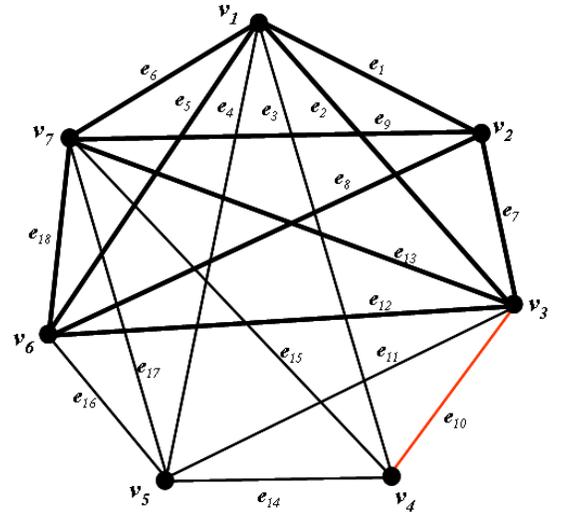

Рис. 2.19. Граф G$_2$.            Рис. 2.20. Клика G$_2$ относительно ребра e$_1$.

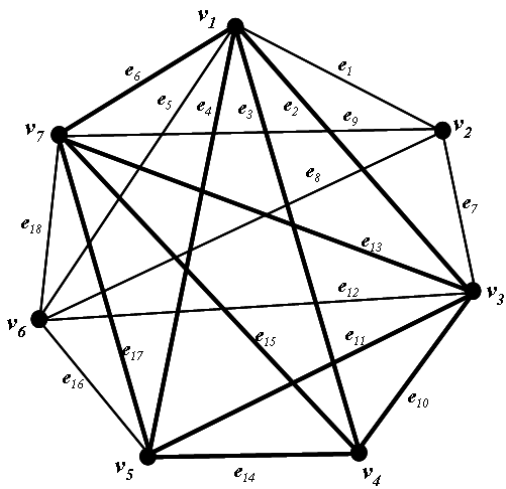
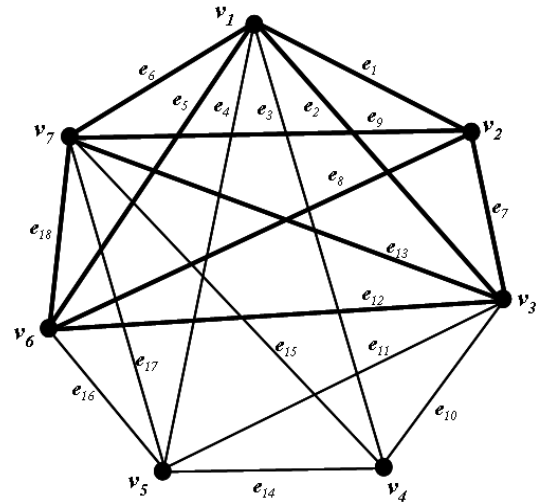

Рис. 2.21. Клика G$_2$ относительно ребра e$_3$.      Рис. 2.22. Клика G$_2$ относительно ребра e$_7$.

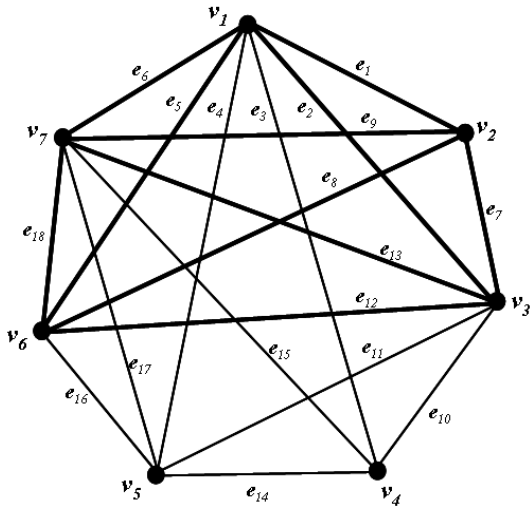
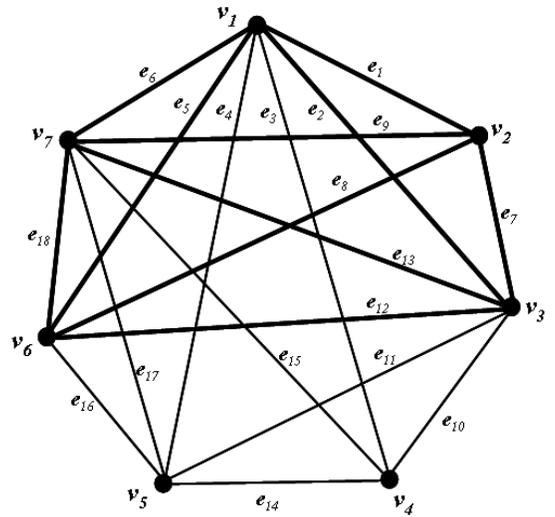

Рис. 2.23. Клика $G_2$ относительно ребра $e_8$.  Рис. 2.24. Клика $G_2$ относительно ребра $e_9$.

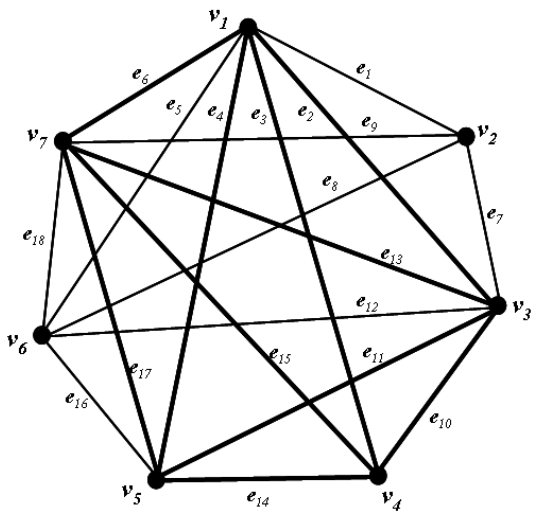
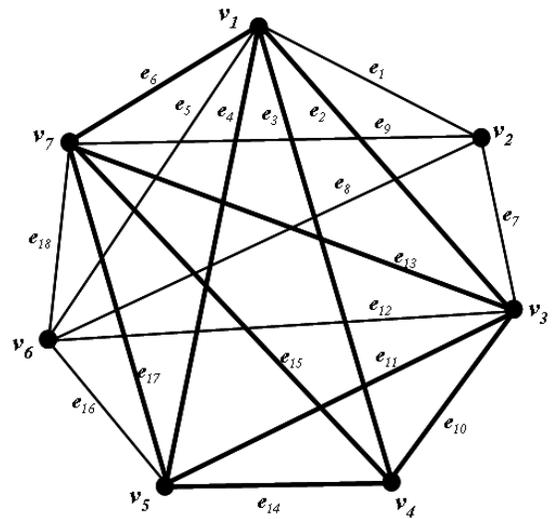

Рис. 2.25. Клика $G_2$ относительно ребра $e_{10}$.  Рис. 2.26. Клика $G_2$ относительно ребра $e_{14}$.

Варианты (a), (c), (d), (e) порождают клику $\{v_1,v_2,v_3,v_6,v_7\}$. Варианты (b), (g), (h) порождают клику $\{v_1,v_3,v_4,v_5,v_7\}$. Вариант (f) порождает клику $\{v_1,v_3,v_4,v_6,v_7\}$. Вариант (i) порождает клику $\{v_1,v_3,v_5,v_6,v_7\}$.

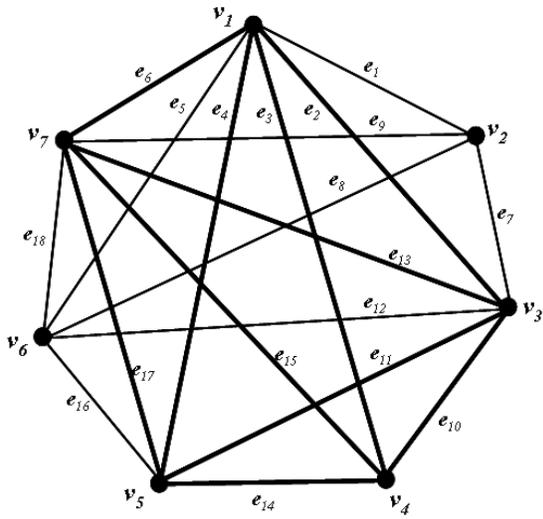 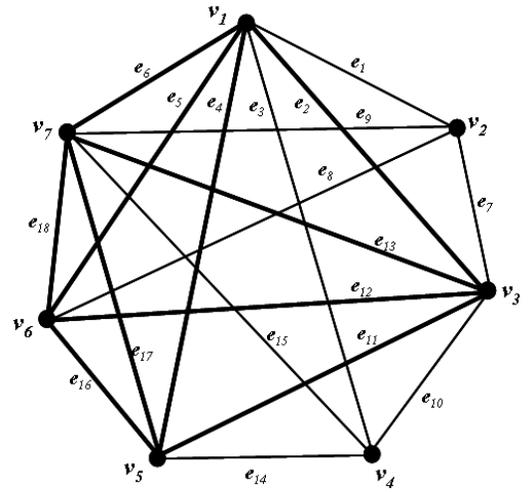

Рис. 2.27. Клика $G_2$ относительно ребра $e_{15}$. Рис. 2.28. Клика $G_2$ относительно ребра $e_{16}$.

# Глава 3. Выделение максимальной клики

## 3.1. Выделение максимальной клики в графе $G_3$

Рассмотрим алгоритм выделения клики максимальной длины в несепарабельном графе. Под длиной клики будем подразумевать количество её вершин. Начнём с примера графа, представленного на рис. 3.1.

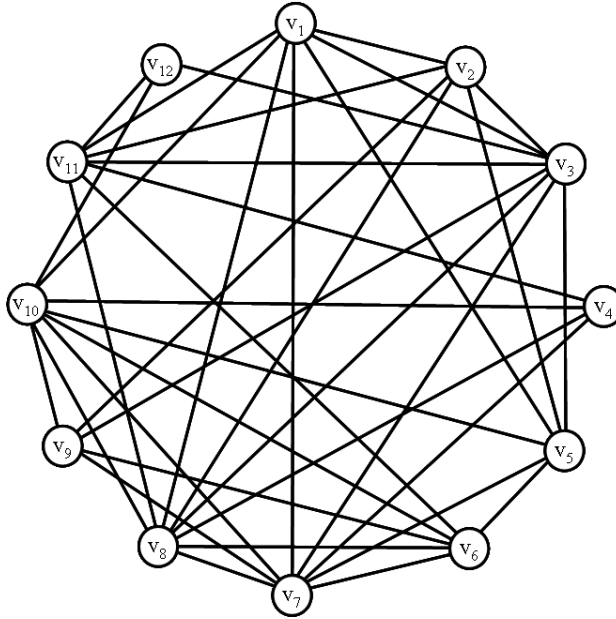

Рис. 3.1. Граф $G_3$.

Количество вершин графа = 12, количество рёбер графа = 38, количество треугольных циклов графа = 39.

Смежность графа:
вершина $v_1$: $\{v_2,v_3,v_5,v_7,v_8,v_{10},v_{11}\}$;
вершина $v_2$: $\{v_1,v_3,v_5,v_8,v_9,v_{11}\}$;
вершина $v_3$: $\{v_1,v_2,v_5,v_7,v_8,v_9,v_{11},v_{12}\}$;
вершина $v_4$: $\{v_7,v_8,v_{10},v_{11}\}$;
вершина $v_5$: $\{v_1,v_2,v_3,v_6,v_7,v_{10}\}$;
вершина $v_6$: $\{v_5,v_7,v_8,v_9,v_{10},v_{11}\}$;
вершина $v_7$: $\{v_1,v_3,v_4,v_5,v_6,v_8,v_9,v_{10}\}$;
вершина $v_8$: $\{v_1,v_2,v_3,v_4,v_6,v_7,v_{10},v_{11}\}$;
вершина $v_9$: $\{v_2,v_3,v_6,v_7,v_{10}\}$;
вершина $v_{10}$: $\{v_1,v_4,v_5,v_6,v_7,v_8,v_9,v_{12}\}$;
вершина $v_{11}$: $\{v_1,v_2,v_3,v_4,v_6,v_8,v_{12}\}$;
вершина $v_{12}$: $\{v_3,v_{10},v_{11}\}$.
Инцидентность графа:
вершина $v_1$: $\{e_1,e_2,e_3,e_4,e_5,e_6,e_7\}$;
вершина $v_2$: $\{e_1,e_8,e_9,e_{10},e_{11},e_{12}\}$;
вершина $v_3$: $\{e_2,e_8,e_{13},e_{14},e_{15},e_{16},e_{17},e_{18}\}$;
вершина $v_4$: $\{e_{19},e_{20},e_{21},e_{22}\}$;

вершина $v_5$: $\{e_3,e_9,e_{13},e_{23},e_{24},e_{25}\}$;
вершина $v_6$: $\{e_{23},e_{26},e_{27},e_{28},e_{29},e_{30}\}$;
вершина $v_7$: $\{e_4,e_{14},e_{19},e_{24},e_{26},e_{31},e_{32},e_{33}\}$;
вершина $v_8$: $\{e_5,e_{10},e_{15},e_{20},e_{27},e_{31},e_{34},e_{35}\}$;
вершина $v_9$: $\{e_{11},e_{16},e_{28},e_{32},e_{36}\}$;
вершина $v_{10}$: $\{e_6,e_{21},e_{25},e_{29},e_{33},e_{34},e_{36},e_{37}\}$;
вершина $v_{11}$: $\{e_7,e_{12},e_{17},e_{22},e_{30},e_{35},e_{38}\}$;
вершина $v_{12}$: $\{e_{18},e_{37},e_{38}\}$.

Множество треугольных циклов в графе:

$c_1=\{e_1,e_2,e_8\} \leftrightarrow \{v_2,v_3,v_1\}$;  $c_2=\{e_1,e_3,e_9\} \leftrightarrow \{v_2,v_5,v_1\}$;
$c_3=\{e_1,e_5,e_{10}\} \leftrightarrow \{v_2,v_8,v_1\}$;  $c_4=\{e_1,e_7,e_{12}\} \leftrightarrow \{v_2,v_{11},v_1\}$;
$c_5=\{e_2,e_3,e_{13}\} \leftrightarrow \{v_3,v_5,v_1\}$;  $c_6=\{e_2,e_4,e_{14}\} \leftrightarrow \{v_3,v_7,v_1\}$;
$c_7=\{e_2,e_5,e_{15}\} \leftrightarrow \{v_3,v_8,v_1\}$;  $c_8=\{e_2,e_7,e_{17}\} \leftrightarrow \{v_3,v_{11},v_1\}$;
$c_9=\{e_3,e_4,e_{24}\} \leftrightarrow \{v_5,v_7,v_1\}$;  $c_{10}=\{e_3,e_6,e_{25}\} \leftrightarrow \{v_5,v_{10},v_1\}$;
$c_{11}=\{e_4,e_5,e_{31}\} \leftrightarrow \{v_7,v_8,v_1\}$;  $c_{12}=\{e_4,e_6,e_{33}\} \leftrightarrow \{v_7,v_{10},v_1\}$;
$c_{13}=\{e_5,e_6,e_{34}\} \leftrightarrow \{v_8,v_{10},v_1\}$;  $c_{14}=\{e_5,e_7,e_{35}\} \leftrightarrow \{v_8,v_{11},v_1\}$;
$c_{15}=\{e_8,e_9,e_{13}\} \leftrightarrow \{v_3,v_5,v_2\}$;  $c_{16}=\{e_8,e_{10},e_{15}\} \leftrightarrow \{v_3,v_8,v_2\}$;
$c_{17}=\{e_8,e_{11},e_{16}\} \leftrightarrow \{v_3,v_9,v_2\}$;  $c_{18}=\{e_8,e_{12},e_{17}\} \leftrightarrow \{v_3,v_{11},v_2\}$;
$c_{19}=\{e_{10},e_{12},e_{35}\} \leftrightarrow \{v_8,v_{11},v_2\}$;  $c_{20}=\{e_{13},e_{14},e_{24}\} \leftrightarrow \{v_5,v_7,v_3\}$;
$c_{21}=\{e_{14},e_{15},e_{31}\} \leftrightarrow \{v_7,v_8,v_3\}$;  $c_{22}=\{e_{14},e_{16},e_{32}\} \leftrightarrow \{v_7,v_9,v_3\}$;
$c_{23}=\{e_{15},e_{17},e_{35}\} \leftrightarrow \{v_8,v_{11},v_3\}$;  $c_{24}=\{e_{17},e_{18},e_{38}\} \leftrightarrow \{v_{11},v_{12},v_3\}$;
$c_{25}=\{e_{19},e_{20},e_{31}\} \leftrightarrow \{v_7,v_8,v_4\}$;  $c_{26}=\{e_{19},e_{21},e_{33}\} \leftrightarrow \{v_7,v_{10},v_4\}$;
$c_{27}=\{e_{20},e_{21},e_{34}\} \leftrightarrow \{v_8,v_{10},v_4\}$;  $c_{28}=\{e_{20},e_{22},e_{35}\} \leftrightarrow \{v_8,v_{11},v_4\}$;
$c_{29}=\{e_{23},e_{24},e_{26}\} \leftrightarrow \{v_6,v_7,v_5\}$;  $c_{30}=\{e_{23},e_{25},e_{29}\} \leftrightarrow \{v_6,v_{10},v_5\}$;
$c_{31}=\{e_{24},e_{25},e_{33}\} \leftrightarrow \{v_7,v_{10},v_5\}$;  $c_{32}=\{e_{26},e_{27},e_{31}\} \leftrightarrow \{v_7,v_8,v_6\}$;
$c_{33}=\{e_{26},e_{28},e_{32}\} \leftrightarrow \{v_7,v_9,v_6\}$;  $c_{34}=\{e_{26},e_{29},e_{33}\} \leftrightarrow \{v_7,v_{10},v_6\}$;
$c_{35}=\{e_{27},e_{29},e_{34}\} \leftrightarrow \{v_8,v_{10},v_6\}$;  $c_{36}=\{e_{27},e_{30},e_{35}\} \leftrightarrow \{v_8,v_{11},v_6\}$;
$c_{37}=\{e_{28},e_{29},e_{36}\} \leftrightarrow \{v_9,v_{10},v_6\}$;  $c_{38}=\{e_{31},e_{33},e_{34}\} \leftrightarrow \{v_8,v_{10},v_7\}$;
$c_{39}=\{e_{32},e_{33},e_{36}\} \leftrightarrow \{v_9,v_{10},v_7\}$.

Определяем количество треугольных циклов, проходящих по ребру, то есть, строим вектор $P_e$ по ребру.

**Итерация 1**

$P_e = <4,5,4,4,5,3,3,5,2,3,1,3,3,4,4,2,4,1,2,3,2,1,2,4,3,4,3,2,4,1,5,3,6,4,5,2,0,1>$;

В векторе циклов по ребру, по первому ребру проходит 4 треугольных цикла, по второму ребру проходит 5 циклов, по третьему – 4 цикла и т.д. Вычисляем максимальное и минимальное значения элементов в векторе по ребрам.

MAX = 6; MIN = 1

Определяем ребра, по которым проходит минимальное количество треугольных циклов равное MIN. Это ребра $e_{11}, e_{18}, e_{22}, e_{30}, e_{38}$. Удаляем из множества циклов те, которые имеют в своем составе выбранные ребра:

$c_{17}=\{e_8,e_{11},e_{16}\} \leftrightarrow \{v_3,v_9,v_2\}$;

$c_{24}=\{e_{17},e_{18},e_{38}\} \leftrightarrow \{v_{11},v_{12},v_3\}$;
$c_{28}=\{e_{20},e_{22},e_{35}\} \leftrightarrow \{v_8,v_{11},v_4\}$;
$c_{36}=\{e_{27},e_{30},e_{35}\} \leftrightarrow \{v_8,v_{11},v_6\}$.

Множество треугольных циклов в графе:

$c_1=\{e_1,e_2,e_8\} \leftrightarrow \{v_2,v_3,v_1\}$;       $c_2=\{e_1,e_3,e_9\} \leftrightarrow \{v_2,v_5,v_1\}$;
$c_3=\{e_1,e_5,e_{10}\} \leftrightarrow \{v_2,v_8,v_1\}$;    $c_4=\{e_1,e_7,e_{12}\} \leftrightarrow \{v_2,v_{11},v_1\}$;
$c_5=\{e_2,e_3,e_{13}\} \leftrightarrow \{v_3,v_5,v_1\}$;    $c_6=\{e_2,e_4,e_{14}\} \leftrightarrow \{v_3,v_7,v_1\}$;
$c_7=\{e_2,e_5,e_{15}\} \leftrightarrow \{v_3,v_8,v_1\}$;    $c_8=\{e_2,e_7,e_{17}\} \leftrightarrow \{v_3,v_{11},v_1\}$;
$c_9=\{e_3,e_4,e_{24}\} \leftrightarrow \{v_5,v_7,v_1\}$;    $c_{10}=\{e_3,e_6,e_{25}\} \leftrightarrow \{v_5,v_{10},v_1\}$;
$c_{11}=\{e_4,e_5,e_{31}\} \leftrightarrow \{v_7,v_8,v_1\}$; $c_{12}=\{e_4,e_6,e_{33}\} \leftrightarrow \{v_7,v_{10},v_1\}$;
$c_{13}=\{e_5,e_6,e_{34}\} \leftrightarrow \{v_8,v_{10},v_1\}$; $c_{14}=\{e_5,e_7,e_{35}\} \leftrightarrow \{v_8,v_{11},v_1\}$;
$c_{15}=\{e_8,e_9,e_{13}\} \leftrightarrow \{v_3,v_5,v_2\}$; $c_{16}=\{e_8,e_{10},e_{15}\} \leftrightarrow \{v_3,v_8,v_2\}$;
$c_{18}=\{e_8,e_{12},e_{17}\} \leftrightarrow \{v_3,v_{11},v_2\}$; $c_{19}=\{e_{10},e_{12},e_{35}\} \leftrightarrow \{v_8,v_{11},v_2\}$;
$c_{20}=\{e_{13},e_{14},e_{24}\} \leftrightarrow \{v_5,v_7,v_3\}$; $c_{21}=\{e_{14},e_{15},e_{31}\} \leftrightarrow \{v_7,v_8,v_3\}$;
$c_{22}=\{e_{14},e_{16},e_{32}\} \leftrightarrow \{v_7,v_9,v_3\}$; $c_{23}=\{e_{15},e_{17},e_{35}\} \leftrightarrow \{v_8,v_{11},v_3\}$;
$c_{25}=\{e_{19},e_{20},e_{31}\} \leftrightarrow \{v_7,v_8,v_4\}$; $c_{26}=\{e_{19},e_{21},e_{33}\} \leftrightarrow \{v_7,v_{10},v_4\}$;
$c_{27}=\{e_{20},e_{21},e_{34}\} \leftrightarrow \{v_8,v_{10},v_4\}$; $c_{29}=\{e_{23},e_{24},e_{26}\} \leftrightarrow \{v_6,v_7,v_5\}$;
$c_{30}=\{e_{23},e_{25},e_{29}\} \leftrightarrow \{v_6,v_{10},v_5\}$; $c_{31}=\{e_{24},e_{25},e_{33}\} \leftrightarrow \{v_7,v_{10},v_5\}$;
$c_{32}=\{e_{26},e_{27},e_{31}\} \leftrightarrow \{v_7,v_8,v_6\}$; $c_{33}=\{e_{26},e_{28},e_{32}\} \leftrightarrow \{v_7,v_9,v_6\}$;
$c_{34}=\{e_{26},e_{29},e_{33}\} \leftrightarrow \{v_7,v_{10},v_6\}$; $c_{35}=\{e_{27},e_{29},e_{34}\} \leftrightarrow \{v_8,v_{10},v_6\}$;
$c_{37}=\{e_{28},e_{29},e_{36}\} \leftrightarrow \{v_9,v_{10},v_6\}$; $c_{38}=\{e_{31},e_{33},e_{34}\} \leftrightarrow \{v_8,v_{10},v_7\}$;
$c_{39}=\{e_{32},e_{33},e_{36}\} \rightarrow \{v_9,v_{10},v_7\}$.

**Итерация 2**

Строим вектор по ребру $P_e =$

$= <4,5,4,4,5,3,3,4,2,3,0,3,3,4,4,1,3,0,2,2,2,0,2,4,3,4,2,2,4,0,5,3,6,4,3,2,0,0>$;

MAX = 6; MIN = 1

Определяем ребра, по которым проходит минимальное количество треугольных циклов равное MIN. Это ребро $e_{16}$. Удаляем из множества цикл $c_{22}$ имеющий в своем составе ребро $e_{16}$:

$c_{22}=\{e_{14},e_{16},e_{32}\} \rightarrow \{v_7,v_9,v_3\}$;

После удаления множество треугольных циклов имеет вид:

$c_1=\{e_1,e_2,e_8\} \leftrightarrow \{v_2,v_3,v_1\}$;       $c_2=\{e_1,e_3,e_9\} \leftrightarrow \{v_2,v_5,v_1\}$;
$c_3=\{e_1,e_5,e_{10}\} \leftrightarrow \{v_2,v_8,v_1\}$;    $c_4=\{e_1,e_7,e_{12}\} \leftrightarrow \{v_2,v_{11},v_1\}$;
$c_5=\{e_2,e_3,e_{13}\} \leftrightarrow \{v_3,v_5,v_1\}$;    $c_6=\{e_2,e_4,e_{14}\} \leftrightarrow \{v_3,v_7,v_1\}$;
$c_7=\{e_2,e_5,e_{15}\} \leftrightarrow \{v_3,v_8,v_1\}$;    $c_8=\{e_2,e_7,e_{17}\} \leftrightarrow \{v_3,v_{11},v_1\}$;
$c_9=\{e_3,e_4,e_{24}\} \leftrightarrow \{v_5,v_7,v_1\}$;    $c_{10}=\{e_3,e_6,e_{25}\} \leftrightarrow \{v_5,v_{10},v_1\}$;
$c_{11}=\{e_4,e_5,e_{31}\} \leftrightarrow \{v_7,v_8,v_1\}$; $c_{12}=\{e_4,e_6,e_{33}\} \leftrightarrow \{v_7,v_{10},v_1\}$;
$c_{13}=\{e_5,e_6,e_{34}\} \leftrightarrow \{v_8,v_{10},v_1\}$; $c_{14}=\{e_5,e_7,e_{35}\} \leftrightarrow \{v_8,v_{11},v_1\}$;
$c_{15}=\{e_8,e_9,e_{13}\} \leftrightarrow \{v_3,v_5,v_2\}$; $c_{16}=\{e_8,e_{10},e_{15}\} \leftrightarrow \{v_3,v_8,v_2\}$;
$c_{18}=\{e_8,e_{12},e_{17}\} \leftrightarrow \{v_3,v_{11},v_2\}$; $c_{19}=\{e_{10},e_{12},e_{35}\} \leftrightarrow \{v_8,v_{11},v_2\}$;
$c_{20}=\{e_{13},e_{14},e_{24}\} \leftrightarrow \{v_5,v_7,v_3\}$; $c_{21}=\{e_{14},e_{15},e_{31}\} \leftrightarrow \{v_7,v_8,v_3\}$;
$c_{23}=\{e_{15},e_{17},e_{35}\} \leftrightarrow \{v_8,v_{11},v_3\}$; $c_{25}=\{e_{19},e_{20},e_{31}\} \leftrightarrow \{v_7,v_8,v_4\}$;
$c_{26}=\{e_{19},e_{21},e_{33}\} \leftrightarrow \{v_7,v_{10},v_4\}$; $c_{27}=\{e_{20},e_{21},e_{34}\} \leftrightarrow \{v_8,v_{10},v_4\}$;

$c_{29}=\{e_{23},e_{24},e_{26}\} \leftrightarrow \{v_6,v_7,v_5\};$   $c_{30}=\{e_{23},e_{25},e_{29}\} \leftrightarrow \{v_6,v_{10},v_5\};$
$c_{31}=\{e_{24},e_{25},e_{33}\} \leftrightarrow \{v_7,v_{10},v_5\};$   $c_{32}=\{e_{26},e_{27},e_{31}\} \leftrightarrow \{v_7,v_8,v_6\};$
$c_{33}=\{e_{26},e_{28},e_{32}\} \leftrightarrow \{v_7,v_9,v_6\};$   $c_{34}=\{e_{26},e_{29},e_{33}\} \leftrightarrow \{v_7,v_{10},v_6\};$
$c_{35}=\{e_{27},e_{29},e_{34}\} \leftrightarrow \{v_8,v_{10},v_6\};$   $c_{37}=\{e_{28},e_{29},e_{36}\} \leftrightarrow \{v_9,v_{10},v_6\};$
$c_{38}=\{e_{31},e_{33},e_{34}\} \leftrightarrow \{v_8,v_{10},v_7\};$   $c_{39}=\{e_{32},e_{33},e_{36}\} \leftrightarrow \{v_9,v_{10},v_7\}.$

**Итерация 3**

$P_e$ = <4,5,4,4,5,3,3,4,2,3,0,3,3,3,4,0,3,0,2,2,2,0,2,4,3,4,2,2,4,0,5,2,6,4,3,2,0,0>;

MAX = 6; MIN = 2.

Определяем ребра, по которым проходит минимальное количество треугольных циклов равное MIN. Это ребра $e_9,e_{19},e_{20},e_{21},e_{23},e_{27},e_{28},e_{32},e_{36}$. Удаляем из множества циклов те, которые имеют в своем составе выбранные ребра:

$c_2=\{e_1,e_3,e_9\} \leftrightarrow \{v_2,v_5,v_1\};$
$c_{15}=\{e_8,e_9,e_{13}\} \leftrightarrow \{v_3,v_5,v_2\};$
$c_{25}=\{e_{19},e_{20},e_{31}\} \leftrightarrow \{v_7,v_8,v_4\};$
$c_{26}=\{e_{19},e_{21},e_{33}\} \leftrightarrow \{v_7,v_{10},v_4\};$
$c_{27}=\{e_{20},e_{21},e_{34}\} \leftrightarrow \{v_8,v_{10},v_4\};$
$c_{29}=\{e_{23},e_{24},e_{26}\} \leftrightarrow \{v_6,v_7,v_5\};$
$c_{30}=\{e_{23},e_{25},e_{29}\} \leftrightarrow \{v_6,v_{10},v_5\};$
$c_{32}=\{e_{26},e_{27},e_{31}\} \leftrightarrow \{v_7,v_8,v_6\};$
$c_{33}=\{e_{26},e_{28},e_{32}\} \leftrightarrow \{v_7,v_9,v_6\};$
$c_{35}=\{e_{27},e_{29},e_{34}\} \leftrightarrow \{v_8,v_{10},v_6\};$
$c_{37}=\{e_{28},e_{29},e_{36}\} \leftrightarrow \{v_9,v_{10},v_6\};$
$c_{39}=\{e_{32},e_{33},e_{36}\} \leftrightarrow \{v_9,v_{10},v_7\}.$

После удаления множество треугольных циклов имеет вид:

$c_1=\{e_1,e_2,e_8\} \leftrightarrow \{v_2,v_3,v_1\};$   $c_3=\{e_1,e_5,e_{10}\} \leftrightarrow \{v_2,v_8,v_1\};$
$c_4=\{e_1,e_7,e_{12}\} \leftrightarrow \{v_2,v_{11},v_1\};$   $c_5=\{e_2,e_3,e_{13}\} \leftrightarrow \{v_3,v_5,v_1\};$
$c_6=\{e_2,e_4,e_{14}\} \leftrightarrow \{v_3,v_7,v_1\};$   $c_7=\{e_2,e_5,e_{15}\} \leftrightarrow \{v_3,v_8,v_1\};$
$c_8=\{e_2,e_7,e_{17}\} \leftrightarrow \{v_3,v_{11},v_1\};$   $c_9=\{e_3,e_4,e_{24}\} \leftrightarrow \{v_5,v_7,v_1\};$
$c_{10}=\{e_3,e_6,e_{25}\} \leftrightarrow \{v_5,v_{10},v_1\};$   $c_{11}=\{e_4,e_5,e_{31}\} \leftrightarrow \{v_7,v_8,v_1\};$
$c_{12}=\{e_4,e_6,e_{33}\} \leftrightarrow \{v_7,v_{10},v_1\};$   $c_{13}=\{e_5,e_6,e_{34}\} \leftrightarrow \{v_8,v_{10},v_1\};$
$c_{14}=\{e_5,e_7,e_{35}\} \leftrightarrow \{v_8,v_{11},v_1\};$   $c_{16}=\{e_8,e_{10},e_{15}\} \leftrightarrow \{v_3,v_8,v_2\};$
$c_{18}=\{e_8,e_{12},e_{17}\} \leftrightarrow \{v_3,v_{11},v_2\};$   $c_{19}=\{e_{10},e_{12},e_{35}\} \leftrightarrow \{v_8,v_{11},v_2\};$
$c_{20}=\{e_{13},e_{14},e_{24}\} \leftrightarrow \{v_5,v_7,v_3\};$   $c_{21}=\{e_{14},e_{15},e_{31}\} \leftrightarrow \{v_7,v_8,v_3\};$
$c_{23}=\{e_{15},e_{17},e_{35}\} \leftrightarrow \{v_8,v_{11},v_3\};$   $c_{31}=\{e_{24},e_{25},e_{33}\} \leftrightarrow \{v_7,v_{10},v_5\};$
$c_{34}=\{e_{26},e_{29},e_{33}\} \leftrightarrow \{v_7,v_{10},v_6\};$   $c_{38}=\{e_{31},e_{33},e_{34}\} \leftrightarrow \{v_8,v_{10},v_7\}.$

**Итерация 4**

$P_e$ = <3,5,3,4,5,3,3,3,0,3,0,3,2,3,4,0,3,0,0,0,0,0,0,3,2,1,0,0,1,0,3,0,4,2,3,0,0,0>;

MAX = 5; MIN = 1.

Определяем ребра, по которым проходит минимальное количество треугольных циклов равное MIN. Это ребра $e_{26},c_{29}$. Удаляем из множества цикл $c_{34}$ имеющий в своем составе ребра $e_{26},c_{29}$:

$c_{34}=\{e_{26},e_{29},e_{33}\} \rightarrow \{v_7,v_{10},v_6\}.$

После удаления множество треугольных циклов имеет вид:

$c_1=\{e_1,e_2,e_8\} \leftrightarrow \{v_2,v_3,v_1\}$;   $c_3=\{e_1,e_5,e_{10}\} \leftrightarrow \{v_2,v_8,v_1\}$;
$c_4=\{e_1,e_7,e_{12}\} \leftrightarrow \{v_2,v_{11},v_1\}$;   $c_5=\{e_2,e_3,e_{13}\} \leftrightarrow \{v_3,v_5,v_1\}$;
$c_6=\{e_2,e_4,e_{14}\} \leftrightarrow \{v_3,v_7,v_1\}$;   $c_7=\{e_2,e_5,e_{15}\} \leftrightarrow \{v_3,v_8,v_1\}$;
$c_8=\{e_2,e_7,e_{17}\} \leftrightarrow \{v_3,v_{11},v_1\}$;   $c_9=\{e_3,e_4,e_{24}\} \leftrightarrow \{v_5,v_7,v_1\}$;
$c_{10}=\{e_3,e_6,e_{25}\} \leftrightarrow \{v_5,v_{10},v_1\}$;   $c_{11}=\{e_4,e_5,e_{31}\} \leftrightarrow \{v_7,v_8,v_1\}$;
$c_{12}=\{e_4,e_6,e_{33}\} \leftrightarrow \{v_7,v_{10},v_1\}$;   $c_{13}=\{e_5,e_6,e_{34}\} \leftrightarrow \{v_8,v_{10},v_1\}$;
$c_{14}=\{e_5,e_7,e_{35}\} \leftrightarrow \{v_8,v_{11},v_1\}$;   $c_{16}=\{e_8,e_{10},e_{15}\} \leftrightarrow \{v_3,v_8,v_2\}$;
$c_{18}=\{e_8,e_{12},e_{17}\} \leftrightarrow \{v_3,v_{11},v_2\}$;   $c_{19}=\{e_{10},e_{12},e_{35}\} \leftrightarrow \{v_8,v_{11},v_2\}$;
$c_{20}=\{e_{13},e_{14},e_{24}\} \leftrightarrow \{v_5,v_7,v_3\}$;   $c_{21}=\{e_{14},e_{15},e_{31}\} \leftrightarrow \{v_7,v_8,v_3\}$;
$c_{23}=\{e_{15},e_{17},e_{35}\} \leftrightarrow \{v_8,v_{11},v_3\}$;   $c_{31}=\{e_{24},e_{25},e_{33}\} \leftrightarrow \{v_7,v_{10},v_5\}$;
$c_{38}=\{e_{31},e_{33},e_{34}\} \leftrightarrow \{v_8,v_{10},v_7\}$.

**Итерация 5**

$P_e$ = <3,5,3,4,5,3,3,3,0,3,0,3,2,3,4,0,3,0,0,0,0,0,0,3,2,0,0,0,0,0,3,0,3,2,3,0,0,0>;

MAX = 5; MIN = 2.

Определяем ребра, по которым проходит минимальное количество треугольных циклов равное MIN. Это ребра $e_{13},e_{25},e_{34}$. Удаляем из множества циклов те, которые имеют в своем составе выбранные ребра:

$c_5=\{e_2,e_3,e_{13}\} \leftrightarrow \{v_3,v_5,v_1\}$;
$c_{10}=\{e_3,e_6,e_{25}\} \leftrightarrow \{v_5,v_{10},v_1\}$;
$c_{13}=\{e_5,e_6,e_{34}\} \leftrightarrow \{v_8,v_{10},v_1\}$;
$c_{20}=\{e_{13},e_{14},e_{24}\} \leftrightarrow \{v_5,v_7,v_3\}$;
$c_{31}=\{e_{24},e_{25},e_{33}\} \leftrightarrow \{v_7,v_{10},v_5\}$;
$c_{38}=\{e_{31},e_{33},e_{34}\} \leftrightarrow \{v_8,v_{10},v_7\}$.

После удаления множество треугольных циклов имеет вид:

$c_1=\{e_1,e_2,e_8\} \leftrightarrow \{v_2,v_3,v_1\}$;   $c_3=\{e_1,e_5,e_{10}\} \leftrightarrow \{v_2,v_8,v_1\}$;
$c_4=\{e_1,e_7,e_{12}\} \leftrightarrow \{v_2,v_{11},v_1\}$;   $c_6=\{e_2,e_4,e_{14}\} \leftrightarrow \{v_3,v_7,v_1\}$;
$c_7=\{e_2,e_5,e_{15}\} \leftrightarrow \{v_3,v_8,v_1\}$;   $c_8=\{e_2,e_7,e_{17}\} \leftrightarrow \{v_3,v_{11},v_1\}$;
$c_9=\{e_3,e_4,e_{24}\} \leftrightarrow \{v_5,v_7,v_1\}$;   $c_{11}=\{e_4,e_5,e_{31}\} \leftrightarrow \{v_7,v_8,v_1\}$;
$c_{12}=\{e_4,e_6,e_{33}\} \leftrightarrow \{v_7,v_{10},v_1\}$;   $c_{14}=\{e_5,e_7,e_{35}\} \leftrightarrow \{v_8,v_{11},v_1\}$;
$c_{16}=\{e_8,e_{10},e_{15}\} \leftrightarrow \{v_3,v_8,v_2\}$;   $c_{18}=\{e_8,e_{12},e_{17}\} \leftrightarrow \{v_3,v_{11},v_2\}$;
$c_{19}=\{e_{10},e_{12},e_{35}\} \leftrightarrow \{v_8,v_{11},v_2\}$;   $c_{21}=\{e_{14},e_{15},e_{31}\} \leftrightarrow \{v_7,v_8,v_3\}$;
$c_{23}=\{e_{15},e_{17},e_{35}\} \leftrightarrow \{v_8,v_{11},v_3\}$.

**Итерация 6**

$P_e$ = <3,4,1,4,4,1,3,3,0,3,0,3,0,2,4,0,3,0,0,0,0,0,0,1,0,0,0,0,0,0,2,0,1,0,3,0,0,0>;

MAX = 4; MIN = 1.

Определяем ребра, по которым проходит минимальное количество треугольных циклов равное MIN. Это ребра $e_3,e_6,e_{24},e_{33}$. Удаляем из множества циклов те, которые имеют в своем составе выбранные ребра:

$c_9=\{e_3,e_4,e_{24}\} \leftrightarrow \{v_5,v_7,v_1\}$;

$c_{12}=\{e_4,e_6,e_{33}\} \leftrightarrow \{v_7,v_{10},v_1\}$.

После удаления множество треугольных циклов имеет вид:

$c_1=\{e_1,e_2,e_8\} \leftrightarrow \{v_2,v_3,v_1\}$;   $c_3=\{e_1,e_5,e_{10}\} \leftrightarrow \{v_2,v_8,v_1\}$;
$c_4=\{e_1,e_7,e_{12}\} \leftrightarrow \leftrightarrow \{v_2,v_{11},v_1\}$;   $c_6=\{e_2,e_4,e_{14}\} \leftrightarrow \{v_3,v_7,v_1\}$;
$c_7=\{e_2,e_5,e_{15}\} \leftrightarrow \{v_3,v_8,v_1\}$;   $c_8=\{e_2,e_7,e_{17}\} \leftrightarrow \{v_3,v_{11},v_1\}$;
$c_{11}=\{e_4,e_5,e_{31}\} \leftrightarrow \{v_7,v_8,v_1\}$;   $c_{14}=\{e_5,e_7,e_{35}\} \leftrightarrow \{v_8,v_{11},v_1\}$;
$c_{16}=\{e_8,e_{10},e_{15}\} \leftrightarrow \{v_3,v_8,v_2\}$;   $c_{18}=\{e_8,e_{12},e_{17}\} \leftrightarrow \{v_3,v_{11},v_2\}$;
$c_{19}=\{e_{10},e_{12},e_{35}\} \leftrightarrow \{v_8,v_{11},v_2\}$;   $c_{21}=\{e_{14},e_{15},e_{31}\} \leftrightarrow \{v_7,v_8,v_3\}$;
$c_{23}=\{e_{15},e_{17},e_{35}\} \leftrightarrow \{v_8,v_{11},v_3\}$.

**Итерация 7**

$P_e$ = <3,4,0,2,4,0,3,3,0,3,0,3,0,2,4,0,3,0,0,0,0,0,0,0,0,0,0,2,0,0,0,3,0,0,0>;

MAX = 4; MIN = 2.

Определяем ребра, по которым проходит минимальное количество треугольных циклов равное MIN. Это ребра $e_4,e_{14},e_{31}$. Удаляем из множества циклов те, которые имеют в своем составе выбранные ребра:

$c_6=\{e_2,e_4,e_{14}\} \leftrightarrow \{v_3,v_7,v_1\}$;
$c_{11}=\{e_4,e_5,e_{31}\} \leftrightarrow \{v_7,v_8,v_1\}$;
$c_{21}=\{e_{14},e_{15},e_{31}\} \leftrightarrow \{v_7,v_8,v_3\}$.

После удаления множество треугольных циклов имеет вид:

$c_1=\{e_1,e_2,e_8\} \leftrightarrow \{v_2,v_3,v_1\}$;   $c_3=\{e_1,e_5,e_{10}\} \leftrightarrow \{v_2,v_8,v_1\}$;
$c_4=\{e_1,e_7,e_{12}\} \leftrightarrow \{v_2,v_{11},v_1\}$;   $c_7=\{e_2,e_5,e_{15}\} \leftrightarrow \{v_3,v_8,v_1\}$;
$c_8=\{e_2,e_7,e_{17}\} \leftrightarrow \{v_3,v_{11},v_1\}$;   $c_{14}=\{e_5,e_7,e_{35}\} \leftrightarrow \{v_8,v_{11},v_1\}$;
$c_{16}=\{e_8,e_{10},e_{15}\} \leftrightarrow \{v_3,v_8,v_2\}$;   $c_{18}=\{e_8,e_{12},e_{17}\} \leftrightarrow \{v_3,v_{11},v_2\}$;
$c_{19}=\{e_{10},e_{12},e_{35}\} \leftrightarrow \{v_8,v_{11},v_2\}$;   $c_{23}=\{e_{15},e_{17},e_{35}\} \leftrightarrow \{v_8,v_{11},v_3\}$.

**Итерация 8**

$P_e$ = <3,3,0,0,3,0,3,3,0,3,0,3,0,0,3,0,3,0,0,0,0,0,0,0,0,0,0,0,0,0,0,3,0,0,0>;

Max = 3; Min = 3;

Max = Min (условие окончания работы алгоритма).

Объединив все вершины, входящие в выделенные циклы, получим множество вершин клики $K_5=\{v_1,v_2,v_3,v_8,v_{11}\}$.

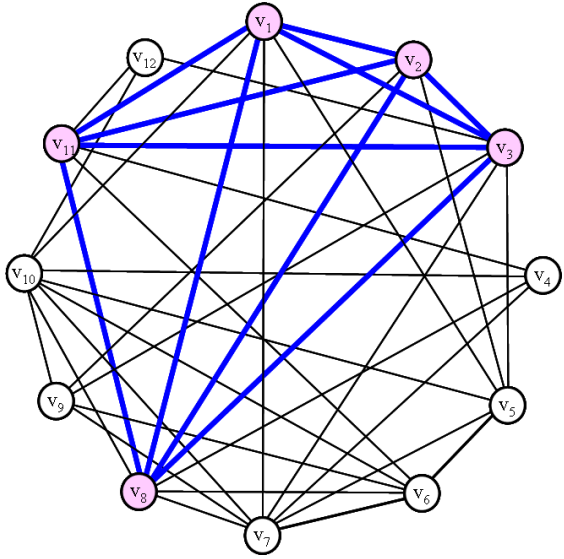

Рис. 3.2. Клика К$_5$ графа G$_3$.

### 3.2. Выделение максимальной клики в графе G$_4$

Рассмотрим случай наличия нескольких максимальных клик в графе G$_3$. Предполагается, что все клики одинаковой длины и эта длина максимальна.

Количество вершин графа = 27, количество ребер графа = 138, количество треугольных циклов графа = 173.

Смежность графа:
вершина $v_1$: $\{v_2,v_4,v_5,v_9,v_{14},v_{16},v_{17},v_{20},v_{21},v_{23},v_{24},v_{25},v_{27}\}$;
вершина $v_2$: $\{v_1,v_8,v_9,v_{13},v_{18},v_{19},v_{20},v_{22},v_{24}\}$;
вершина $v_3$: $\{v_6,v_{15},v_{19},v_{21},v_{23},v_{26}\}$;
вершина $v_4$: $\{v_1,v_7,v_8,v_9,v_{12},v_{14},v_{15},v_{18},v_{21}\}$;
вершина $v_5$: $\{v_1,v_6,v_8,v_{10},v_{13},v_{14},v_{16},v_{19},v_{20},v_{22},v_{25},v_{27}\}$;
вершина $v_6$: $\{v_3,v_5,v_9,v_{11},v_{13},v_{21},v_{27}\}$;
вершина $v_7$: $\{v_4,v_8,v_{11},v_{13},v_{16},v_{17},v_{18},v_{19},v_{20},v_{22},v_{23}\}$;
вершина $v_8$: $\{v_2,v_4,v_5,v_7,v_{10},v_{17},v_{19},v_{20},v_{21},v_{22},v_{23},v_{24},v_{27}\}$;
вершина $v_9$: $\{v_1,v_2,v_4,v_6,v_{13},v_{16},v_{18},v_{20},v_{23},v_{24},v_{25},v_{26}\}$;
вершина $v_{10}$: $\{v_5,v_8,v_{13},v_{20}\}$;
вершина $v_{11}$: $\{v_6,v_7,v_{14},v_{15},v_{16},v_{17},v_{19},v_{22},v_{24},v_{27}\}$;
вершина $v_{12}$: $\{v_4,v_{17},v_{20},v_{21},v_{23},v_{27}\}$;
вершина $v_{13}$: $\{v_2,v_5,v_6,v_7,v_9,v_{10},v_{15},v_{16},v_{20},v_{21},v_{22},v_{26},v_{27}\}$;
вершина $v_{14}$: $\{v_1,v_4,v_5,v_{11},v_{15},v_{17},v_{18},v_{19},v_{22},v_{27}\}$;
вершина $v_{15}$: $\{v_3,v_4,v_{11},v_{13},v_{14},v_{16},v_{17},v_{18},v_{20},v_{22},v_{23},v_{25},v_{26},v_{27}\}$;
вершина $v_{16}$: $\{v_1,v_5,v_7,v_9,v_{11},v_{13},v_{15},v_{17},v_{21},v_{22},v_{25}\}$;
вершина $v_{17}$: $\{v_1,v_7,v_8,v_{11},v_{12},v_{14},v_{15},v_{16},v_{19},v_{22}\}$;
вершина $v_{18}$: $\{v_2,v_4,v_7,v_9,v_{14},v_{15},v_{20},v_{22},v_{23},v_{26}\}$;
вершина $v_{19}$: $\{v_2,v_3,v_5,v_7,v_8,v_{11},v_{14},v_{17},v_{20},v_{21},v_{25}\}$;
вершина $v_{20}$: $\{v_1,v_2,v_5,v_7,v_8,v_9,v_{10},v_{12},v_{13},v_{15},v_{18},v_{19},v_{22},v_{24}\}$;

вершина v₂₁: {v₁,v₃,v₄,v₆,v₈,v₁₂,v₁₃,v₁₆,v₁₉,v₂₂,v₂₅,v₂₆};
вершина v₂₂: {v₂,v₅,v₇,v₈,v₁₁,v₁₃,v₁₄,v₁₅,v₁₆,v₁₇,v₁₈,v₂₀,v₂₁,v₂₅};
вершина v₂₃: {v₁,v₃,v₇,v₈,v₉,v₁₂,v₁₅,v₁₈,v₂₄,v₂₅};
вершина v₂₄: {v₁,v₂,v₈,v₉,v₁₁,v₂₀,v₂₃};
вершина v₂₅: {v₁,v₅,v₉,v₁₅,v₁₆,v₁₉,v₂₁,v₂₂,v₂₃,v₂₇};
вершина v₂₆: {v₃,v₉,v₁₃,v₁₅,v₁₈,v₂₁,v₂₇};
вершина v₂₇: {v₁,v₅,v₆,v₈,v₁₁,v₁₂,v₁₃,v₁₄,v₁₅,v₂₅,v₂₆}.

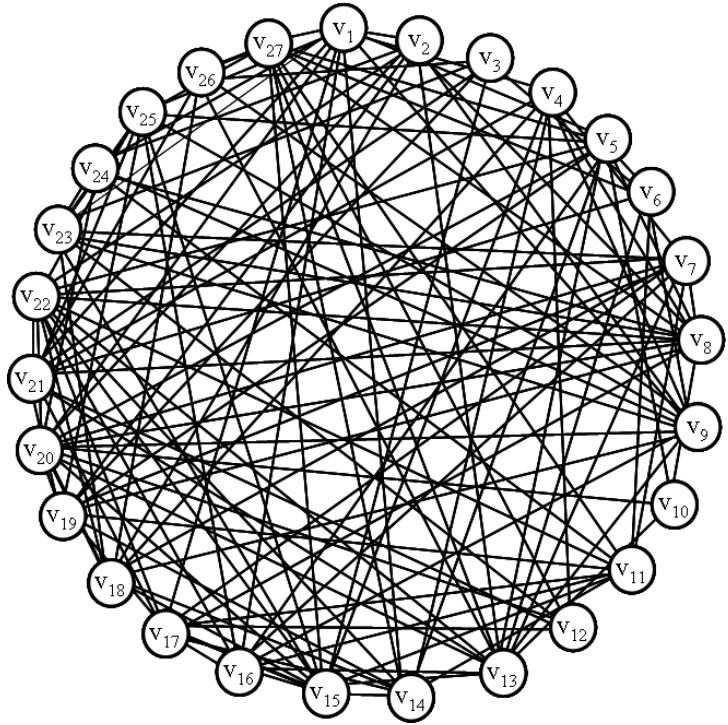

Рис. 3.3. Граф $G_4$.

Инцидентность графа:

вершина $v_1$: {$e_1,e_2,e_3,e_4,e_5,e_6,e_7,e_8,e_9,e_{10},e_{11},e_{12},e_{13}$};
вершина $v_2$: {$e_1,e_{14},e_{15},e_{16},e_{17},e_{18},e_{19},e_{20},e_{21}$};
вершина $v_3$: {$e_{22},e_{23},e_{24},e_{25},e_{26},e_{27}$};
вершина $v_4$: {$e_2,e_{28},e_{29},e_{30},e_{31},e_{32},e_{33},e_{34},e_{35}$};
вершина $v_5$: {$e_3,e_{36},e_{37},e_{38},e_{39},e_{40},e_{41},e_{42},e_{43},e_{44},e_{45},e_{46}$};
вершина $v_6$: {$e_{22},e_{36},e_{47},e_{48},e_{49},e_{50},e_{51}$};
вершина $v_7$: {$e_{28},e_{52},e_{53},e_{54},e_{55},e_{56},e_{57},e_{58},e_{59},e_{60},e_{61}$};
вершина $v_8$: {$e_{14},e_{29},e_{37},e_{52},e_{62},e_{63},e_{64},e_{65},e_{66},e_{67},e_{68},e_{69},e_{70}$};
вершина $v_9$: {$e_4,e_{15},e_{30},e_{47},e_{71},e_{72},e_{73},e_{74},e_{75},e_{76},e_{77},e_{78}$};
вершина $v_{10}$: {$e_{38},e_{62},e_{79},e_{80}$};
вершина $v_{11}$: {$e_{48},e_{53},e_{81},e_{82},e_{83},e_{84},e_{85},e_{86},e_{87},e_{88}$};
вершина $v_{12}$: {$e_{31},e_{89},e_{90},e_{91},e_{92},e_{93}$};
вершина $v_{13}$: {$e_{16},e_{39},e_{49},e_{54},e_{71},e_{79},e_{94},e_{95},e_{96},e_{97},e_{98},e_{99},e_{100}$};
вершина $v_{14}$: {$e_5,e_{32},e_{40},e_{81},e_{101},e_{102},e_{103},e_{104},e_{105},e_{106}$};
вершина $v_{15}$: {$e_{23},e_{33},e_{82},e_{94},e_{101},e_{107},e_{108},e_{109},e_{110},e_{111},e_{112},e_{113},e_{114},e_{115}$};
вершина $v_{16}$: {$e_6,e_{41},e_{55},e_{72},e_{83},e_{95},e_{107},e_{116},e_{117},e_{118},e_{119}$};
вершина $v_{17}$: {$e_7,e_{56},e_{63},e_{84},e_{89},e_{102},e_{108},e_{116},e_{120},e_{121}$};
вершина $v_{18}$: {$e_{17},e_{34},e_{57},e_{73},e_{103},e_{109},e_{122},e_{123},e_{124},e_{125}$};

вершина $v_{19}$: $\{e_{18},e_{24},e_{42},e_{58},e_{64},e_{85},e_{104},e_{120},e_{126},e_{127},e_{128}\}$;
вершина $v_{20}$: $\{e_{8},e_{19},e_{43},e_{59},e_{65},e_{74},e_{80},e_{90},e_{96},e_{110},e_{122},e_{126},e_{129},e_{130}\}$;
вершина $v_{21}$: $\{e_{9},e_{25},e_{35},e_{50},e_{66},e_{91},e_{97},e_{117},e_{127},e_{131},e_{132},e_{133}\}$;
вершина $v_{22}$: $\{e_{20},e_{44},e_{60},e_{67},e_{86},e_{98},e_{105},e_{111},e_{118},e_{121},e_{123},e_{129},e_{131},e_{134}\}$;
вершина $v_{23}$: $\{e_{10},e_{26},e_{61},e_{68},e_{75},e_{92},e_{112},e_{124},e_{135},e_{136}\}$;
вершина $v_{24}$: $\{e_{11},e_{21},e_{69},e_{76},e_{87},e_{130},e_{135}\}$;
вершина $v_{25}$: $\{e_{12},e_{45},e_{77},e_{113},e_{119},e_{128},e_{132},e_{134},e_{136},e_{137}\}$;
вершина $v_{26}$: $\{e_{27},e_{78},e_{99},e_{114},e_{125},e_{133},e_{138}\}$;
вершина $v_{27}$: $\{e_{13},e_{46},e_{51},e_{70},e_{88},e_{93},e_{100},e_{106},e_{115},e_{137},e_{138}\}$.

Множество треугольных циклов в графе:

$c_1=\{e_1,e_4,e_{15}\} \leftrightarrow \{v_2,v_9,v_1\}$; $\quad$ $c_2=\{e_1,e_8,e_{19}\} \leftrightarrow \{v_2,v_{20},v_1\}$;
$c_3=\{e_1,e_{11},e_{21}\} \leftrightarrow \{v_2,v_{24},v_1\}$; $\quad$ $c_{13}=\{e_4,e_8,e_{74}\} \leftrightarrow \{v_9,v_{20},v_1\}$;
$c_{15}=\{e_4,e_{11},e_{76}\} \leftrightarrow \{v_9,v_{24},v_1\}$; $\quad$ $c_{22}=\{e_8,e_{11},e_{130}\} \leftrightarrow \{v_{20},v_{24},v_1\}$;
$c_{33}=\{e_{15},e_{19},e_{74}\} \leftrightarrow \{v_9,v_{20},v_2\}$; $\quad$ $c_{34}=\{e_{15},e_{21},e_{76}\} \leftrightarrow \{v_9,v_{24},v_2\}$;
$c_{41}=\{e_{19},e_{21},e_{130}\} \leftrightarrow \{v_{20},v_{24},v_2\}$; $\quad$ $c_{87}=\{e_{53},e_{55},e_{83}\} \leftrightarrow \{v_{11},v_{16},v_7\}$;
$c_{88}=\{e_{53},e_{56},e_{84}\} \leftrightarrow \{v_{11},v_{17},v_7\}$; $\quad$ $c_{90}=\{e_{53},e_{60},e_{86}\} \leftrightarrow \{v_{11},v_{22},v_7\}$;
$c_{94}=\{e_{55},e_{56},e_{116}\} \leftrightarrow \{v_{16},v_{17},v_7\}$; $\quad$ $c_{95}=\{e_{55},e_{60},e_{118}\} \leftrightarrow \{v_{16},v_{22},v_7\}$;
$c_{97}\{e_{56},e_{60},e_{121}\} \leftrightarrow \{v_{17},v_{22},v_7\}$; $\quad$ $c_{119}=\{e_{74},e_{76},e_{130}\} \leftrightarrow \{v_{20},v_{24},v_9\}$;
$c_{123}=\{e_{81},e_{82},e_{101}\} \leftrightarrow \{v_{14},v_{15},v_{11}\}$; $\quad$ $c_{124}=\{e_{81},e_{84},e_{102}\} \leftrightarrow \{v_{14},v_{17},v_{11}\}$;
$c_{126}=\{e_{81},e_{86},e_{105}\} \leftrightarrow \{v_{14},v_{22},v_{11}\}$; $\quad$ $c_{128}=\{e_{82},e_{83},e_{107}\} \leftrightarrow \{v_{15},v_{16},v_{11}\}$;
$c_{129}=\{e_{82},e_{84},e_{108}\} \leftrightarrow \{v_{15},v_{17},v_{11}\}$; $\quad$ $c_{130}=\{e_{82},e_{86},e_{111}\} \leftrightarrow \{v_{15},v_{22},v_{11}\}$;
$c_{132}=\{e_{83},e_{84},e_{116}\} \leftrightarrow \{v_{16},v_{17},v_{11}\}$; $\quad$ $c_{133}=\{e_{83},e_{86},e_{118}\} \leftrightarrow \{v_{16},v_{22},v_{11}\}$;
$c_{135}=\{e_{84},e_{86},e_{121}\} \leftrightarrow \{v_{17},v_{22},v_{11}\}$; $\quad$ $c_{147}=\{e_{101},e_{102},e_{108}\} \leftrightarrow \{v_{15},v_{17},v_{14}\}$;
$c_{149}=\{e_{101},e_{105},e_{111}\} \leftrightarrow \{v_{15},v_{22},v_{14}\}$; $\quad$ $c_{152}=\{e_{102},e_{105},e_{121}\} \leftrightarrow \{v_{17},v_{22},v_{14}\}$;
$c_{154}=\{e_{107},e_{108},e_{116}\} \leftrightarrow \{v_{16},v_{17},v_{15}\}$; $\quad$ $c_{155}=\{e_{107},e_{111},e_{118}\} \leftrightarrow \{v_{16},v_{22},v_{15}\}$;
$c_{157}=\{e_{108},e_{111},e_{121}\} \leftrightarrow \{v_{17},v_{22},v_{15}\}$; $\quad$ $c_{167}=\{e_{116},e_{118},e_{121}\} \leftrightarrow \{v_{17},v_{22},v_{16}\}$.

Начальный вектор по рёбрам имеет вид (для облегчения подсчёта местоположения элементов разобьём вектор $P_e$ на двадцатки):

$P_e$ = <3,3,5,7,4,5,2,4,3,3,4,6,3,4,5,3,3,2,8,4,
4,0,1,0,1,0,2,2,2,2,0,3,2,4,2,2,4,3,6,4,
4,4,6,6,5,5,0,0,3,1,2,6,4,3,4,5,4,4,5,7,
2,2,3,6,7,3,6,2,3,0,4,3,5,5,4,4,3,2,2,3,
5,5,4,6,3,5,0,2,0,0,0,0,0,5,6,7,4,7,4,4,
6,5,3,3,5,4,5,4,6,3,8,2,4,4,5,5,4,8,6,4,
6,5,5,3,2,4,2,2,7,4,4,4,2,4,3,3,3,2>;
для MAX = 8, MIN = 1.

Окончательный вид вектора по рёбрам:

$P_e$ = <3,0,0,3,0,0,0,3,0,0,3,0,0,0,3,0,0,0,3,0,
3,0,0,0,0,0,0,0,0,0,0,0,0,0,0,0,0,0,0,0,
0,0,0,0,0,0,0,0,0,0,0,3,0,3,3,0,0,0,0,3,
0,0,0,0,0,0,0,0,0,0,0,0,3,0,3,0,0,0,0,0,
3,4,4,5,0,5,0,0,0,0,0,0,0,0,0,0,0,0,0,0,
3,3,0,0,3,0,3,4,0,0,4,0,0,0,0,4,0,4,0,0,
5,0,0,0,0,0,0,0,3,0,0,0,0,0,0,0,0,0>;

для MAX = 5, MIN = 3.

Далее будем искать максимальные клики. Множество циклов, принадлежащих первой клике:

$c_1=\{e_1,e_4,e_{15}\} \leftrightarrow \{v_2,v_9,v_1\}$; $\qquad c_2=\{e_1,e_8,e_{19}\} \leftrightarrow \{v_2,v_{20},v_1\}$;
$c_3=\{e_1,e_{11},e_{21}\} \leftrightarrow \{v_2,v_{24},v_1\}$; $\qquad c_{13}=\{e_4,e_8,e_{74}\} \leftrightarrow \{v_9,v_{20},v_1\}$;
$c_{15}=\{e_4,e_{11},e_{76}\} \leftrightarrow \{v_9,v_{24},v_1\}$; $\qquad c_{22}=\{e_8,e_{11},e_{130}\} \leftrightarrow \{v_{20},v_{24},v_1\}$;
$c_{33}=\{e_{15},e_{19},e_{74}\} \leftrightarrow \{v_9,v_{20},v_2\}$; $\qquad c_{34}=\{e_{15},e_{21},e_{76}\} \leftrightarrow \{v_9,v_{24},v_2\}$;
$c_{41}=\{e_{19},e_{21},e_{130}\} \leftrightarrow \{v_{20},v_{24},v_2\}$; $\qquad c_{119}=\{e_{74},e_{76},e_{130}\} \leftrightarrow \{v_{20},v_{24},v_9\}$.

Множество вершин первой клики: $K_1=\{v_1,v_2,v_9,v_{20},v_{24}\}$.

Множество циклов, принадлежащих второй клике:

$c_{87}=\{e_{53},e_{55},e_{83}\} \leftrightarrow \{v_{11},v_{16},v_7\}$; $\qquad c_{88}=\{e_{53},e_{56},e_{84}\} \leftrightarrow \{v_{11},v_{17},v_7\}$;
$c_{90}=\{e_{53},e_{60},e_{86}\} \leftrightarrow \{v_{11},v_{22},v_7\}$; $\qquad c_{94}=\{e_{55},e_{56},e_{116}\} \leftrightarrow \{v_{16},v_{17},v_7\}$;
$c_{95}=\{e_{55},e_{60},e_{118}\} \leftrightarrow \{v_{16},v_{22},v_7\}$; $\qquad c_{97}\{e_{56},e_{60},e_{121}\} \leftrightarrow \{v_{17},v_{22},v_7\}$;
$c_{132}=\{e_{83},e_{84},e_{116}\} \leftrightarrow \{v_{16},v_{17},v_{11}\}$; $\qquad c_{133}=\{e_{83},e_{86},e_{118}\} \leftrightarrow \{v_{16},v_{22},v_{11}\}$;
$c_{135}=\{e_{84},e_{86},e_{121}\} \leftrightarrow \{v_{17},v_{22},v_{11}\}$; $\qquad c_{167}=\{e_{116},e_{118},e_{121}\} \leftrightarrow \{v_{17},v_{22},v_{16}\}$.

Множество вершин второй клики: $K_2=\{v_7,v_{11},v_{16},v_{17},v_{22}\}$.

Множество циклов, принадлежащих третьей клике:

$c_{123}=\{e_{81},e_{82},e_{101}\} \leftrightarrow \{v_{14},v_{15},v_{11}\}$; $\qquad c_{124}=\{e_{81},e_{84},e_{102}\} \leftrightarrow \{v_{14},v_{17},v_{11}\}$;
$c_{126}=\{e_{81},e_{86},e_{105}\} \leftrightarrow \{v_{14},v_{22},v_{11}\}$; $\qquad c_{129}=\{e_{82},e_{84},e_{108}\} \leftrightarrow \{v_{15},v_{17},v_{11}\}$;
$c_{130}=\{e_{82},e_{86},e_{111}\} \leftrightarrow \{v_{15},v_{22},v_{11}\}$; $\qquad c_{135}=\{e_{84},e_{86},e_{121}\} \leftrightarrow \{v_{17},v_{22},v_{11}\}$;
$c_{147}=\{e_{101},e_{102},e_{108}\} \leftrightarrow \{v_{15},v_{17},v_{14}\}$; $\qquad c_{149}=\{e_{101},e_{105},e_{111}\} \leftrightarrow \{v_{15},v_{22},v_{14}\}$;
$c_{152}=\{e_{102},e_{105},e_{121}\} \leftrightarrow \{v_{17},v_{22},v_{14}\}$; $\qquad c_{157}=\{e_{108},e_{111},e_{121}\} \leftrightarrow \{v_{17},v_{22},v_{15}\}$.

Множество вершин третьей клики: $K_3=\{v_{11},v_{14},v_{15},v_{17},v_{22}\}$.

Множество циклов, принадлежащих четвертой клике:

$c_{128}=\{e_{82},e_{83},e_{107}\} \leftrightarrow \{v_{15},v_{16},v_{11}\}$; $\qquad c_{129}=\{e_{82},e_{84},e_{108}\} \leftrightarrow \{v_{15},v_{17},v_{11}\}$;
$c_{130}=\{e_{82},e_{86},e_{111}\} \leftrightarrow \{v_{15},v_{22},v_{11}\}$; $\qquad c_{132}=\{e_{83},e_{84},e_{116}\} \leftrightarrow \{v_{16},v_{17},v_{11}\}$;
$c_{133}=\{e_{83},e_{86},e_{118}\} \leftrightarrow \{v_{16},v_{22},v_{11}\}$; $\qquad c_{135}=\{e_{84},e_{86},e_{121}\} \leftrightarrow \{v_{17},v_{22},v_{11}$
$c_{154}=\{e_{107},e_{108},e_{116}\} \leftrightarrow \{v_{16},v_{17},v_{15}\}$; $\qquad c_{155}=\{e_{107},e_{111},e_{118}\} \leftrightarrow \{v_{16},v_{22},v_{15}\}$;
$c_{157}=\{e_{108},e_{111},e_{121}\} \leftrightarrow \{v_{17},v_{22},v_{15}\}$; $\qquad c_{167}=\{e_{116},e_{118},e_{121}\} \leftrightarrow \{v_{17},v_{22},v_{16}\}$.

Множество вершин четвертой клики: $K_4=\{v_{11},v_{15},v_{16},v_{17},v_{22}\}$.

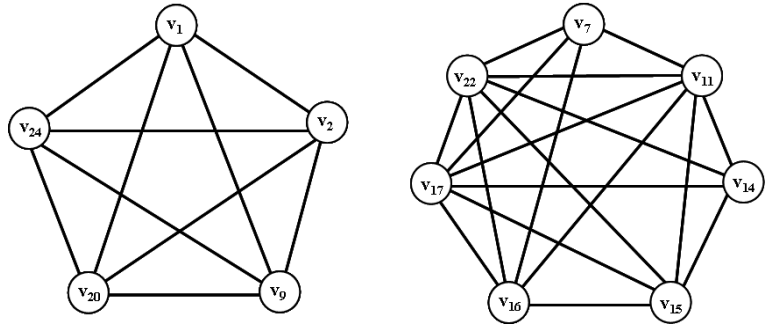

Рис. 3.4. Подграфы графа $G_4$.

В этом случае возникает неоднозначность решения, требующая перебора различных вариантов, так как циклы $c_{129}, c_{130}, c_{132}, c_{133}, c_{157}, c_{167}$ участвуют в формировании двух клик, а цикл $c_{135}$ – даже в трех.

### 3.3. Выделение максимальной клики в графе Турана

Если рассматривать графы Муна-Мозера, то имеется некоторая особенность в построении начального вектора циклов по ребрам. Здесь значения MAX и MIN совпадают уже на первом начальном шаге.

Начальное значение вектора по ребрам графа Муна-Мозера на 9 вершин – MAX=3, MIN=3, на 12 вершин – MAX=6, MIN=6, на 15 вершин – MAX=9, MIN=9, на 18 вершин – MAX=12, MIN=12, и так далее.

В графе Турана величины в начальном векторе циклов по ребрам $P_e$ могут отличаться. Рассмотрим граф Турана (рис. 3.5).

Количество вершин графа = 13, количество ребер графа = 63, количество изометрических циклов графа = 216, количество циклов графа длиной три = 135.

Смежность графа:

вершина  1:  $<v_4,v_5,v_6,v_7,v_8,v_9,v_{10},v_{11},v_{12},v_{13}>$
вершина  2:  $<v_4,v_5,v_6,v_7,v_8,v_9,v_{10},v_{11},v_{12},v_{13}>$
вершина  3:  $<v_4,v_5,v_6,v_7,v_8,v_9,v_{10},v_{11},v_{12},v_{13}>$
вершина  4:  $<v_1,v_2,v_3,v_7,v_8,v_9,v_{10},v_{11},v_{12},v_{13}>$
вершина  5:  $<v_1,v_2,v_3,v_7,v_8,v_9,v_{10},v_{11},v_{12},v_{13}>$
вершина  6:  $<v_1,v_2,v_3,v_7,v_8,v_9,v_{10},v_{11},v_{12},v_{13}>$
вершина  7:  $<v_1,v_2,v_3,v_4,v_5,v_6,v_{10},v_{11},v_{12},v_{13}>$
вершина  8:  $<v_1,v_2,v_3,v_4,v_5,v_6,v_{10},v_{11},v_{12},v_{13}>$
вершина  9:  $<v_1,v_2,v_3,v_4,v_5,v_6,v_{10},v_{11},v_{12},v_{13}>$
вершина  10: $<v_1,v_2,v_3,v_4,v_5,v_6,v_7,v_8,v_9>$
вершина  11: $<v_1,v_2,v_3,v_4,v_5,v_6,v_7,v_8,v_9>$
вершина  12: $<v_1,v_2,v_3,v_4,v_5,v_6,v_7,v_8,v_9>$
вершина  13: $<v_1,v_2,v_3,v_4,v_5,v_6,v_7,v_8,v_9>$

Инцидентность графа:

вершина  1:  $<e_1,e_2,e_3,e_4,e_5,e_6,e_7,e_8,e_9,e_{10}>$
вершина  2:  $<e_{11},e_{12},e_{13},e_{14},e_{15},e_{16},e_{17},e_{18},e_{19},e_{20}>$
вершина  3:  $<e_{21},e_{22},e_{23},e_{24},e_{25},e_{26},e_{27},e_{28},e_{29},e_{30}>$
вершина  4:  $<e_1,e_{11},e_{21},e_{31},e_{32},e_{33},e_{34},e_{35},e_{36},e_{37}>$
вершина  5:  $<e_2,e_{12},e_{22},e_{38},e_{39},e_{40},e_{41},e_{42},e_{43},e_{44}>$
вершина  6:  $<e_3,e_{13},e_{23},e_{45},e_{46},e_{47},e_{48},e_{49},e_{50},e_{51}>$
вершина  7:  $<e_4,e_{14},e_{24},e_{31},e_{38},e_{45},e_{52},e_{53},e_{54},e_{55}>$
вершина  8:  $<e_5,e_{15},e_{25},e_{32},e_{39},e_{46},e_{56},e_{57},e_{58},e_{59}>$
вершина  9:  $<e_6,e_{16},e_{26},e_{33},e_{40},e_{47},e_{60},e_{61},e_{62},e_{63}>$
вершина  10: $<e_7,e_{17},e_{27},e_{34},e_{41},e_{48},e_{52},e_{56},e_{60}>$

вершина 11: <$e_8, e_{18}, e_{28}, e_{35}, e_{42}, e_{49}, e_{53}, e_{57}, e_{61}$>
вершина 12: <$e_9, e_{19}, e_{29}, e_{36}, e_{43}, e_{50}, e_{54}, e_{58}, e_{62}$>
вершина 13: <$e_{10}, e_{20}, e_{30}, e_{37}, e_{44}, e_{51}, e_{55}, e_{59}, e_{63}$>

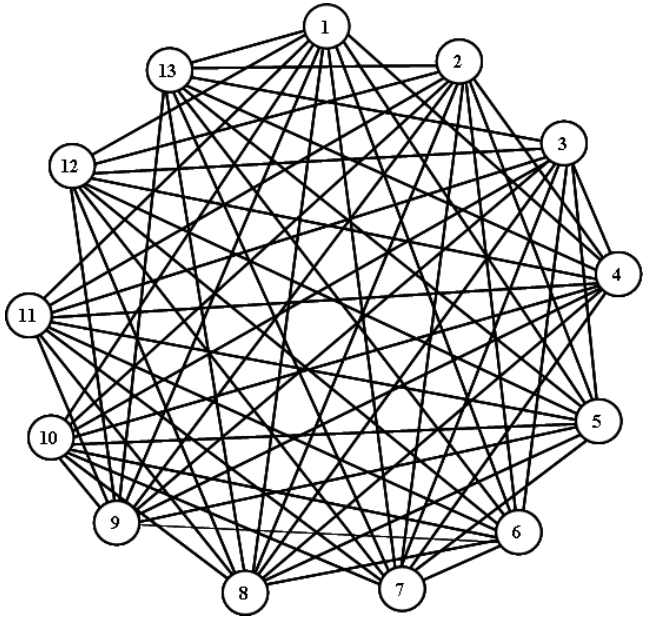

Рис. 3.5. Граф Турана.

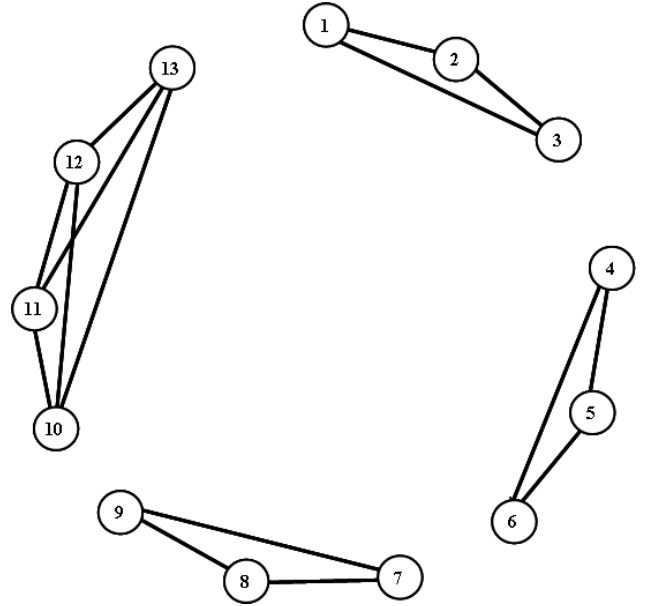

Рис. 3.6. Дополнительный граф Турана.

Множество циклов длиной три:

$c_1 = \{e_1, e_4, e_{31}\} \leftrightarrow \{v_4, v_7, v_1\}$;  $c_2 = \{e_1, e_5, e_{32}\} \leftrightarrow \{v_4, v_8, v_1\}$;
$c_3 = \{e_1, e_6, e_{33}\} \leftrightarrow \{v_4, v_9, v_1\}$;  $c_4 = \{e_1, e_7, e_{34}\} \leftrightarrow \{v_4, v_{10}, v_1\}$;
$c_5 = \{e_1, e_8, e_{35}\} \leftrightarrow \{v_4, v_{11}, v_1\}$;  $c_6 = \{e_1, e_9, e_{36}\} \leftrightarrow \{v_4, v_{12}, v_1\}$;
$c_7 = \{e_1, e_{10}, e_{37}\} \leftrightarrow \{v_4, v_{13}, v_1\}$;  $c_8 = \{e_2, e_4, e_{38}\} \leftrightarrow \{v_5, v_7, v_1\}$;
$c_9 = \{e_2, e_5, e_{39}\} \leftrightarrow \{v_5, v_8, v_1\}$;  $c_{10} = \{e_2, e_6, e_{40}\} \leftrightarrow \{v_5, v_9, v_1\}$;
$c_{11} = \{e_2, e_7, e_{41}\} \leftrightarrow \{v_5, v_{10}, v_1\}$;  $c_{12} = \{e_2, e_8, e_{42}\} \leftrightarrow \{v_5, v_{11}, v_1\}$;
$c_{13} = \{e_2, e_9, e_{43}\} \leftrightarrow \{v_5, v_{12}, v_1\}$;  $c_{14} = \{e_2, e_{10}, e_{44}\} \leftrightarrow \{v_5, v_{13}, v_1\}$;
$c_{15} = \{e_3, e_4, e_{45}\} \leftrightarrow \{v_6, v_7, v_1\}$;  $c_{16} = \{e_3, e_5, e_{46}\} \leftrightarrow \{v_6, v_8, v_1\}$;
$c_{17} = \{e_3, e_6, e_{47}\} \leftrightarrow \{v_6, v_9, v_1\}$;  $c_{18} = \{e_3, e_7, e_{48}\} \leftrightarrow \{v_6, v_{10}, v_1\}$;
$c_{19} = \{e_3, e_8, e_{49}\} \leftrightarrow \{v_6, v_{11}, v_1\}$;  $c_{20} = \{e_3, e_9, e_{50}\} \leftrightarrow \{v_6, v_{12}, v_1\}$;
$c_{21} = \{e_3, e_{10}, e_{51}\} \leftrightarrow \{v_6, v_{13}, v_1\}$;  $c_{22} = \{e_4, e_7, e_{52}\} \leftrightarrow \{v_7, v_{10}, v_1\}$;
$c_{23} = \{e_4, e_8, e_{53}\} \leftrightarrow \{v_7, v_{11}, v_1\}$;  $c_{24} = \{e_4, e_9, e_{54}\} \leftrightarrow \{v_7, v_{12}, v_1\}$;
$c_{25} = \{e_4, e_{10}, e_{55}\} \leftrightarrow \{v_7, v_{13}, v_1\}$;  $c_{26} = \{e_5, e_7, e_{56}\} \leftrightarrow \{v_8, v_{10}, v_1\}$;
$c_{27} = \{e_5, e_8, e_{57}\} \leftrightarrow \{v_8, v_{11}, v_1\}$;  $c_{28} = \{e_5, e_9, e_{58}\} \leftrightarrow \{v_8, v_{12}, v_1\}$;
$c_{29} = \{e_5, e_{10}, e_{59}\} \leftrightarrow \{v_8, v_{13}, v_1\}$;  $c_{30} = \{e_6, e_7, e_{60}\} \leftrightarrow \{v_9, v_{10}, v_1\}$;
$c_{31} = \{e_6, e_8, e_{61}\} \leftrightarrow \{v_9, v_{11}, v_1\}$;  $c_{32} = \{e_6, e_9, e_{62}\} \leftrightarrow \{v_9, v_{12}, v_1\}$;
$c_{33} = \{e_6, e_{10}, e_{63}\} \leftrightarrow \{v_9, v_{13}, v_1\}$;  $c_{34} = \{e_{11}, e_{14}, e_{31}\} \leftrightarrow \{v_4, v_7, v_2\}$;
$c_{35} = \{e_{11}, e_{15}, e_{32}\} \leftrightarrow \{v_4, v_8, v_2\}$;  $c_{36} = \{e_{11}, e_{16}, e_{33}\} \leftrightarrow \{v_4, v_9, v_2\}$;
$c_{37} = \{e_{11}, e_{17}, e_{34}\} \leftrightarrow \{v_4, v_{10}, v_2\}$;  $c_{38} = \{e_{11}, e_{18}, e_{35}\} \leftrightarrow \{v_4, v_{11}, v_2\}$;
$c_{39} = \{e_{11}, e_{19}, e_{36}\} \leftrightarrow \{v_4, v_{12}, v_2\}$;  $c_{40} = \{e_{11}, e_{20}, e_{37}\} \leftrightarrow \{v_4, v_{13}, v_2\}$;
$c_{41} = \{e_{12}, e_{14}, e_{38}\} \leftrightarrow \{v_5, v_7, v_2\}$;  $c_{42} = \{e_{12}, e_{15}, e_{39}\} \leftrightarrow \{v_5, v_8, v_2\}$;
$c_{43} = \{e_{12}, e_{16}, e_{40}\} \leftrightarrow \{v_5, v_9, v_2\}$;  $c_{44} = \{e_{12}, e_{17}, e_{41}\} \leftrightarrow \{v_5, v_{10}, v_2\}$;

$c_{45} = \{e_{12},e_{18},e_{42}\} \leftrightarrow \{v_5,v_{11},v_2\};$    $c_{46} = \{e_{12},e_{19},e_{43}\} \leftrightarrow \{v_5,v_{12},v_2\};$

$c_{47} = \{e_{12},e_{20},e_{44}\} \leftrightarrow \{v_5,v_{13},v_2\};$    $c_{48} = \{e_{13},e_{14},e_{45}\} \leftrightarrow \{v_6,v_7,v_2\};$

$c_{49} = \{e_{13},e_{15},e_{46}\} \leftrightarrow \{v_6,v_8,v_2\};$    $c_{50} = \{e_{13},e_{16},e_{47}\} \leftrightarrow \{v_6,v_9,v_2\};$

$c_{51} = \{e_{13},e_{17},e_{48}\} \leftrightarrow \{v_6,v_{10},v_2\};$    $c_{52} = \{e_{13},e_{18},e_{49}\} \leftrightarrow \{v_6,v_{11},v_2\};$

$c_{53} = \{e_{13},e_{19},e_{50}\} \leftrightarrow \{v_6,v_{12},v_2\};$    $c_{54} = \{e_{13},e_{20},e_{51}\} \leftrightarrow \{v_6,v_{13},v_2\};$

$c_{55} = \{e_{14},e_{17},e_{52}\} \leftrightarrow \{v_7,v_{10},v_2\};$    $c_{56} = \{e_{14},e_{18},e_{53}\} \leftrightarrow \{v_7,v_{11},v_2\};$

$c_{57} = \{e_{14},e_{19},e_{54}\} \leftrightarrow \{v_7,v_{12},v_2\};$    $c_{58} = \{e_{14},e_{20},e_{55}\} \leftrightarrow \{v_7,v_{13},v_2\};$

$c_{59} = \{e_{15},e_{17},e_{56}\} \leftrightarrow \{v_8,v_{10},v_2\};$    $c_{60} = \{e_{15},e_{18},e_{57}\} \leftrightarrow \{v_8,v_{11},v_2\};$

$c_{61} = \{e_{15},e_{19},e_{58}\} \leftrightarrow \{v_8,v_{12},v_2\};$    $c_{62} = \{e_{15},e_{20},e_{59}\} \leftrightarrow \{v_8,v_{13},v_2\};$

$c_{63} = \{e_{16},e_{17},e_{60}\} \leftrightarrow \{v_9,v_{10},v_2\};$    $c_{64} = \{e_{16},e_{18},e_{61}\} \leftrightarrow \{v_9,v_{11},v_2\};$

$c_{65} = \{e_{16},e_{19},e_{62}\} \leftrightarrow \{v_9,v_{12},v_2\};$    $c_{66} = \{e_{16},e_{20},e_{63}\} \leftrightarrow \{v_9,v_{13},v_2\};$

$c_{67} = \{e_{21},e_{24},e_{31}\} \leftrightarrow \{v_4,v_7,v_3\};$    $c_{68} = \{e_{21},e_{25},e_{32}\} \leftrightarrow \{v_4,v_8,v_3\};$

$c_{69} = \{e_{21},e_{26},e_{33}\} \leftrightarrow \{v_4,v_9,v_3\};$    $c_{70} = \{e_{21},e_{27},e_{34}\} \leftrightarrow \{v_4,v_{10},v_3\};$

$c_{71} = \{e_{21},e_{28},e_{35}\} \leftrightarrow \{v_4,v_{11},v_3\};$    $c_{72} = \{e_{21},e_{29},e_{36}\} \leftrightarrow \{v_4,v_{12},v_3\};$

$c_{73} = \{e_{21},e_{30},e_{37}\} \leftrightarrow \{v_4,v_{13},v_3\};$    $c_{74} = \{e_{22},e_{24},e_{38}\} \leftrightarrow \{v_5,v_7,v_3\};$

$c_{75} = \{e_{22},e_{25},e_{39}\} \leftrightarrow \{v_5,v_8,v_3\};$    $c_{76} = \{e_{22},e_{26},e_{40}\} \leftrightarrow \{v_5,v_9,v_3\};$

$c_{77} = \{e_{22},e_{27},e_{41}\} \leftrightarrow \{v_5,v_{10},v_3\};$    $c_{78} = \{e_{22},e_{28},e_{42}\} \leftrightarrow \{v_5,v_{11},v_3\};$

$c_{79} = \{e_{22},e_{29},e_{43}\} \leftrightarrow \{v_5,v_{12},v_3\};$    $c_{80} = \{e_{22},e_{30},e_{44}\} \leftrightarrow \{v_5,v_{13},v_3\};$

$c_{81} = \{e_{23},e_{24},e_{45}\} \leftrightarrow \{v_6,v_7,v_3\};$    $c_{82} = \{e_{23},e_{25},e_{46}\} \leftrightarrow \{v_6,v_8,v_3\};$

$c_{83} = \{e_{23},e_{26},e_{47}\} \leftrightarrow \{v_6,v_9,v_3\};$    $c_{84} = \{e_{23},e_{27},e_{48}\} \leftrightarrow \{v_6,v_{10},v_3\};$

$c_{85} = \{e_{23},e_{28},e_{49}\} \leftrightarrow \{v_6,v_{11},v_3\};$    $c_{86} = \{e_{23},e_{29},e_{50}\} \leftrightarrow \{v_6,v_{12},v_3\};$

$c_{87} = \{e_{23},e_{30},e_{51}\} \leftrightarrow \{v_6,v_{13},v_3\};$    $c_{88} = \{e_{24},e_{27},e_{52}\} \leftrightarrow \{v_7,v_{10},v_3\};$

$c_{89} = \{e_{24},e_{28},e_{53}\} \leftrightarrow \{v_7,v_{11},v_3\};$    $c_{90} = \{e_{24},e_{29},e_{54}\} \leftrightarrow \{v_7,v_{12},v_3\};$

$c_{91} = \{e_{24},e_{30},e_{55}\} \leftrightarrow \{v_7,v_{13},v_3\};$    $c_{92} = \{e_{25},e_{27},e_{56}\} \leftrightarrow \{v_8,v_{10},v_3\};$

$c_{93} = \{e_{25},e_{28},e_{57}\} \leftrightarrow \{v_8,v_{11},v_3\};$    $c_{94} = \{e_{25},e_{29},e_{58}\} \leftrightarrow \{v_8,v_{12},v_3\};$

$c_{95} = \{e_{25},e_{30},e_{59}\} \leftrightarrow \{v_8,v_{13},v_3\};$    $c_{96} = \{e_{26},e_{27},e_{60}\} \leftrightarrow \{v_9,v_{10},v_3\};$

$c_{97} = \{e_{26},e_{28},e_{61}\} \leftrightarrow \{v_9,v_{11},v_3\};$    $c_{98} = \{e_{26},e_{29},e_{62}\} \leftrightarrow \{v_9,v_{12},v_3\};$

$c_{99} = \{e_{26},e_{30},e_{63}\} \leftrightarrow \{v_9,v_{13},v_3\};$    $c_{100} = \{e_{31},e_{34},e_{52}\} \leftrightarrow \{v_7,v_{10},v_4\};$

$c_{101} = \{e_{31},e_{35},e_{53}\} \leftrightarrow \{v_7,v_{11},v_4\};$    $c_{102} = \{e_{31},e_{36},e_{54}\} \leftrightarrow \{v_7,v_{12},v_4\};$

$c_{103} = \{e_{31},e_{37},e_{55}\} \leftrightarrow \{v_7,v_{13},v_4\};$    $c_{104} = \{e_{32},e_{34},e_{56}\} \leftrightarrow \{v_8,v_{10},v_4\};$

$c_{105} = \{e_{32},e_{35},e_{57}\} \leftrightarrow \{v_8,v_{11},v_4\};$    $c_{106} = \{e_{32},e_{36},e_{58}\} \leftrightarrow \{v_8,v_{12},v_4\};$

$c_{107} = \{e_{32},e_{37},e_{59}\} \leftrightarrow \{v_8,v_{13},v_4\};$    $c_{108} = \{e_{33},e_{34},e_{60}\} \leftrightarrow \{v_9,v_{10},v_4\};$

$c_{109} = \{e_{33},e_{35},e_{61}\} \leftrightarrow \{v_9,v_{11},v_4\};$    $c_{110} = \{e_{33},e_{36},e_{62}\} \leftrightarrow \{v_9,v_{12},v_4\};$

$c_{111} = \{e_{33},e_{37},e_{63}\} \leftrightarrow \{v_9,v_{13},v_4\};$    $c_{112} = \{e_{38},e_{41},e_{52}\} \leftrightarrow \{v_7,v_{10},v_5\};$

$c_{113} = \{e_{38},e_{42},e_{53}\} \leftrightarrow \{v_7,v_{11},v_5\};$    $c_{114} = \{e_{38},e_{43},e_{54}\} \leftrightarrow \{v_7,v_{12},v_5\};$

$c_{115} = \{e_{38},e_{44},e_{55}\} \leftrightarrow \{v_7,v_{13},v_5\};$    $c_{116} = \{e_{39},e_{41},e_{56}\} \leftrightarrow \{v_8,v_{10},v_5\};$

$c_{117} = \{e_{39},e_{42},e_{57}\} \leftrightarrow \{v_8,v_{11},v_5\};$    $c_{118} = \{e_{39},e_{43},e_{58}\} \leftrightarrow \{v_8,v_{12},v_5\};$

$c_{119} = \{e_{39},e_{44},e_{59}\} \leftrightarrow \{v_8,v_{13},v_5\};$    $c_{120} = \{e_{40},e_{41},e_{60}\} \leftrightarrow \{v_9,v_{10},v_5\};$

$c_{121} = \{e_{40},e_{42},e_{61}\} \leftrightarrow \{v_9,v_{11},v_5\};$    $c_{122} = \{e_{40},e_{43},e_{62}\} \leftrightarrow \{v_9,v_{12},v_5\};$

$c_{123} = \{e_{40},e_{44},e_{63}\} \leftrightarrow \{v_9,v_{13},v_5\};$    $c_{124} = \{e_{45},e_{48},e_{52}\} \leftrightarrow \{v_7,v_{10},v_6\};$

$c_{125} = \{e_{45},e_{49},e_{53}\} \leftrightarrow \{v_7,v_{11},v_6\};$    $c_{126} = \{e_{45},e_{50},e_{54}\} \leftrightarrow \{v_7,v_{12},v_6\};$

$c_{127} = \{e_{45},e_{51},e_{55}\} \leftrightarrow \{v_7,v_{13},v_6\};$    $c_{128} = \{e_{46},e_{48},e_{56}\} \leftrightarrow \{v_8,v_{10},v_6\};$

$c_{129} = \{e_{46},e_{49},e_{57}\} \leftrightarrow \{v_8,v_{11},v_6\};$    $c_{130} = \{e_{46},e_{50},e_{58}\} \leftrightarrow \{v_8,v_{12},v_6\};$

$c_{131} = \{e_{46},e_{51},e_{59}\} \leftrightarrow \{v_8,v_{13},v_6\};$    $c_{132} = \{e_{47},e_{48},e_{60}\} \leftrightarrow \{v_9,v_{10},v_6\};$

$c_{133} = \{e_{47},e_{49},e_{61}\} \leftrightarrow \{v_9,v_{11},v_6\};$    $c_{134} = \{e_{47},e_{50},e_{62}\} \leftrightarrow \{v_9,v_{12},v_6\};$

$c_{135} = \{e_{47},e_{51},e_{63}\} \leftrightarrow \{v_9,v_{13},v_6\}.$

**Итерация 0**

Max = 7, Min = 6

$P_e = \langle 7\,7\,7\,7\,7\,7\,7\,6\,6\,6\,7\,7\,7\,7\,7\,7\,6\,6\,6$
$\qquad 7\,7\,7\,7\,7\,7\,6\,6\,6\,7\,7\,7\,6\,6\,6\,7\,7\,7$
$\qquad 6\,6\,6\,6\,7\,7\,7\,6\,6\,6\,6\,6\,6\,6\,6\,6\,6\,6$
$\qquad 6\,6\,6\rangle$

Выбираем ребро $e_1$; строим на базе выбранных циклов и вершин подграф (рис. 3.7).

$c_1 = \{e_1,e_4,e_{31}\} \leftrightarrow \{v_4,v_7,v_1\};$  $\qquad c_2 = \{e_1,e_5,e_{32}\} \leftrightarrow \{v_4,v_8,v_1\};$
$c_3 = \{e_1,e_6,e_{33}\} \leftrightarrow \{v_4,v_9,v_1\};$  $\qquad c_4 = \{e_1,e_7,e_{34}\} \leftrightarrow \{v_4,v_{10},v_1\};$
$c_5 = \{e_1,e_8,e_{35}\} \leftrightarrow \{v_4,v_{11},v_1\};$  $\qquad c_6 = \{e_1,e_9,e_{36}\} \leftrightarrow \{v_4,v_{12},v_1\};$
$c_7 = \{e_1,e_{10},e_{37}\} \leftrightarrow \{v_4,v_{13},v_1\}.$

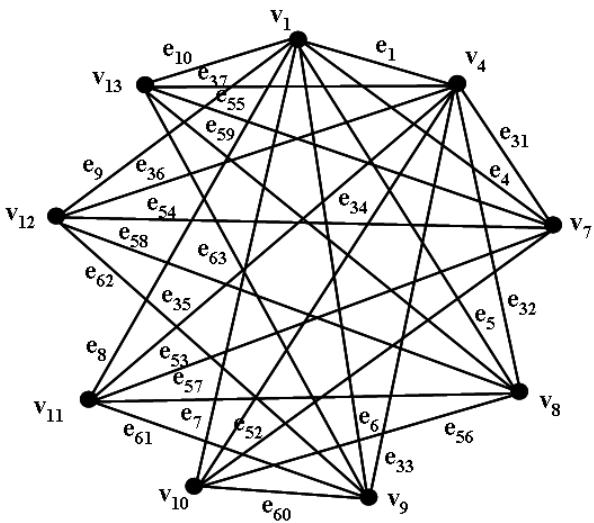   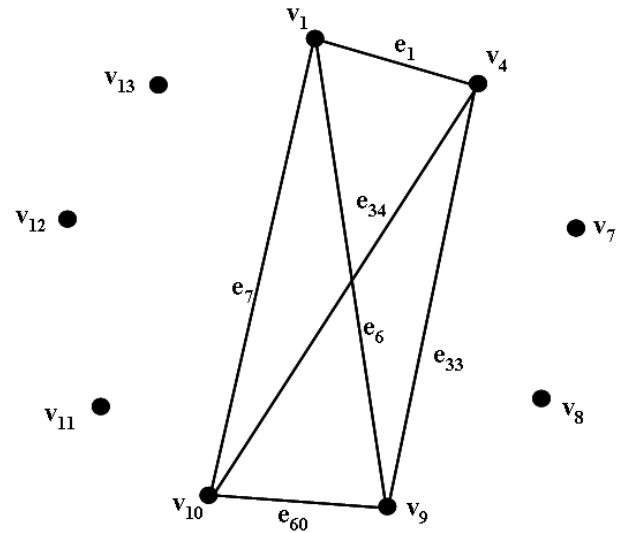

3.7. Подграф относительно ребра $e_1$.  $\qquad$  3.8. Клика относительно рёбер $e_1$ и $e_{60}$.

Множество циклов длиной три для 1-й итерации:

$c_1 = \{e_1,e_4,e_{31}\} \leftrightarrow \{v_4,v_7,v_1\};$  $\qquad c_2 = \{e_1,e_5,e_{32}\} \leftrightarrow \{v_4,v_8,v_1\};$
$c_3 = \{e_1,e_6,e_{33}\} \leftrightarrow \{v_4,v_9,v_1\};$  $\qquad c_4 = \{e_1,e_7,e_{34}\} \leftrightarrow \{v_4,v_{10},v_1\};$
$c_5 = \{e_1,e_8,e_{35}\} \leftrightarrow \{v_4,v_{11},v_1\};$  $\qquad c_6 = \{e_1,e_9,e_{36}\} \leftrightarrow \{v_4,v_{12},v_1\};$
$c_7 = \{e_1,e_{10},e_{37}\} \leftrightarrow \{v_4,v_{13},v_1\};$  $\qquad c_{22} = \{e_4,e_7,e_{52}\} \leftrightarrow \{v_7,v_{10},v_1\};$
$c_{23} = \{e_4,e_8,e_{53}\} \leftrightarrow \{v_7,v_{11},v_1\};$  $\qquad c_{24} = \{e_4,e_9,e_{54}\} \leftrightarrow \{v_7,v_{12},v_1\};$
$c_{25} = \{e_4,e_{10},e_{55}\} \leftrightarrow \{v_7,v_{13},v_1\};$  $\qquad c_{26} = \{e_5,e_7,e_{56}\} \leftrightarrow \{v_8,v_{10},v_1\};$
$c_{27} = \{e_5,e_8,e_{57}\} \leftrightarrow \{v_8,v_{11},v_1\};$  $\qquad c_{28} = \{e_5,e_9,e_{58}\} \leftrightarrow \{v_8,v_{12},v_1\};$
$c_{29} = \{e_5,e_{10},e_{59}\} \leftrightarrow \{v_8,v_{13},v_1\};$  $\qquad c_{30} = \{e_6,e_7,e_{60}\} \leftrightarrow \{v_9,v_{10},v_1\};$
$c_{31} = \{e_6,e_8,e_{61}\} \leftrightarrow \{v_9,v_{11},v_1\};$  $\qquad c_{32} = \{e_6,e_9,e_{62}\} \leftrightarrow \{v_9,v_{12},v_1\};$
$c_{33} = \{e_6,e_{10},e_{63}\} \leftrightarrow \{v_9,v_{13},v_1\};$  $\qquad c_{100} = \{e_{31},e_{34},e_{52}\} \leftrightarrow \{v_7,v_{10},v_4\};$
$c_{101} = \{e_{31},e_{35},e_{53}\} \leftrightarrow \{v_7,v_{11},v_4\};$  $\qquad c_{102} = \{e_{31},e_{36},e_{54}\} \leftrightarrow \{v_7,v_{12},v_4\};$
$c_{103} = \{e_{31},e_{37},e_{55}\} \leftrightarrow \{v_7,v_{13},v_4\};$  $\qquad c_{104} = \{e_{32},e_{34},e_{56}\} \leftrightarrow \{v_8,v_{10},v_4\};$
$c_{105} = \{e_{32},e_{35},e_{57}\} \leftrightarrow \{v_8,v_{11},v_4\};$  $\qquad c_{106} = \{e_{32},e_{36},e_{58}\} \leftrightarrow \{v_8,v_{12},v_4\};$
$c_{107} = \{e_{32},e_{37},e_{59}\} \leftrightarrow \{v_8,v_{13},v_4\};$  $\qquad c_{108} = \{e_{33},e_{34},e_{60}\} \leftrightarrow \{v_9,v_{10},v_4\};$
$c_{109} = \{e_{33},e_{35},e_{61}\} \leftrightarrow \{v_9,v_{11},v_4\};$  $\qquad c_{110} = \{e_{33},e_{36},e_{62}\} \leftrightarrow \{v_9,v_{12},v_4\};$

$c_{111} = \{e_{33}, e_{37}, e_{63}\} \leftrightarrow \{v_9, v_{13}, v_4\}$.

**Итерация 1**

Max = 7, Min = 2

$P_e$ = <7 0 0 5 5 5 5 5 5 5 0 0 0 0 0 0 0 0 0 0

0 0 0 0 0 0 0 0 0 0 5 5 5 5 5 5 5 0 0 0

0 0 0 0 0 0 0 0 0 0 0 2 2 2 2 2 2 2 2 2

2 2 2>

Выбираем ребро $e_{60}$ имеющее минимальный вес. Строим подграф на вершинах выбранных циклов, проходящих по рёбрам $e_1$ и $e_{60}$ (рис. 3.8):

$c_{30} = \{e_6, e_7, e_{60}\} \leftrightarrow \{v_9, v_{10}, v_1\}$;  $c_{108} = \{e_{33}, e_{34}, e_{60}\} \leftrightarrow \{v_9, v_{10}, v_4\}$;

# Выводы

Задача выделения максимальной клики графа является задачей с полиномиальной вычислительной сложностью. В свою очередь, задача перечисления всех максимальных клик графа является задачей с экспоненциальной вычислительной сложностью.

Основная идея предложенного в данной работе метода заключается в определении вектора количества циклов, проходящих по ребрам $P_e$ основной итерации. Построение подграфа с участием циклов, проходящих по ребру с минимальным весом, индуцирует (порождает) максимальную клику графа.

Основу представленного в работе алгоритма составляет выделение множества треугольных циклов графа. Такое выделение позволяет находить решение, используя как подпространство циклов $C(G)$, так и подпространство разрезов графа $S(G)$.

# Список используемой литературы